\documentclass{aa}
\usepackage{txfonts}
\usepackage{pictex}
\usepackage{latexsym}
\usepackage{comment}
\usepackage{graphics}
\usepackage{amsmath}
\usepackage{afterpage}
\usepackage[]{hyperref}
\usepackage{color}
\usepackage{xcolor}
\usepackage{booktabs}
\usepackage{array}
\usepackage{longtable}
\usepackage[figuresright]{rotating}
\usepackage{caption}
\usepackage{subcaption}
\usepackage{upgreek}
\usepackage{ulem}
\usepackage{wasysym}
\usepackage{enumitem}
\usepackage{units}
\usepackage{ifthen}
\setlist[itemize]{label=\textbullet}
\usepackage{threeparttable}
\usepackage{ragged2e}
\usepackage{float}

\newcommand{\nup}{\nu_{\rm p}}
\newcommand{\tauff}{\tau_{\rm ff}}
\newcommand{\tauFFA}{\tau_{\rm FFA}}
\newcommand{\tausyn}{\tau_{\rm syn}}
\newcommand{\tauSSA}{\tau_{\rm SSA}}

\begin{document} 

   \title{Broadband radio continuum and host galaxy properties of extremely inverted spectrum compact radio galaxies}
\author{
        Mukul Mhaskey \inst{1,2}\fnmsep\thanks{E-mail: mukul@tls-tautenburg.de (MM); aritra.basu@dzastro.de (AB)}, Barenya Dev \inst{3}, Aritra Basu\inst{4,5,1 \star}, Helmut Meusinger \inst{1}, Surajit Paul\inst{6}, \and Matthias Hoeft\inst{1} 
        }
   \institute{Th\"uringer Landessternwarte, Sternwarte 5, 07778 Tautenburg, Germany
         \and
             Homi Bhabha Centre for Science Education (HBCSE-TIFR), V.N.Purav Marg, Mankhurd, Mumbai 400088, India
          \and
             Indian Institute of Technology Indore, Simrol, Indore, Madhya Pradesh 453552, India
          \and
             Deutsches Zentrum f\"ur Astrophysik, Postplatz 1, 02826 G\"orlitz, Germany         
          \and
             Max-Planck-Institut f\"ur Radioastronomie, Auf dem H\"ugel 69, 53121 Bonn, Germany
           \and
             Manipal Centre for Natural Sciences, Centre of Excellence, Manipal Academy of Higher Education, Manipal, Karnataka 576104, India
             }

\abstract{We present broadband radio continuum spectra spanning $0.15$--$8$\,GHz for a sample of ten Extremely Inverted Spectrum Extragalactic Radio Sources (EISERS), selected from narrowband, low-frequency surveys below 1.4\,GHz. In these surveys, EISERS show unusually steep low-frequency spectral indices, approaching or exceeding the synchrotron self-absorption (SSA) limit of $\alpha \approx +2.5$ ($S_\nu \propto \nu^\alpha$), expected for homogeneous synchrotron sources. We have performed follow-up quasi-simultaneous radio observations of these sources using uGMRT (0.150--1.4 GHz) and VLA (1.0--8.0 GHz). The new observations provide improved frequency coverage and consistent flux calibration across both the optically thick and optically thin spectral regimes. We find that the spectral slopes of EISERS in optically thick regime are flatter than the two-frequency spectral indices obtained using TGSS and WENSS, and lie in the range +1.6 to +3, suggesting that heterogeneous survey data can introduce biases through flux-scale offsets, intrinsic variability of these compact sources, and diverse angular resolutions between surveys. We modelled the broadband radio spectra as synchrotron emission affected by synchrotron self-absorption (SSA) and subsequently modified by free-free absorption (FFA) in an external ionised medium. The estimated physical parameters based on the best-fits indicate that the magnetic fields ($B_{\rm SSA}$) and thermal electron densities ($n_{\rm e}$) are similar to the general class of peak spectrum sources, with $B_{\rm SSA}$ spanning over a large range of values from 20\,mG to 8\,G and $n_{\rm e} \sim 29-128\,{\rm cm}^{-3}$. Optical and infrared analysis further indicates that the EISERS are preferentially associated with dusty starburst galaxies, composite starburst--AGN systems, or obscured AGN embedded in dense circumnuclear environments, supporting a scenario in which at least a fraction of these sources represent young or environmentally confined radio AGN.}

   \keywords{}

  \titlerunning{EISERS2026}
  \authorrunning{M. Mhaskey}

   \maketitle
  
\nolinenumbers
\section{Introduction}

Sensitive large-area radio surveys covering a wide frequency range between 0.15 and 5\,GHz have enabled the assembly of large and diverse samples of extragalactic radio sources to study their radio continuum spectrum in detail.
These samples encompass radio galaxies across different evolutionary stages, ranging from compact ($< 10$\,kpc) to large ($>100$\,kpc) and recurrent systems, thus providing key insights into the life cycle of active galactic nuclei, AGN \citep{Odea_2021, Hardcastle2020}. Particularly, compact radio galaxies ($<10$\,kpc) represent a vast majority of the radio galaxy population \citep[]{Baldi2018}. 

Of particular interest is a class of compact AGN whose radio continuum emission shows a characteristic spectral peak and/or turnover at frequencies typically below $\sim 1.5$\,GHz, referred to as peak spectrum (PS) sources. It is believed that the turnover is due to synchrotron self-absorption (SSA), occurring within the compact emitting region, or by free–free absorption (FFA), caused by foreground ionised gas in the circumnuclear environment and/or in the interstellar medium of the host galaxy \citep{Odea1998}. High-angular resolution observations of PS sources through very long baseline interferometry (VLBI; milliarcsecond, typically 1--10\,mas angular resolution) have shown that they are compact and have sizes $\lesssim 10$\,kpc \citep{PhillipsMutel1982, Fanti1990, Dallacasa2013}. Since the radio morphology and the spectral shape of PS sources are determined by physical properties, such as emission volume, magnetic field strength, energy distribution of relativistic electrons, jet power and the density of the ambient medium, they can also provide insights into the environment of AGN. 

\begin{table*}
\centering
\caption{Source name, positions, VLBI angular sizes, and spectral indices ($\alpha_{0.150\rm ~GHz}^{0.325\rm ~GHz}$) for the EISERS sample.}
\label{table:specindex}
\begin{tabular}{ccccccccc}
\hline
Source &\multicolumn{2}{c}{Coordinates (J2000)} &\multicolumn{2}{c}{uGMRT bands} & \multicolumn{1}{c}{VLA bands}& \multicolumn{2}{c}{VLBI S-band (2.3 GHz)} & $\alpha^{0.325\,{\rm GHz}}_{0.150\,{\rm GHz}}$ \\
 &RA&Dec& Epoch\,1 & Epoch\,2&  & LAS& Compact & This work\\
    & & & & & & &component & \\
 & [hh mm ss] & [dd mm ss]& & &  & [mas] &[mas] & \\
\hline
J0614+6108 & 06 14 50.46 & 61 08 04.8 & $-$& 2,3,4 & L,S,C&  09$^{\star}$ & 05 & $2.01^{+0.20}_{-0.20}$ \\
J0649+5947 & 06 49 09.69 & 59 47 53.6 & $-$&2,3 & L,S,C& 40 & 14 & $1.79^{+0.20}_{-0.20}$ \\
J0740+7129 & 07 40 56.90 & 71 29 53.9 & 3,4,5&2,3 & L,S,C&13  &  08 & $2.46^{+0.28}_{-0.28}$ \\
J0847+5723 & 08 47 27.86 & 57 23 35.5 &3,5 &2,3,4 &  L,S,C&32 & 12 & $1.80^{+0.21}_{-0.21}$ \\
J0853+6722 & 08 53 34.38 & 67 22 15.5 &$-$ &2,3,4 &  L,S,C&15 & 09 & $2.04^{+0.21}_{-0.21}$ \\
J0858+7501 & 08 58 33.43 & 75 01 18.5 & 3,5&2,3,4 &  L,S,C&54 & 15 & $3.11^{+0.20}_{-0.20}$ \\
J1326+5712 & 13 26 50.47 & 57 12 06.8 &2,3,4 &2,3,4,5 &L,S,C&  10 & 07 & $2.47^{+0.20}_{-0.20}$ \\
J1536+8154 & 15 36 59.83 & 81 54 31.3 &3,5 &2,3 &L,S,C&  24 & 10 & $1.67^{+0.21}_{-0.21}$ \\
J1549+5038 & 15 49 17.46 & 50 38 05.3 &2,3 &2,3,4 &  L,S,C&28 & 12 & $2.53^{+0.21}_{-0.21}$ \\
J1658+4732 & 16 58 26.54 & 47 32 14.9 & 3,4,5&2,3,4 & L,S,C& 08$^{\star}$ & 05 & $2.26^{+0.20}_{-0.20}$ \\
\hline
\end{tabular}
\tablefoot{The VLBI S-band (2.3~GHz) angular sizes 
are listed for both the largest angular scale (LAS) and the most compact VLBI component. uGMRT Epochs\,1 and 2 represent the observations that were carried out during Jan--Feb 2019 and Nov 2021--Feb 2022, respectively. VLA observations were carried out in a single epoch during Mar--July 2022.\\
$^\star$ Measured at C-band (4.3 GHz).}
\end{table*}

At least some PS sources appear to be young, only a few hundred years old, as demonstrated, for example, by VLBI measurements of the movements of hotspots \citep{Polatidis1999, Stanghellini2001, Polatidis2003, An2012}, and age estimates based on synchrotron radiation \citep{Murgia2002}. This suggests that the PS sources are likely associated with recent AGN activity, such as the launching of relativistic jets and the injection of energy into the surrounding medium. That is, they are ``young and evolving''. However, in an alternative scenario, jet evolution models infer that the abundance of compact ($< 10$\,kpc) and young ($<1000$\,yr) PS sources in radio surveys are related to their confinement in their host galaxies in which the jets cannot propagate further and that their ages are comparable to those determined for large sources \citep{An2012}. In this alternative scenario, some AGN are not powerful enough to propagate the jet into the intergalactic medium (IGM). That is, they are ``old and frustrated'' \citep[e.g.,][]{Callingham2017}. Thus, when complemented by their large-scale old radio emission, and information on the properties of their host galaxies, PS sources are ideal candidates to study AGN evolution \citep{Odea1998,Odea_2021}.

In addition to the radio spectral properties, the characteristics of the host galaxy play a crucial role in understanding the nature and evolution of compact radio sources. The presence of dense gas and dust in the circumnuclear region or interstellar medium can significantly influence the propagation of radio jets, potentially leading to confinement or strong jet–ISM interactions \citep{Holt2008, Morganti2013, Tadhunter2016}. Such environments are also expected to contribute to external free–free absorption, thereby modifying the observed radio spectra \citep{Bicknell1997,Callingham2017}. Therefore, investigating the host-galaxy properties—such as star formation activity, dust content, and AGN obscuration—is essential for distinguishing between young, evolving sources and those that are frustrated by a dense ambient medium.

In this context, we focus on a rare class of compact radio sources known as Extremely Inverted Spectrum Extragalactic Radio Sources \citep[EISERS;][]{GopalKrishna2014,Mhaskey2019a,Mhaskey2019b}, characterised by exceptionally steep low-frequency (0.150--0.325 GHz) spectral indices that exceed the classical synchrotron self-absorption limit ($\alpha >$  +2.5  for the power law spectrum $S \propto \nu^{\alpha}$). Such extreme spectral behaviour cannot be readily explained within the framework of homogeneous SSA alone, and therefore provides a unique opportunity to investigate the role of source inhomogeneity and external absorption processes. Studying EISERS allows us to place strong constraints on the physical conditions in the emitting and absorbing regions, including the magnetic field strength and the density of the surrounding ionised medium. Moreover, these sources are likely to represent compact radio AGN embedded in dense, gas-rich environments, making them ideal laboratories to probe the interplay between jet evolution and the host galaxy environment.

To this end, we have constructed a sample of 
EISERS to build a well-populated spectrum covering a wide range of frequencies that sample both sides of the peak frequency. 
In Section~\ref{sec:sample} we describe the selection procedure for the ten EISERS. Section~\ref{sec:radio_obs} contains the details of their quasi-simultaneous radio observations and the analysis procedure. Radiation absorption models and the spectral fitting results on the EISERS candidates are discussed in Section~\ref{sec:modelling}. Host galaxy properties are discussed in Section~\ref{sec:hosts}. 
We discuss and conclude our results in Sections~\ref{sec:discussion} and \ref{sec:summary}, respectively.

In this work, we use $\Lambda$CDM cosmology with $H_0 = 73 \,{\rm km\,s^{-1}\,Mpc^{-1}}$,
$\Omega_{\rm \Lambda} =0.73$ and $\Omega_{\rm M}=0.27$ \citep{Spergel2007}.

\section{Sample selection of the EISERS} \label{sec:sample}
The frequency of the peak radio continuum emission in PS sources, $\nup$, has been observed to lie over a broad frequency range, $\sim 0.1$ to $\sim10$\,GHz. In this work, we carefully select a sample of PS sources based on their spectral slope $\alpha$  at frequencies below $\nup$, i.e., in the optically thick part of the emission \citep{GopalKrishna2014,Mhaskey2019a}. A reliable measurement of $\alpha$ in the low frequency regime, especially below about 0.5\,GHz was hampered owing to challenging calibrations, limited sensitivity and angular resolution. However, with the advent of sensitive telescopes with broad frequency coverage operating below 0.5\,GHz, such as, the Low Frequency Array (LOFAR), the upgraded Giant Metrewave Radio Telescope (uGMRT), and the Murchison Wide-field Array (MWA), studying the optically thick regime of AGN emission has become feasible. These telescope facilities offer complementary frequency coverages that allow us to seamlessly study the radio spectrum of PS sources from 0.1 to 1\,GHz.

Our parent EISERS sample was selected using three major radio surveys: the TIFR GMRT Sky Survey Alternate Data Release~1 \citep[TGSS-ADR1;][]{Intema2017} at 0.15\,GHz, the Westerbork Northern Sky Survey \citep[WENSS;][]{Rengelink1997} at 0.325\,GHz, and the NRAO VLA Sky Survey \citep[NVSS;][]{Condon1998} at 1.4\,GHz. In this parent sample, morphologically single-component sources from the WENSS catalogue (denoted as `S'-type) with flux densities greater than 150\,mJy at 0.325\,GHz were chosen and cross-matched with the TGSS-ADR1 catalogue using a matching radius of 20\,arcsec. The flux-density limit was chosen such that sources with spectral indices up to $+1.75$ at frequencies below $\sim0.5$\,GHz remain detectable in TGSS, given its typical rms noise of 5--7\,mJy\,beam$^{-1}$. The two-point spectral index ($\alpha_{0.150\,\mathrm{GHz}}^{0.325\,\mathrm{GHz}}$) was then computed, including lower limits for sources undetected in TGSS. Candidate EISERS were selected by requiring: (i) a single-component/unresolved morphology; (ii) exclusion of sources within $10^\circ$ of the Galactic plane or identified as H\,\textsc{ii} regions in the NASA Extragalactic Database (NED);\footnote{\url{https://ned.ipac.caltech.edu/}} (iii) clean detections in both the WENSS and NVSS, verified through visual inspection of the radio images; and (iv) an ultra-steep low-frequency spectrum, defined as $\alpha^{0.325\,\mathrm{GHz}}_{0.150\,\mathrm{GHz}}>+2.75$ for TGSS-detected sources or $\alpha^{0.325\,\mathrm{GHz}}_{0.150\,\mathrm{GHz}}>+2.5$ for TGSS non-detections \citep[see][for details]{GopalKrishna2014,Mhaskey2019b}. The more conservative threshold of $+2.75$ for TGSS detections was adopted to account for possible flux-density variability in compact radio sources \citep{Bell2018,Chhetri2018}. This selection yielded 29 EISERS, all of which were subsequently observed with the uGMRT using its Band\,2, 3, 4 and 5 receivers covering 0.12 to 1.45\,GHz allowing us to characterise their spectra below the turnover frequency.

Of these 29 EISERS, 10 sources with the steepest low-frequency 
spectrum ($\alpha^{0.325\,\mathrm{GHz}}_{0.150\,\mathrm{GHz}}>+1.5$) 
were subsequently observed with the VLA between 1 and 8\,GHz using its L-, S- and C-band receivers.
These ten EISERS therefore have a broad frequency coverage from 0.15 to 8\,GHz enabling consistent modelling of the radio spectra covering both the optically thick and thin regimes.
Table~\ref{table:specindex} lists the source positions, observing epochs, archival VLBI angular sizes, and low-frequency spectral indices for the ten EISERS analysed in this work. There, the spectral indices $\alpha^{0.325\,\mathrm{GHz}}_{0.150\,\mathrm{GHz}}$ obtained using the new data of these sources presented in Section~\ref{sec:radio_obs} are listed. Within errors, the spectral index-based selection criteria applied to TGSS and WENSS data are confirmed by our datasets.
The remaining 19 candidates for which uGMRT observations are available, will be presented in another study.

\begin{figure}
\centering      
\includegraphics[trim={0cm 0.2cm 0cm 0.4cm},clip,width=1\columnwidth]{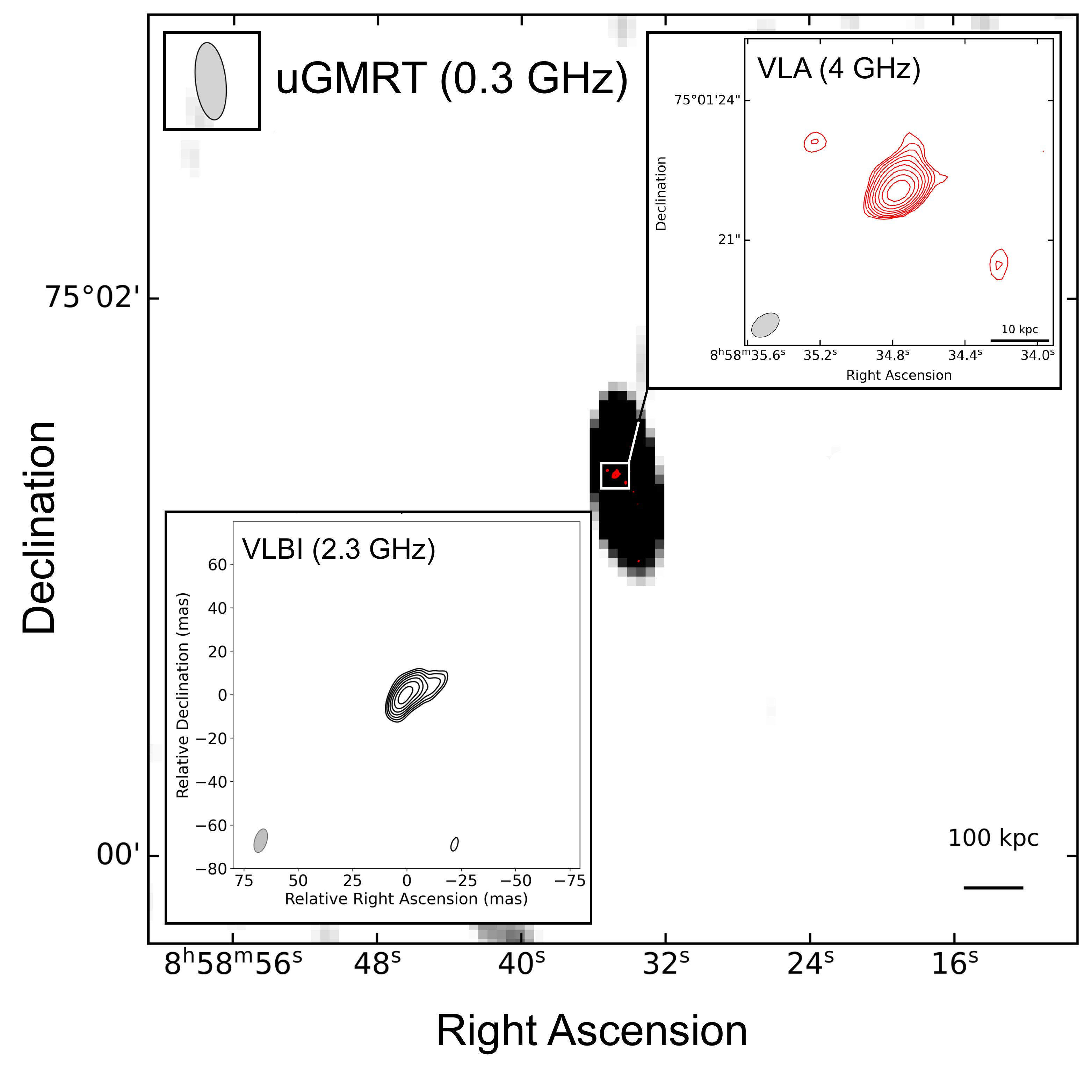}
\caption{Composite radio image of J0858+7501. The VLA C-band (4 GHz; beam = $0.6''\times0.4''$) contours (in red) over-plotted on the uGMRT Band-3 (0.3 GHz; beam = $17''\times6.5''$) total intensity map is in the raster. In the inset we have the 2.3\,GHz map of the same source at an angular resolution of 10.8\,mas, obtained from the VLBI calibrator catalogue \citep{Petrov2011,Petrov2021}. All contours are plotted starting at the respective $5\sigma$ and subsequent levels increase by factor of $\sqrt 2$.}
\label{fig:Radiomap_J0858+7501}  
\end{figure}

Furthermore, archival VLBI observations obtained via the Astrogeo VLBI calibrator list and image database (and the associated Radio Fundamental catalogue, e.g., \citet{Petrov2011, Petrov2021}) were used to estimate the angular sizes of the compact radio-emitting regions required for the physical parameter calculations later in this paper. Depending on the availability of archival data, the angular sizes were obtained from observations with typical resolutions of a few milliarcseconds (approximately 2--10 mas). As the apparent source size depends on synchrotron opacity, we consistently adopt the size measured at the lowest available frequency in the optically thin regime. From the Astrogeo VLBI database, we use source sizes at 2.3\,GHz wherever available. Sources for which 2.3\,GHz is unavailable, we use the sizes at 4.3\,GHz. In Table~\ref{table:specindex} we list the largest angular scale (LAS) and the extent of the brightest compact component for our sample. In Fig.~\ref{fig:Radiomap_J0858+7501}, we present the uGMRT Band-3 image of J0858+7501 at 0.325~GHz, with the VLA C-band (4~GHz) contours overlaid. The inset shows the 2.3~GHz VLBI contour map of the same source at an angular resolution of 10.8~mas. The Radio-on-Radio images of the remaining EISERS are shown in the Fig.\ref{fig:RoR} of Appendix~\ref{appendix:ror}.

\section{Radio observations} \label{sec:radio_obs}
EISERS are compact, and their emission is likely to evolve rapidly on timescales of months to years, making spectral indices derived from non-simultaneous survey data susceptible to variability-induced biases. Hence, it is critical to obtain quasi-simultaneous observations across a wide range of frequencies to robustly constrain the spectrum below $\rm \nu_p$ and probe the underlying absorption mechanism(s). TGSS-ADR1 and WENSS were conducted nearly a decade apart, and therefore, $\alpha^{0.325\,\mathrm{GHz}}_{0.150\,\mathrm{GHz}}$ could be affected by flux density variability intrinsic to EISERS, or refractive interstellar scintillation because of their compact sizes \citep{Bell2018}. Besides flux density variability, systematic calibration uncertainties could further significantly affect the spectral slope towards low frequencies \cite[e.g.,][]{Rengelink1997, Intema2017, Hardcastle2016}. To accurately characterise the low-frequency spectral properties of EISERS, sensitive, quasi-simultaneous observations with high angular resolution ($\lesssim 20$\,arcsec) are required to minimise uncertainties due to variability and source confusion. To this end, our new uGMRT observations presented in this work are suitable to study the low frequency spectrum of EISERS.

\subsection{uGMRT} \label{sec:ugmrt}
The uGMRT observing campaign of the ten EISERS was carried out over two epochs. In the first epoch, between January and February 2019 (Proposal code 35\_010, ddtC077, PI: Mukul Mhaskey), seven of them were observed, while in the second epoch, between November 2021 and February 2022 (Proposal code 39\_049, ddtC168, PI: Mukul Mhaskey), all 10 were observed. Depending on the scheduling constraints during each epoch, sources were observed at different uGMRT frequency bands covered with Band\,2 (0.12 to 0.25\,GHz), Band\,3 (0.25 to 0.5\,GHz), Band\,4 (0.55 to 0.9\,GHz) and Band\,5 (1 to 1.45\,GHz). The bands at which the sources were observed at each epoch are listed in Table~\ref{table:specindex}. That means, a majority of the EISERS were observed between 0.12 and 1.45\,GHz with the uGMRT, and seven of them have multi-epoch data.
These observations were undertaken in both the 32\,MHz narrow-band mode and the $\sim 200$\,MHz wide-band mode. 
In this paper, these wide-band data are used to study the radio continuum spectra of the ten EISERS.

The uGMRT data were processed using the Source Peeling and Atmospheric Modelling (\texttt{SPAM}) package \citep{Intema2014}. The wide-band uGMRT data at bands 2, 3, 4 and 5 were first divided into 16 or 32\,MHz wide sub-bands, determined by the signal-to-noise-ratio of visibility data. Each of these sub-band data were processed through \texttt{SPAM}, and direction-dependent calibration and subsequent imaging were also performed, thereby providing us a contiguous frequency coverage between 0.15 and 1.4\,GHz.

The flux density scale for the uGMRT observations was set using standard primary calibrators, adopting the Perley-Butler flux density scale \citep{Perley2017}. The calibration was performed independently for each observing band (Bands 2, 3, and 4), taking into account the frequency-dependent flux densities of the calibrators. While the absolute flux scale is expected to be consistent across bands within typical calibration uncertainties, residual systematic differences at the level of $\sim$10–15\% may arise due to band-dependent effects such as primary beam corrections and radio frequency interference (RFI). These uncertainties are incorporated into the flux density error budget used in the spectral analysis.

In our previous studies \citep{Mhaskey2019a}, the flux density uncertainties of individual sources were estimated using  
$\sigma_S = [\sigma_{\rm rms}^2 + (0.10\,S_{\rm peak})^2]^{0.5}$, where the 10\% term accounts for systematic uncertainties, including small calibration errors arising from the elevation dependence of antenna gains \citep{Chandra2004}. Based on our experience with observations using the upgraded Giant Metrewave Radio Telescope (uGMRT), a conservative uncertainty of about 15\% of the integrated flux density is typically appropriate for individual sources.

\subsection{VLA} \label{sec:vla}
The sample of ten EISERS was observed with the VLA between March and July 2022 (Project code 22A-233, PI: Mukul Mhaskey) covering 1 to 8\,GHz using L-, S- and C-band receivers.
These are snapshot observations ($\sim$5 min) for the 10 targets earlier observed with the uGMRT in full polarisation mode. Note that, due to scheduling constraints, each target was observed over multiple observing sessions during the period, but each session covered the entire 1--8\,GHz range.

The VLA data were calibrated using the Common Astronomy Software Applications \texttt{(CASA)} \citep{McMullin2007,CASA2022} based standard NRAO VLA calibration pipeline, which adopts the Perley \& Butler (2017) flux density scale \citep{Perley2017}. This scale is consistent with that used for the uGMRT observations, ensuring that the flux densities across different instruments are on a common scale. Residual cross-instrument differences are expected to be within the typical calibration uncertainties of a few percent. 
The VLA wide-band data were subsequently divided into smaller spectral sub-bands, following the same approach adopted for the GMRT data, and each sub-band was imaged independently using the \texttt{WSClean} package \citep{offringa-wsclean-2014,offringa-wsclean-2017}. For each source, the 1--2 GHz, 2--4 GHz, and 4--8 GHz bands were subdivided into five, three, and two sub-bands, respectively, and the resulting sub-bands were imaged separately.

For the sources observed with the VLA, we adopt a conservative flux density uncertainty of 5\%. This accounts for the systematic uncertainties in the absolute flux density calibration, including residual gain variations and differences in elevation and atmospheric conditions between the calibrator and target sources. Under typical observing conditions, the accuracy of the VLA absolute flux density scale is at the level of $\sim$2–4\% \citep{Perley2017}. Since the sources in our sample are unresolved, additional uncertainties due to imaging or flux recovery are minimal. We therefore adopt a conservative flux density uncertainty of 5\% for the VLA measurements. 

\subsection{Flux density verification of the EISERS}
Flux densities were measured using the source-fitting procedure implemented in \texttt{PyBDSF}\footnote{https://github.com/lofar-astron/pybdsf} \citep{mohan2015_pybdsf}. All EISERS in the sample remained unresolved at the angular resolution of these observations throughout the 0.15 to 8\,GHz coverage. Therefore, the flux densities were determined by modelling these sources at each frequency as a single component 2-dimensional Gaussian. We note that it is possible that
the measurement of flux densities of the compact EISERS for observations carried out using multiple telescope receivers, and over multiple epochs could be affected by intrinsic variability of the sources, especially in the optically thick regime at low frequencies, and/or differences in flux density scale. This might hamper a reliable modelling of the radio continuum spectral energy distribution (SED) performed later in the paper.

To assess the level of flux-density variability at low frequencies, we compared the flux densities of the seven EISERS observed in two uGMRT epochs separated by approximately 3 yr. As a control sample, we selected ten of the brightest unresolved sources from the same seven fields. The measured flux densities are listed in Table~\ref{table:variability}, together with the fractional variability, defined as $\Delta\,f_\nu = |\Delta\,S_\nu|/\langle S_\nu \rangle$, and expressed in percentage.
Here, $|\Delta\,S_\nu|$ is the absolute difference of the flux densities measured between the two epochs, and $\langle S_\nu\rangle$ is the mean. For both EISERS and the field sources, we find a typical $\Delta\,f_\nu \approx 15\%$, which is consistent with the calibration uncertainties. This suggests that the flux density variations in EISERS are not significant over a 3\,yr timescale.
A few individual sources at specific frequencies show substantially larger apparent variations between the two epochs, occasionally exceeding 50\% and, in extreme cases, approaching or exceeding 100\%. These large deviations do not follow the overall frequency-dependent trend and are unlikely to represent genuine intrinsic variability of the radio sources. Instead, they are most plausibly attributed to observational effects, including small mismatches in the central observing frequency between epochs combined with the extremely steep spectral slopes of EISERS near the turnover, differences in uv-coverage, angular-scale sensitivity, residual calibration and ionospheric effects inherent to metre-wavelength interferometry \citep[e.g.,][]{Intema2017,Chhetri2018,Bell2018}. We therefore treat such extreme cases as outliers and do not consider them representative of the intrinsic variability behaviour of the population.

\begin{figure}
\centering      
\includegraphics[trim={0cm 0.3cm 0cm 0cm},clip,width=0.85\columnwidth]{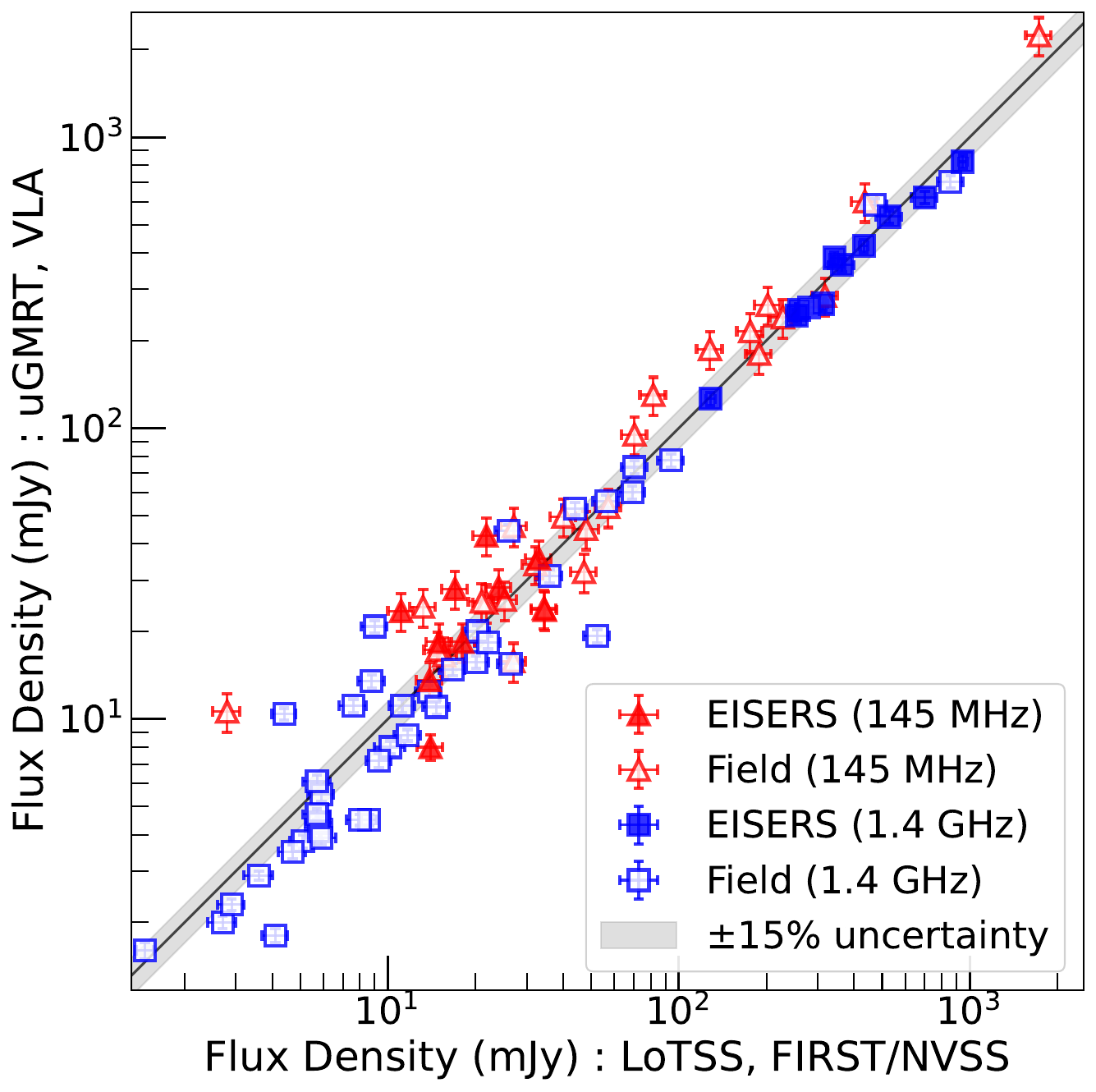}
\caption{Flux density ratio comparison of uGMRT (0.145\,GHz) and VLA (1.4\,GHz) measurements against LoTSS and FIRST/NVSS, respectively; EISERS (filled) and field sources (open) are shown relative to the 1:1 relation, with the shaded band representing the $\pm$15\% flux calibration uncertainty.} 
\label{fig:fluxcorrelation}  
\end{figure}

To evaluate the reliability of the adopted flux scale at both low frequencies and optically thin high frequencies, we compared the flux densities of the ten EISERS and other sources in the field to those from well-established radio surveys. The flux densities are listed in Table.~\ref{tab:data-validation-I}. Specifically, the uGMRT flux densities at $\sim0.15$\,GHz are compared with the LOFAR Two-metre Sky Survey (LoTSS; \citet{Shimwell2026}) at 0.14 GHz, while the 1.4~GHz measurements from the VLA are compared with corresponding values from the FIRST and NVSS (whichever is available for a given source) of observed flux densities to different surveys. Fig.~\ref{fig:fluxcorrelation} shows the comparison between the measured and survey flux densities for both EISERS and the field sources. The solid line indicates the 1:1 relation and the shaded region marks the expected $\pm15\%$ flux calibration uncertainty.

At 1.4~GHz, the VLA/uGMRT flux density measurements from our data are in excellent agreement with those obtained from FIRST/NVSS, with a median flux density ratio of 0.98 for the EISERS and 0.83 for the field sources. All EISERS lie within the $\pm15\%$ uncertainty envelope, indicating that the flux density scale at GHz frequencies is robust.

At 0.145\,GHz, however, the uGMRT flux densities are systematically higher than LoTSS by $\sim$20–25\% on average, with median ratios of $\sim$1.17 for field sources and $\sim1.13$ for EISERS. The comparable behaviour of field sources and EISERS indicates that this is a systematic effect rather than source-specific variability.
The LoTSS data are calibrated on the \citet{ScaifeHeald2012} flux-density scale, while the uGMRT and VLA measurements are tied to the \citet{Perley2017} scale. 
At 1.4 GHz, our VLA measurements are in excellent agreement with the FIRST/NVSS flux densities, confirming the consistency of the adopted GHz flux-density scale. In Table \ref{tab:fluxJ0858} and in Fig.\,\ref{fig:SED-J0858+7501} we show a typical spectrum of the EISERS J0858+7501 obtained with our data. Except for a few outliers at narrow frequencies discussed above, we do not find any drastic discontinuity between different receiver bands. The spectra for the other EISERS are presented in Fig.~\ref{fig:SED-EISERS}.\\

\section{Optical and infrared counterparts} \label{sec:hosts}    

In order to identify the hosts of the EISERS at optical and/or infrared wavebands, we make use of publicly available surveys.
The inspection of the images from the Sloan Digital Sky Survey Data Release 16 \citep[SDSS DR16,][]{Ahumada_2020} reveals that the optical counterparts of the EISERS are generally weak. The only exception is J1549+5038.\footnote{This is the only source from our list for which an SDSS spectrum is available. It shows the emission lines and blue continuum typical of quasars.} The images from the DESI Legacy Imaging Surveys \citep[][hereafter the Legacy Surveys, LS]{Dey_2019}, are 1 to 2 mag deeper in the $g$, $r$, and $z$ bands and cover a larger sky area than SDSS.  As of the Data Release 10 (LS DR10)\footnote{https://www.legacysurvey.org/dr10/description/}, the area is expanded to $> 20\,000$ square degrees  of the extragalactic sky and the publicly available data also include images and magnitudes in the $i$ band. Thanks to the very wide areal coverage, all objects from our sample are located within the LS footprint area. In addition to the LS images, we used the image cutouts and the photometry from the Wide-field Infrared Survey Explorer (WISE) satellite \citep{Wright_2010} in the four passbands W1 (3.4\,$\upmu$m), W2 (4.6\,$\upmu$m), W3 (12\,$\upmu$m), and W4  (22\,$\upmu$m). In the optical and infrared bands, the counterparts of the EISERS are usually faint, but recognisable for about two-thirds of our sources. Remarkably, the detection rate in the near and mid infrared (NIR, MIR)  is higher than in the optical. The co-added images from the LS bands and those from the WISE bands are shown in Figs. \ref{fig:MBSED} \& \ref{fig:EISERS_wo_MBSED}. 
In Section~\ref{sec:optical_SED}, we describe how the photometric data in the optical and infrared regions were used to estimate the photometric redshifts where no spectroscopic redshifts are available.

\section{Modelling the radio continuum spectra of EISERS} \label{sec:modelling}

Here, we model the broadband radio continuum SED of the ten EISERS using our flux density measurements between 0.15 and 8\,GHz. To this end, through radiative transfer, we use a combination of SSA and FFA mechanisms to model the observed spectra. Optically thin emission has $\alpha < 0$, while optically thick emission has $\alpha > 0$.

\subsection{Parametrised SSA and FFA} \label{sec:SSA+FFA}

Under SSA, synchrotron emission from compact radio sources originates from relativistic electrons spiralling in magnetic fields. The optical thickness of synchrotron emission ($\tau_{\rm SSA}$) increases with decreasing frequency so that at sufficiently low frequencies, the emission becomes self-absorbed as the same population of electrons that emits the radiation also re-absorb it. The SSA mechanism leads to a turnover in the spectrum, with the emission transitioning from an optically thin regime at high frequencies to an optically thick regime at low frequencies \citep{Rybicki1979,Longair2011}.

For a homogeneous synchrotron emitting plasma, the optical depth ($\tausyn$)
is determined by the physical properties, such as, the magnetic field strength ($B$), density of the cosmic ray electrons ($n_{\rm CRE}$), their energy distribution given as {$n_{\rm CRE}(E)\,{\rm d}E = n_0\,E^{\gamma}\,{\rm d}E$},
pathlength through the medium ($L$), and the frequency. Here, the energy index ($\gamma$) is related to the spectral index ($\alpha_{\rm thin}$) of the optically thin synchrotron emission as $\gamma = 2\,\alpha_{\rm thin} - 1$, for our convention where $\alpha_{\rm thin} < 0$. $\tausyn$ is related to them as,
\begin{equation}
\tausyn(\nu) \propto
n_0\,
B^{\left(3-2\alpha_{\rm thin}\right)/2}\,
\nu^{\alpha_{\rm thin}-2.5}\,
L.
\label{eq:tau_syn}
\end{equation}
For the purpose of modelling the SSA component of radio continuum SED, we adopt a simple parametrisation of the optical depth ($\tauSSA$) following \citet{Tingay2003} as,
\begin{equation}
\tau_{\rm SSA}(\nu) =
k_1 \left(\frac{\nu}{\nu_0}\right)^{\alpha_{\rm thin} - \alpha_{\rm thick}}.
\label{eq:tau_SSA}
\end{equation}
Here, $\nu_0$ is a fixed pivot frequency (taken to be 1 GHz), $k_1$ is a dimensionless normalisation constant corresponding to the SSA optical depth at $\nu_0$, and $\alpha_{\rm thick}$ is the optically thick spectral index. Note that, as per Eq.~\ref{eq:tau_syn}, $\alpha_{\rm thick}$ is $\approx +2.5$ for a homogeneous synchrotron emitting source \citep{Kellermann1969}. Deviations from this value are expected in inhomogeneous or partially resolved sources \citep{Bicknell1997}.
We allow $\alpha_{\rm thick}$ to vary in our modelling, such that $\tauSSA$ in Eq.~\ref{eq:tau_SSA} effectively captures the integrated spectral signature of an inhomogeneous synchrotron source, where variations in $B$, $n_0$, or $L$ lead to a superposition of multiple self-absorbed regions with different local turnover frequencies, resulting in a flattened optically thick spectrum ($\alpha_{\mathrm{thick}} < 2.5$). We emphasise that, in such cases, $\alpha_{\rm thin}$ is not an adequate indicator of the energy index $\gamma$ of the cosmic ray electron populations in the inhomogeneous medium.

The intrinsic synchrotron spectrum 
is given as,
\begin{equation}
S_{\rm syn
}(\nu) = a_1\,
\left(\frac{\nu}{\nu_0}\right)^{\alpha_{\rm thin}}
\left(\frac{1 - {\rm e}^{-\tau_{\rm SSA}(\nu)}}
{\tau_{\rm SSA}(\nu)}\right),
\label{eq:sync_emission}
\end{equation}
where $a_1$ is the flux density at $\nu_0 = 1$\,GHz for our model.

In addition to SSA occurring internally in the source,
the synchrotron emission can be further modified when propagating through a surrounding ionised medium or an ionised cloud lying in between the observer along the line of sight (LoS). FFA arises when free electrons in an ionised, thermal plasma scatter and absorb background radiation, leading to additional attenuation at low frequencies \citep{Rybicki1979,Condon1992}. This process is particularly important in dense circumnuclear environments, where high electron densities can produce significant optical depths \citep{Bicknell1997,Kameno2000}. The free-free optical depth ($\tauff$) at frequency $\nu$ is related to the emission measure,
$\mathrm{EM}=\int n_{\rm e}^2\,{\rm d}l$, as,
\begin{equation}
\label{eq:tau_ff}
\tauff(\nu) =
0.08235\,
T_{\rm e}^{-1.35}
\left[
\frac{\nu}{\mathrm{GHz}}
\right]^{-2.1}
\left[
\frac{\mathrm{EM}}{\mathrm{cm^{-6}\,pc}}
\right],
\end{equation}
where $T_{\rm e}$ is the electron temperature and $n_{\rm e}$ is the number density of thermal free electrons. 
We parametrise $\tauff$ as,
\begin{equation}\label{eq:ffa-k2}
\tau_{\rm FFA}(\nu) =
k_2 \left(\frac{\nu}{\nu_0}\right)^{-2.1}.
\end{equation}
Here, $k_2$ represents the free–free optical depth at the pivot frequency $\nu_0$, and for an assumed value of $T_{\rm e}$, it provides an estimate for EM.

Thus, the resultant radio continuum emission, $S_{\rm RT}(\nu)$, from a dense, relativistic synchrotron emitting plasma that is undergoing SSA and further attenuated by thermal plasma along the LoS follows from radiative transfer (RT) as,
\begin{equation}
\label{eq:full_model}
\begin{aligned}
S_{\rm RT}(\nu) =\;& S_{\rm syn}(\nu)\,{\rm e}^{-\tauFFA(\nu)} + S_{\rm ff}(\nu),\\
=\;&
a_1
\left(\frac{\nu}{\nu_0}\right)^{\alpha_{\rm thin}}
\left(\frac{1 - {\rm e}^{-\tauSSA(\nu)}}{\tauSSA(\nu)}\right) \,{\rm e}^{-\tauFFA(\nu)}
\\
&+
a_2
\left(\frac{\nu}{\nu_0}\right)^{-0.1}
\left(\frac{1 - {\rm e}^{-\tauFFA(\nu)}}{\tauFFA(\nu)}\right).
\end{aligned}
\end{equation}
The first term represents synchrotron emission modified by SSA and attenuated by FFA, while the second term, $S_{\rm ff}(\nu)$, corresponds to thermal free–free emission from the ionised plasma. We refer to this model as SSA+FFA in the text. We denote the peak frequency of the total SSA+FFA spectrum as $\nu_{\rm p}$, and the peak of the intrinsic synchrotron (SSA) component in Eq.~\ref{eq:sync_emission} as $\nu_{\rm p,SSA}$.

In our adopted formulation of the model, the curvature and position of the spectral turnover are jointly controlled by 
the parameters $a_1$, $k_1$, $k_2$, $\alpha_{\mathrm{thick}}$ and $\alpha_{\mathrm{thin}}$,
leading to partial degeneracy among these quantities. This behaviour is inherent to single-zone, parametric models and reflects the fact that different combinations of intrinsic source properties can yield similar spectral shapes. Nevertheless, the inclusion of an external free-free absorption component allows us to explicitly assess the contribution of foreground ionised material to the observed turnover.

\subsection{Other radio SED models}
\label{sec:ancillary_models}

The SEDs of compact peaked-spectrum radio sources have often been modelled as pure SSA or pure FFA depending on the assumed physical configuration of the emitting and absorbing media \citep[e.g.][]{Bicknell1997, Tingay2003, Callingham2015, Mhaskey2019b}. While we consider a geometrical configuration in which the EISERS SEDs are primarily modelled using the SSA+FFA mechanism (Eq.~\ref{eq:full_model}), we also consider other possible SED models for our sample. 
They are --- pure SSA given by Eq.~\ref{eq:sync_emission};  
and an FFA+SSA model. The FFA+SSA is a model where the thermal free--free emission is seen through a synchrotron emitting plasma, i.e., an embedded ionised region viewed through, e.g., frustrated synchrotron lobe. The SED for such a scenario is given by,
\begin{equation}
\label{eq:FFA+SSA}
\begin{aligned}
S_{\rm RT}(\nu) =\;& S_{\rm ff}(\nu)\,{\rm e}^{-\tauSSA(\nu)} + S_{\rm syn}(\nu),\\
=\;&
a_2
\left(\frac{\nu}{\nu_0}\right)^{-0.1}
\left(\frac{1 - {\rm e}^{-\tauFFA(\nu)}}{\tauFFA(\nu)}\right)\,\,{\rm e}^{-\tauSSA(\nu)}
\\
&+
a_1
\left(\frac{\nu}{\nu_0}\right)^{\alpha_{\rm thin}}
\left(\frac{1 - {\rm e}^{-\tauSSA(\nu)}}{\tauSSA(\nu)}\right) .
\end{aligned}
\end{equation}

We emphasise that although radio continuum spectra shed light onto the emission mechanism(s) and plausible geometrical orientation of these unresolved EISERS, they alone cannot uniquely distinguish between different geometrical arrangements of the emitting and absorbing plasma.  
Consequently, the adopted model should be regarded as a phenomenological description rather than a unique physical solution.
 
\begin{figure*}
\centering
\begin{minipage}{0.48\textwidth}
\centering
\includegraphics[trim={0cm 0cm 0cm 3cm},clip,width=\columnwidth]{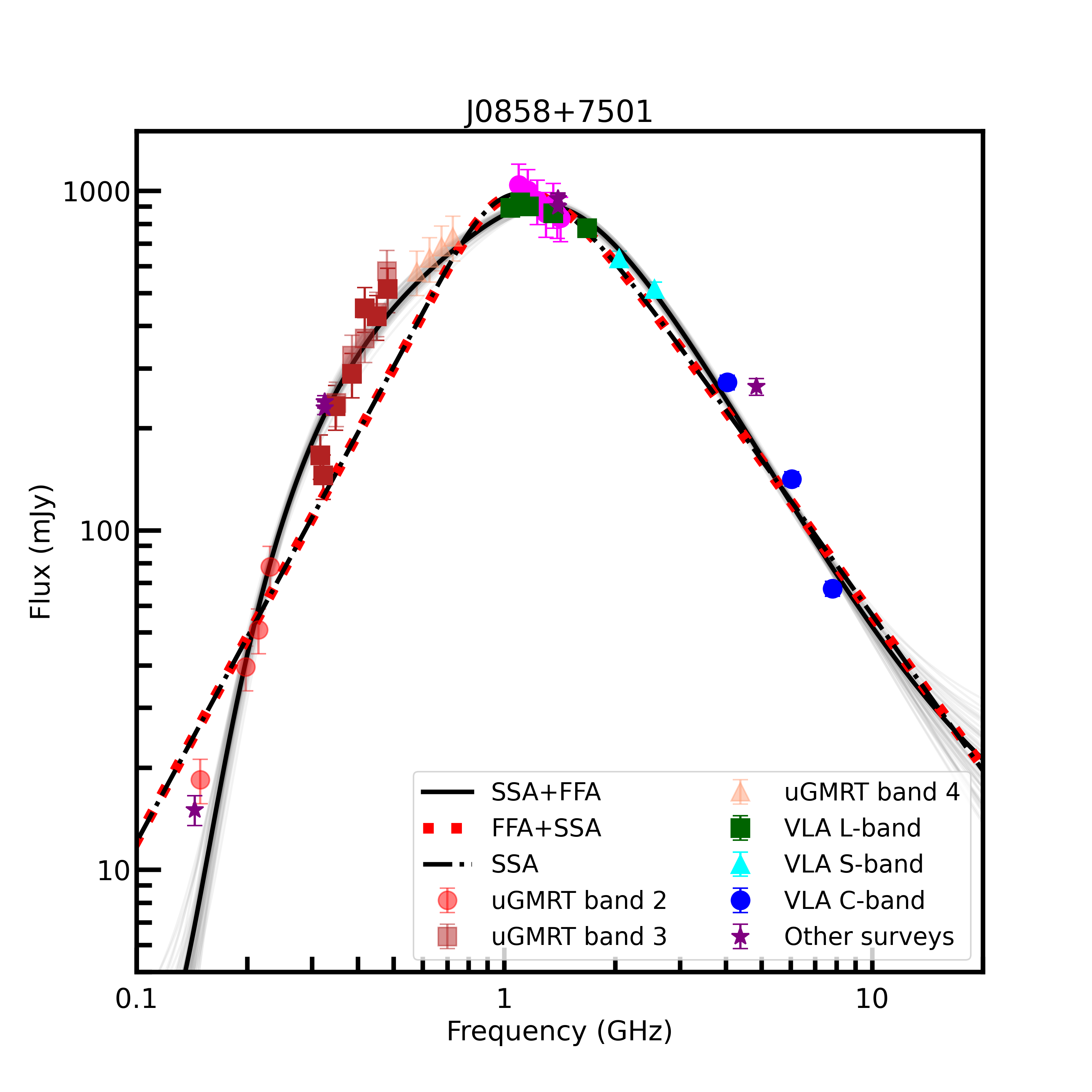}

\end{minipage}
\hfill
\begin{minipage}{0.48\textwidth}
\centering
\includegraphics[trim={0cm 0cm 0cm 0cm},clip,width=\columnwidth]{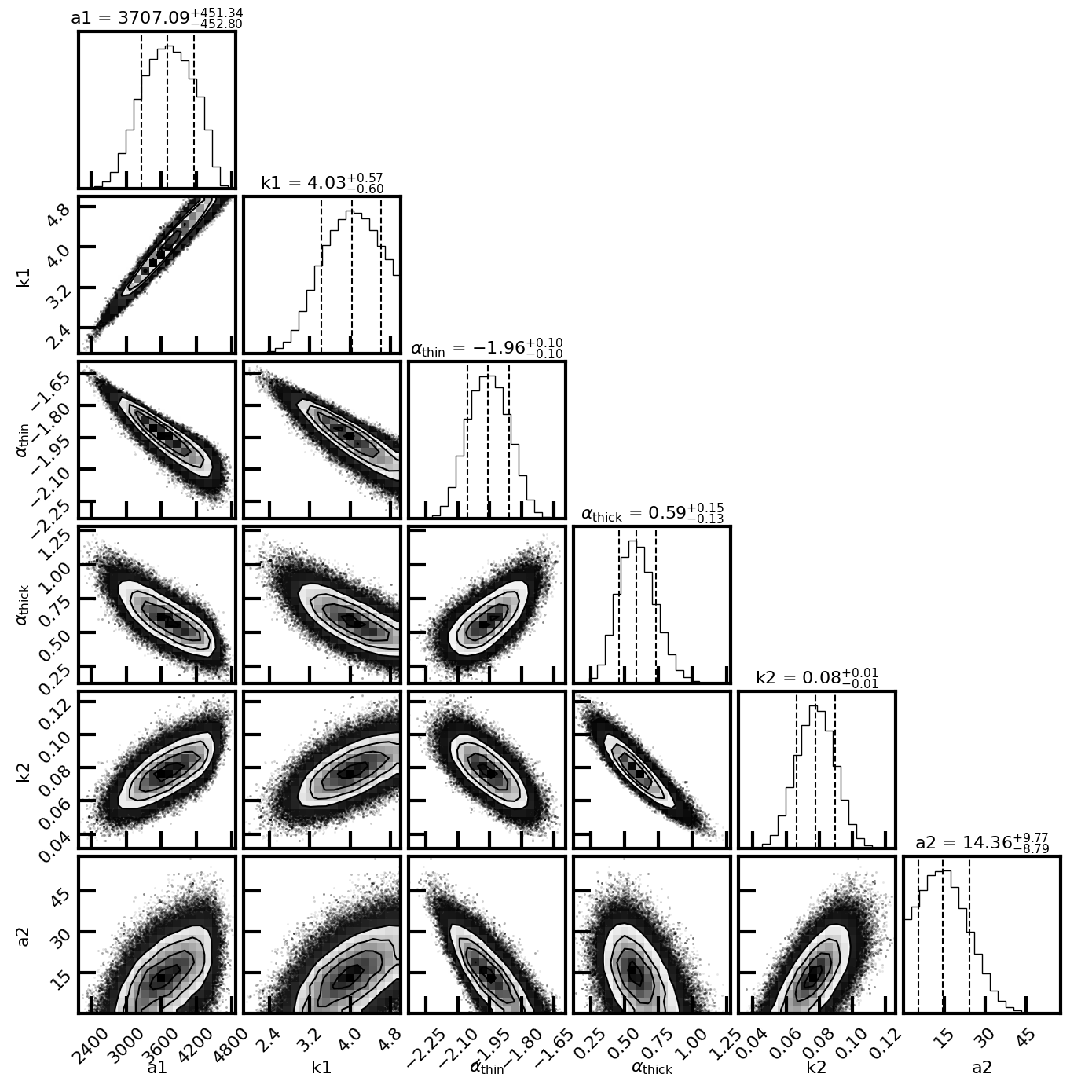}

\end{minipage}

\caption{Left: The radio continuum SED between 0.15 and 8\,GHz of J0858+7501. 
The different frequency bands covered by the uGMRT and VLA are marked as different colours. The solid and shaded uGMRT datapoints correspond to Epochs\,1 and 2, respectively. The other available radio surveys (eg., LoTSS, FIRST, NVSS, WENSS etc.) are shown under the legend \lq other surveys', which are not included for the fitting. The solid, dotted and dash-dotted lines represent the best fit SSA+FFA, FFA+SSA and pure SSA emission models discussed in Section~\ref{sec:modelling}. 
Right: Corner plot showing the marginalized posterior probability distributions and parameter covariances from the SSA+FFA model MCMC fit. Contours indicate the 1$\sigma$, 2$\sigma$, and 3$\sigma$ confidence regions.}
\label{fig:SED-J0858+7501}
\end{figure*}

\subsection{Parameter estimation using MCMC}
We use Bayesian inference to estimate the best-fitting values of the model parameters of a composite SSA+FFA model given in Eq.~\ref{eq:full_model}. The parameters are estimated using the Markov-Chain Monte Carlo (MCMC) approach implemented in the \texttt{emcee} package \citep{Foreman_Mackey_2013}. The observed spectra $S_{\rm obs}(\nu)$ is fitted by maximising the Gaussian log-likelihood function,
\begin{equation}
\ln \mathcal{L}(\vartheta) = -\frac{1}{2} \sum_{i} \left\{ \frac{\left[S_{\rm obs}(\nu_i) - S_{\rm mod}(\vartheta)\right]^2}{\sigma_i^2} + \ln(2\pi\sigma_i^2) \right\},
\end{equation}
where, $S_{\rm obs}(\nu_i)$ is the observed flux density at frequencies $\nu_i$, $S_{\rm mod}(\vartheta)$ is the corresponding model flux density computed from Eqs.~\ref{eq:sync_emission}, \ref{eq:full_model} and \ref{eq:FFA+SSA},
and $\sigma_i$ represents the total uncertainty at each frequency, including both measurement noise and systematic calibration uncertainties. These uncertainties are assumed to be Gaussian and uncorrelated.
$\vartheta$ is the parameter vector over which the posterior distributions are explored.

The model in Eq.~\ref{eq:full_model} has six free parameters, such that,
\begin{equation}
\vartheta = \{a_1,\ k_1,\ \alpha_{\rm thin},\ \alpha_{\rm thick},\ k_2,\ a_2\}.
\end{equation}
We adopt uniform priors for all the six free parameters, each restricted to a physically motivated range; 
\begin{equation}
\begin{aligned}
a_1 &\in [0,\ 10000], \quad a_2 \in [0,\ 10000],\\
k_1 &\in [0,\ 10], \quad k_2 \in [0,\ 10],\\
\alpha_{\rm thin} &\in [-2.5,\ 0],\\
\alpha_{\rm thick} &\in [0,\ 3.5].
\end{aligned}
\label{eq:prior}
\end{equation} 

For the SSA+FFA and FFA+SSA models, $\alpha_{\rm thick}$ was allowed to vary between zero  up to $+3.5$ to account for deviations from the canonical homogeneous SSA spectrum due to additional absorption. For the pure SSA it was varied up to its theoretical limit of $+2.5$.
The posterior distribution is sampled using an ensemble of 50 walkers evolved over 5000 steps. The walkers are initialised in a small Gaussian ball around initial guess values to ensure efficient convergence. The first 20\% of the chain is discarded as burn-in to minimise the influence of initial conditions.
The final parameter estimates are taken as the median (50$^{\rm th}$ percentile) of the posterior distributions, with uncertainties derived from the 16$^{\rm th}$ and 84$^{\rm th}$ percentiles, corresponding to a 68\% credible interval. In Fig.~\ref{fig:SED-J0858+7501}, we present the fit for J0858+7501 as a typical example, where the solid black line is the best-fit SSA+FFA model, and the gray lines represent posterior realisations. The fits for the other nine EISERS are presented in Fig.~\ref{fig:SED-EISERS}. In Table~\ref{tab:best_fit_parameters}, we present the best-fit parameters for the ten EISERS, together with the Bayesian Information Criterion \citep[BIC;][]{Kass&Raftery1995} values for the different absorption models fitted to the broadband radio spectra.

We note that for some sources, the FFA+SSA and SSA-only model fit curves overlap because the contribution of free--free absorption and emission is negligible over the observed frequency range. In such cases, the fitted free--free optical depth remains small ($\tau_{\rm FFA}\ll1$) and the thermal free--free component contributes only weakly compared with the synchrotron emission. Consequently, the FFA+SSA model effectively reduces to the pure SSA form, yielding nearly indistinguishable spectral shapes and fit curves.

From the values of BIC in Table~\ref{tab:best_fit_parameters} and the model SEDs in Fig.~\ref{fig:SED-EISERS}, it is evident that SSA+FFA model well represents the radio spectrum for all EISERS with the exception of J1326+5712. For a few other sources, namely, J0740+7129, J1536+8154 and J1549+5038, a mild preference towards SSA-only and/or FFA+SSA models could not be ruled out. We note that, the physical parameters derived based on the SSA+FFA model in the next section are meaningful only if the underlying assumptions of the adopted homogeneous SSA model are satisfied. In sources with complex or inhomogeneous radio structures, the inferred quantities (e.g., magnetic field strength and electron density) should therefore be regarded as effective parameters of the adopted model rather than unique physical properties of the source. This is particularly relevant for the J1549+5038, whose known core--jet morphology from archival VLBI observations \citep{Xu1995, Fey2000, Helmboldt2007} suggests that the assumption of a single homogeneous synchrotron-emitting region is unlikely to be valid. We therefore do not consider the derived physical parameters for this source in the subsequent discussion of the EISERS population.

\begin{table*}
\centering
\caption{Best-fit parameters ($a_1$, $a_2$, $k_1$, $k_2$, $\alpha_{\rm thin}$, and $\alpha_{\rm thick}$) obtained from fitting the SSA+FFA model described by Eq.~\ref{eq:full_model}.}

\label{tab:best_fit_parameters}
\addtolength{\tabcolsep}{-3pt}
\begin{tabular}{lccccccccccc}

\hline
Source &
$a_1$ &
$a_2$ &
$k_1$ &
$k_2$ &
$\alpha_{\rm thin}$ &
$\alpha_{\rm thick}$ &
$\chi^2_{\rm red}$ &
\multicolumn{3}{c}{${\rm BIC}$} \\
&
&
&
&
&
&
&
&
SSA+FFA&
SSA &
FFA+SSA \\
\hline \addlinespace

J0614+6108 &
$677.4^{+34.5}_{-50.5}$ &
$2.19^{+3.03}_{-1.62}$ &
$1.804^{+0.139}_{-0.231}$ &
$0.039^{+0.006}_{-0.006}$ &
$-1.46^{+0.05}_{-0.05}$ &
$0.41^{+0.12}_{-0.10}$ &
0.97 &247
&270
&278
\\ \addlinespace

J0649+5947 &
$259.2^{+7.1}_{-7.9}$ &
$0.156^{+0.22}_{-0.11}$ &
$0.962^{+0.028}_{-0.059}$ &
$0.031^{+0.005}_{-0.005}$ &
$-1.68^{+0.02}_{-0.02}$ &
$0.305^{+0.10}_{-0.09}$ &
5.56 &174
&210
&217
\\ \addlinespace

J0740+7129 $^\dagger$ &
$403.2^{+30.3}_{-38.5}$ &
$2.88^{+4.08}_{-2.13}$ &
$0.744^{+0.18}_{-0.23}$ &
$0.060^{+0.012}_{-0.014}$ &
$-1.13^{+0.06}_{-0.05}$ &
$0.80^{+0.38}_{-0.22}$ &
2.01 &388
&389
&397
\\ \addlinespace

J0847+5723 &
$1574.8^{+93.4}_{-121.2}$ &
$15.39^{+19.28}_{-11.14}$ &
$4.75^{+0.19}_{-0.35}$ &
$0.034^{+0.006}_{-0.005}$ &
$-1.22^{+0.05}_{-0.05}$ &
$0.320^{+0.056}_{-0.053}$ &
0.48 &374
&445
&454
\\ \addlinespace

J0853+6722 &
$2263.5^{+341.7}_{-428.5}$ &
$67.44^{+25.55}_{-27.12}$ &
$8.26^{+1.24}_{-1.81}$ &
$0.097^{+0.029}_{-0.031}$ &
$-1.43^{+0.16}_{-0.11}$ &
$0.53^{+0.12}_{-0.11}$ &
2.69 &254
&290
&298
\\ \addlinespace

J0858+7501 &
$3480.6^{+470.9}_{-416.6}$ &
$14.60^{+10.33}_{-8.98}$ &
$3.74^{+0.63}_{-0.57}$ &
$0.076^{+0.013}_{-0.012}$ &
$-1.92^{+0.10}_{-0.11}$ &
$0.67^{+0.13}_{-0.12}$ &
2.89 & 275
&349
&356
\\ \addlinespace

J1326+5712 &
$684.8^{+90.7}_{-62.4}$ &
$175.5^{+37.1}_{-60.3}$ &
$0.467^{+0.32}_{-0.22}$ &
$0.247^{+0.05}_{-0.08}$ &
$-1.10^{+0.26}_{-0.24}$ &
$1.20^{+0.99}_{-0.70}$ &
0.33 &315
&318
&321
\\ \addlinespace

J1536+8154 &
$622.3^{+40.3}_{-43.4}$ &
$36.28^{+32.69}_{-24.43}$ &
$0.894^{+0.08}_{-0.14}$ &
$0.026^{+0.012}_{-0.009}$ &
$-0.85^{+0.06}_{-0.10}$ &
$1.07^{+0.15}_{-0.12}$ &
0.74 &183
&183.4
&191
\\ \addlinespace

J1549+5038 &
$744.5^{+95.3}_{-95.2}$ &
$108.6^{+42.6}_{-37.9}$ &
$0.625^{+0.26}_{-0.30}$ &
$0.100^{+0.021}_{-0.023}$ &
$-0.06^{+0.04}_{-0.05}$ &
$0.64^{+0.39}_{-0.30}$ &
2.03 &227
&230
&230.2
\\ \addlinespace

J1658+4732 &
$463.2^{+27.2}_{-35.6}$ &
$18.15^{+13.07}_{-11.64}$ &
$0.845^{+0.11}_{-0.20}$ &
$0.050^{+0.013}_{-0.014}$ &
$-0.95^{+0.07}_{-0.07}$ &
$0.79^{+0.25}_{-0.17}$ &
2.26 & 492 & 494 & 503
\\

\hline
\end{tabular}
\tablefoot{The values of the Bayesian Information Criterion (BIC) for the different SED models are listed for comparison.\\
$^\dagger$ For the fitting of the source J0740+7129, we have combined both the uGMRT epochs to cover the frequency range.}
\vspace{0.1cm}
\end{table*}

\begin{table*}
\centering
\caption{Derived spectral and physical quantities obtained from the SSA+FFA spectral fits.}

\label{tab:derived_parameters}

\addtolength{\tabcolsep}{-3pt}

\begin{tabular}{lccccccccc}

\hline

Source &
$\nu_{\rm p}$ &
$S_{\rm p}$ &
$\nu_{\rm p,SSA}$ &
$S_{\rm p,SSA}$ &
$\tau_{\rm SSA}(\nu_{\rm p})$ &
$\tau_{\rm FFA}(\nu_{\rm p})$ &
$B_{\rm SSA}$ &
EM &
$n_{\rm e}$ \\

&
[GHz] &
[mJy] &
[GHz] &
[mJy] &
&
&
[G] &
[$10^{5}\,\rm cm^{-6}\,pc$] &
[cm$^{-3}$] \\

\hline \addlinespace

J0614+6108 &
$0.89^{+0.03}_{-0.03}$ &
$309.40^{+8.70}_{-7.60}$ &
$0.83^{+0.04}_{-0.04}$ &
$338.60^{+11.30}_{-10.10}$ &
$2.24^{+0.40}_{-0.43}$ &
$0.05^{+0.01}_{-0.01}$ &
$0.04^{+0.01}_{-0.01}$$^{\ddagger}$ &
$1.22^{+0.18}_{-0.18}$ &
$55.20^{+8.60}_{-8.60}$ 
\\ \addlinespace

J0649+5947 &
$0.63^{+0.03}_{-0.03}$ &
$198.40^{+6.50}_{-6.10}$ &
$0.66^{+0.06}_{-0.03}$ &
$211.60^{+15.70}_{-19.50}$ &
$2.41^{+0.38}_{-0.38}$ &
$0.08^{+0.02}_{-0.02}$ &
$1.88^{+0.69}_{-0.62}$$^{\ddagger}$ &
$0.95^{+0.15}_{-0.15}$ &
$29.00^{+2.30}_{-2.30}$ 
\\ \addlinespace

J0740+7129 &
$0.78^{+0.02}_{-0.03}$ &
$285.20^{+10.90}_{-8.90}$ &
$0.72^{+0.02}_{-0.02}$ &
$319.50^{+8.90}_{-8.70}$ &
$1.24^{+0.42}_{-0.38}$ &
$0.10^{+0.02}_{-0.02}$ &
$0.14^{+0.04}_{-0.04}$$^{\ddagger}$&
$1.83^{+0.30}_{-0.30}$ &
$53.40^{+4.50}_{-4.50}$ 
\\ \addlinespace

J0847+5723 &
$1.50^{+0.05}_{-0.05}$ &
$362.20^{+7.50}_{-6.80}$ &
$1.18^{+0.10}_{-0.08}$ &
$367.60^{+7.00}_{-7.00}$ &
$2.57^{+0.35}_{-0.39}$ &
$0.02^{+0.01}_{-0.01}$ &
$5.47^{+1.77}_{-1.61}$ &
$1.04^{+0.15}_{-0.12}$ &
$32.40^{+2.40}_{-1.90}$ 
\\ \addlinespace

J0853+6722 &
$1.98^{+0.08}_{-0.08}$ &
$405.10^{+9.90}_{-10.20}$ &
$1.62^{+0.07}_{-0.07}$ &
$401.80^{+9.90}_{-9.10}$ &
$2.19^{+1.05}_{-0.84}$ &
$0.02^{+0.01}_{-0.01}$ &
$7.92^{+2.30}_{-2.09}$$^{\ddagger}$ &
$2.96^{+0.92}_{-0.92}$ &
$64.10^{+9.90}_{-9.90}$ 
\\ \addlinespace

J0858+7501 &
$1.26^{+0.03}_{-0.03}$ &
$912.50^{+22.70}_{-20.40}$ &
$1.12^{+0.05}_{-0.05}$ &
$986.80^{+24.60}_{-22.10}$ &
$2.29^{+0.92}_{-0.68}$ &
$0.05^{+0.01}_{-0.01}$ &
$1.07^{+0.29}_{-0.27}$ &
$2.44^{+0.31}_{-0.31}$ &
$44.10^{+2.80}_{-2.80}$ 
\\ \addlinespace

J1326+5712 &
$0.98^{+0.04}_{-0.04}$ &
$579.80^{+15.20}_{-14.80}$ &
$1.01^{+0.02}_{-0.02}$ &
$571.90^{+11.50}_{-11.50}$ &
$0.50^{+0.33}_{-0.24}$ &
$0.26^{+0.05}_{-0.08}$ &
$0.22^{+0.04}_{-0.04}$ &
$8.80^{+1.70}_{-2.70}$ &
$128.00^{+12.00}_{-20.00}$ 
\\ \addlinespace

J1536+8154 &
$0.93^{+0.03}_{-0.03}$ &
$444.70^{+11.90}_{-11.00}$ &
$0.91^{+0.05}_{-0.05}$ &
$450.90^{+12.10}_{-12.10}$ &
$1.03^{+0.20}_{-0.22}$ &
$0.03^{+0.02}_{-0.01}$ &
$0.57^{+0.18}_{-0.17}$$^{\ddagger}$ &
$0.79^{+0.30}_{-0.30}$ &
$31.50^{+6.10}_{-6.10}$ 
\\ \addlinespace

J1549+5038$^\dagger$ &
$4.28^{+3.86}_{-1.15}$ &
$697.80^{+16.70}_{-15.30}$ &
$3.52^{+2.43}_{-0.92}$ &
$682.60^{+12.60}_{-11.50}$ &
$0.23^{+0.35}_{-0.20}$ &
$-$ &
$-$ &
$-$ &
$-$ 
\\ \addlinespace

J1658+4732 &
$0.84^{+0.03}_{-0.03}$ &
$325.00^{+7.30}_{-6.60}$ &
$0.76^{+0.03}_{-0.03}$ &
$342.00^{+7.40}_{-7.10}$ &
$1.15^{+0.34}_{-0.34}$ &
$0.07^{+0.02}_{-0.02}$ &
$0.02^{+0.01}_{-0.01}$ &
$1.52^{+0.30}_{-0.30}$ &
$60.30^{+6.00}_{-6.00}$ \\

\hline

\end{tabular}
\tablefoot{Columns list the turnover properties of the composite and synchrotron spectra, the SSA and FFA optical depths evaluated at the turnover frequency of the composite spectrum, the magnetic field strength estimated from the SSA formalism, and the emission measure and electron density derived assuming a uniform external free--free absorbing screen with $T_{\rm e}=10^4$\,K. The frequencies and the flux densities listed here are in the observer's frame.\\
$^\dagger$ J1549+5038 is a known QSO 
with a core-jet morphology. The derived physical parameters based on the best-fit SSA+FFA model, wherein the SSA component is implicitly assumed to be homogeneous, may not be applicable.\\
$^\ddagger$ The redshifts of these sources are not known, and therefore, we have assumed a fiducial value of $z=1$ to compute $B_{\rm SSA}$. Our reported values should be multiplied by $2/(1+z)$ when their redshifts become available.}

\end{table*}

\subsection{Estimating physical properties of EISERS}
\label{sec:derived_par}
To estimate the different physical quantities on which SSA and FFA depend, we determined the turnover frequency $\nu_{\rm p,SSA}$ and the corresponding flux density $S_{\rm p, SSA}$ for each of the SSA and FFA components in Eq.~\ref{eq:full_model} separately from the best-fit model. Under SSA, the peak frequency ($\nu_{\rm p,SSA}$) of the synchrotron emitting component given by Eq.~\ref{eq:sync_emission} is determined by the magnetic field strength ($B_{\rm SSA}$) among other parameters, and are related as \citep[e.g.,][]{KellermannPaulinyToth1981, Readhead1994},
\begin{equation}
\label{eq:B_mag_field}
\left[\frac{B_{\mathrm{SSA}}}{\rm G}\right] \approx
10^{-5}\,
b(\alpha)
\,
\left[\frac{\nu_{\rm p,SSA}}{\rm GHz}\right]^5\,
\left[\frac{\theta_{\rm syn}}{\rm mas}\right]^4\,
\left[\frac{S_{\rm p,SSA}}{\rm Jy}\right]^{-2}
\left(
\frac{\delta}{1+z}
\right).
\end{equation} 

Here, $b(\alpha)$ is a weakly varying function that depends on $\alpha_{\rm thin}$, with typical values in the range $\sim 2-4$, $\theta_{\rm syn}$ is the angular extent of the optically thick synchrotron source, $\delta$ is the Doppler factor and $z$ is the redshift. We adopt the commonly used value of $b(\alpha)=3.5$ \citep[e.g.,][]{MarscherGear1985,Readhead1994}, and use the LAS of the brightest compact component from VLBI observations listed in Table~\ref{table:specindex} as the $\theta_{\rm syn}$. In the absence of independent constraints on relativistic beaming of our sample of EISERS, we assume $\delta=1$, corresponding to negligible Doppler boosting. This is motivated by the fact that EISERS are generally thought to be compact radio galaxies rather than strongly beamed, blazar-like sources, and therefore are not expected to exhibit significant relativistic boosting. In Table~\ref{tab:derived_parameters} we list the estimated strength of $B_{\rm SSA}$, and find them to lie over a range of values between 0.02 -- 8 G for the nine EISERS, excluding J1549+5038. 

Free--free emission arising from scattering of free electrons by ions is a direct tracer of the number density of thermal free electrons ($n_{\rm e}$), such that, $\tauff$ provides an estimate of EM for the external screen assumed in our model. By comparing Eq.~\ref{eq:tau_ff} with its parametrisation in Eq.~\ref{eq:ffa-k2}, EM can be computed as,
\begin{equation}
\mathrm{EM} =
\left[\frac{k_2}
{0.08235\,T_{\rm e}^{-1.35}}\right]\,{\rm cm^{-6}\,pc}.
\label{eq:EM}
\end{equation}
assuming an electron temperature $T_{\rm e} = 10^4$~K, typical of photoionised circumnuclear gas.

We estimate the mean electron density $\langle n_{\rm e}\rangle$ from EM under the assumption of uniform volume filling factor as,
\begin{equation}
\langle n_{\rm e} \rangle \approx
\left(
\frac{\mathrm{EM}}{L}
\right)^{1/2}.
\label{eq:ne}
\end{equation}
Here, $L$ is the path length in parsecs of the free--free absorbing screen, which we assume to be comparable to the angular size $\theta_{\rm syn}$ so that, $L \approx \theta_{\rm syn}\, D_A$. Here, $D_A$ is the angular diameter distance. The estimated EM and $\langle n_{\rm e }\rangle$ are presented in Table~\ref{tab:derived_parameters}.

\section{Optical and infrared properties} \label{sec:optical_SED}
\subsection{SEDs and redshifts} \label{subsec:SED} 
We determined the SED of the EISERS host galaxies in the optical and infrared using the photometric fluxes in the optical LS bands in combination with the fluxes in the WISE near-infrared bands W1 and W2 from unWISE \citep{Schlafly_2019} and W3 and W4 from AllWISE \citep{Cutri_2014}.
Measurements are also available in the SDSS u and i bands for the three relatively bright objects J0847+5723, J1326+5712 and J1549+5038. The quasi-stellar object (QSO) J1549+5038 is the only source from our list that has a flux measurement in the NUV band by the GALEX mission \citep{Morrissey_2007}.

To classify the host galaxies, we compared each SED with the set of galaxy template spectra from the SWIRE library \citep{Lonsdale_2003, Polletta_2007}, which consists of 25 SEDs that contain all major galaxy types, including ellipticals of different ages, early-type and late-type spirals, starburst (SB) galaxies,  type 1 AGN,  type 2 AGN,  bright QSOs, and composite types SB+AGN. In this paper, our approach is the same as in  \citet{Meusinger_2024}, where this method has been applied to 242 galaxies. We found that assigning an SED to a template spectrum becomes highly unreliable or impossible when flux densities are available for only three or fewer photometric bands. We have therefore restricted the analysis to objects for which flux densities are available in at least four bands, at least one of which must be in the infrared range. This limits the sample to the counterparts of the eight EISERS listed in Table\,\ref{tab:redshifts}. 
On average, these sources have measurements in seven bands. The SED-fitting method involves determining the photometric redshift. We used $z_{\rm phot}$ from  \citet{Duncan_2022} as an initial value and then varied the value incrementally, matching all SWIRE templates to the observed SED at each given $z$.  The best and the second best matches were determined using the minimum chi-squared method. 
For J1549+5038 the resulting  $z_{\rm phot}$ is well confirmed by the spectroscopic redshift from SDSS. Fig. \,\ref{fig:MBSED} displays the resulting MBSEDs (black squares with error bars),  the best-fitting template (red), the second best fit (magenta). Table\,\ref{tab:redshifts} summarises the results. 

\begin{figure}[htbp]

\includegraphics[trim={0cm 0.0cm 0cm 0.0cm},clip,width=0.95\columnwidth]{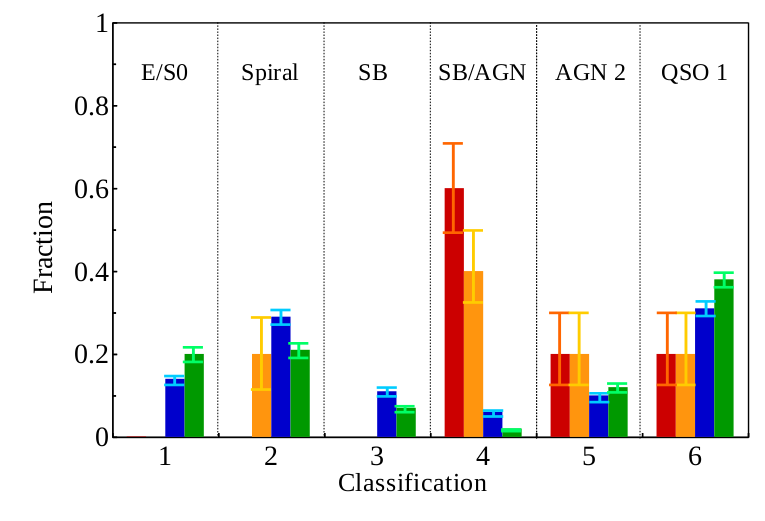}
\caption{Histogram distributions of the SED classes of the EISERS from Table\,\ref{tab:redshifts} (red for best matching templates, orange for second best matches), compared with the PSS-CSO (blue) and the ECS (green) samples from \citet{Meusinger_2024}. The error bars indicate the standard errors of proportions.} 
\label{fig:histogram_tSED}
\end{figure}

Following \citet{Meusinger_2024}, the different templates were grouped into the six different classes: 
1 for E and S0 galaxies,  
2 for spiral galaxies,  
3 for starburst (SB) galaxies, 
4 for SB-AGN composites (or transition objects), 
5 for AGNs of type 1.8 to 2, and 6 for type 1 QSOs.
Figure\,\ref{fig:histogram_tSED} shows the distribution for the five EISERS from Table\,\ref{tab:redshifts}. For comparison, the peak spectrum sources (PSS)/compact symmetrical objects (CSO) and the comparison sample of extended radio galaxies (ECS)
from \citet{Meusinger_2024} are plotted.

The only source where the best and second-best fits result in slightly different classes is J1658+4732. This object illustrates how the classification can be limited by the use of only a small number of photometric bands. The best match, Arp 220, is a strong SB galaxy. The second best match, Sc galaxy, is also characterised by a young stellar population, but is not an SB. The two SEDs do not significantly differ from one another between 0.2 and 3\,$\upmu$m (see Fig.\,\ref{fig:MBSED}). At $z \sim 1.3$, this is the range covered by the LS photometric bands and the first two WISE bands. Arp 220 shows strong emission in the mid and far IR. However, the AllWISE catalogue offers only upper limits for W3 and W4, and no data are available in the far IR for J1658+4732. We consider it unlikely that the near-IR flux of an Sc galaxy at $z = 1.3$ would be high enough to be detected by WISE, but here we have focused exclusively on the SED fit.

\begin{table*}[htbp]\centering                                        \caption{Redshifts and best matching SED templates for the EISERS with photometric measurements in at least four bands in the optical and IR}.                              
\begin{tabular}{cccccccccc}                                             
\hline
 Source 
& $z_{\rm spec}$ 
& \multicolumn{3}{c}{$z_{\rm phot}$}  & & \multicolumn{4}{c}{best matching template} \\
   \cline{3 - 5} 
   \cline{7 - 10} 
   &  & LS DR9 & Duncan (2022)  & This work 
   & & $1^{\rm st}$ & class & $2^{\rm nd}$ & class\\
\hline           

J0847+5723   &   -                    & 1.090 $\pm$ 0.16  & 1.155 $\pm$ 0.145 & 1.15 & & IC\,19254 & 4 & Arp\,220 & 4 \\                       
J0858+7501   &   -                    & -                 & 1.010 $\pm$ 0.190 & 2.00 & & Torus  & 5 & IC\,19254 & 4 \\         
             
J1326+5712   &   -                    & 0.589 $\pm$ 0.085 & 0.713 $\pm$ 0.067 & 0.90 & & Arp\,220 & 4 &  Sey\,2  & 5 \\                          
J1549+5038   & 2.174\tablefootmark{b} & 0.433 $\pm$ 0.202 & 1.518 $\pm$ 0.519 & 2.17 & & TQSO\,1  & 6 & QSO\,1   & 6 \\
J1658+4732   &   -                    & 1.086 $\pm$ 0.234 & 1.011 $\pm$ 0.219 & 1.30 & & Arp\,220 & 4 &  Sc    & 2 \\         
\hline                                                                    
\end{tabular}                                     
\label{tab:redshifts}                                                   
\end{table*}

\subsection{Galaxy mergers}
\begin{figure}[htbp]
\centering

\includegraphics[trim={0cm 0.3cm 0cm 0cm},clip,width=0.85\columnwidth]{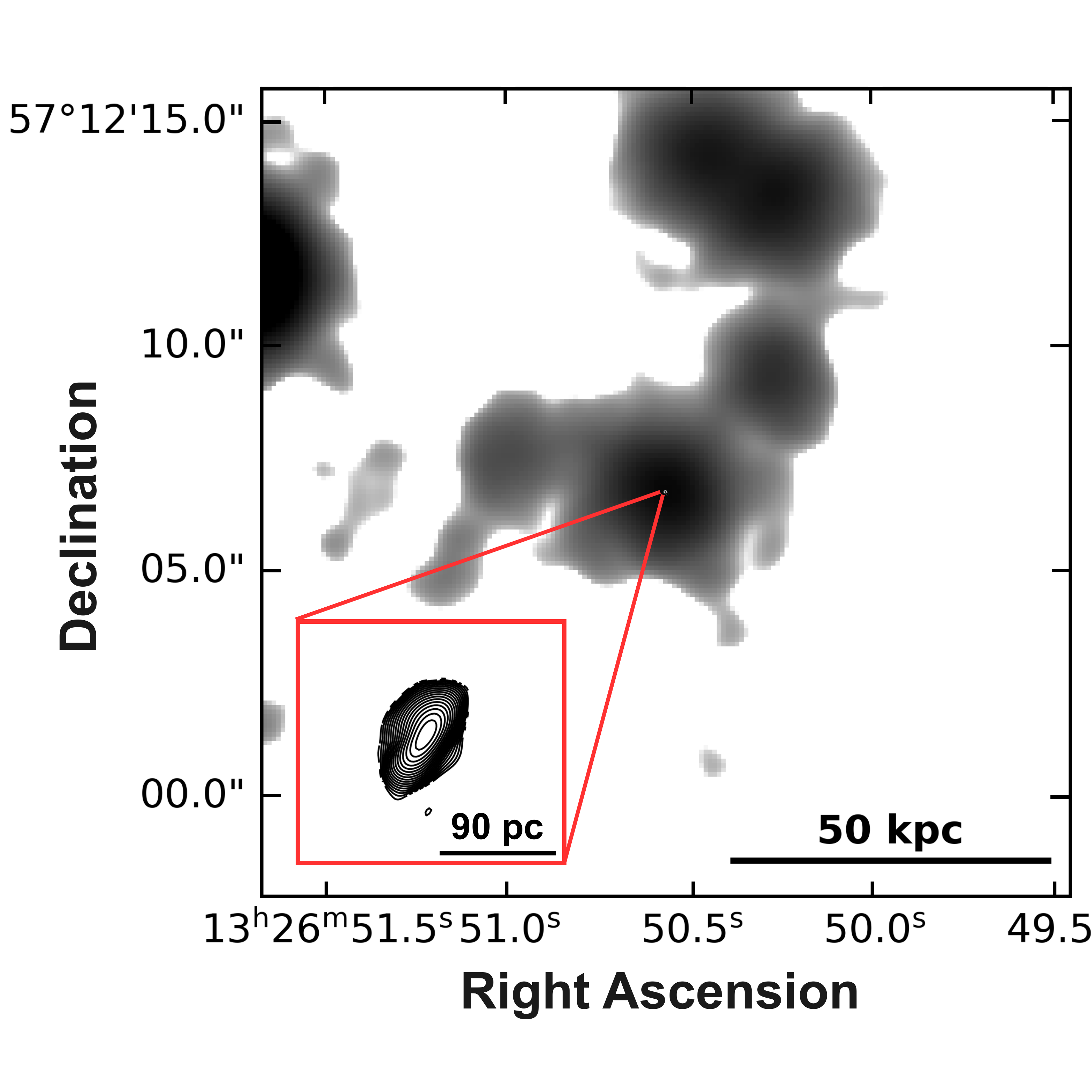}
\caption{Optical legacy surveys image of a possible merger for the source J1326+5712, smoothed (boxcar filter with radius 5) and contrast enhanced. VLBI 2.3\,GHz contours are shown in the inset.}
\label{fig:J1326}
\end{figure}

Fig.\,\ref{fig:J1326} shows the optical images of J1326+5712. It has two nearby neighbours, both at a core distance of $\sim 3.5\arcsec$ (27\,kpc). A connection in the form of a light bridge is indicated, at least for the western galaxy. The other galaxies in Table \ref{tab:redshifts} are too far away ($z > 1$) for morphological peculiarities to be discernible in the LS images.

It has long been suspected that efficient radial gas flows, triggered by galaxy mergers involving at least one gas-rich component, can trigger both AGNs and circumnuclear star formation activity \citep[e.g.][]{Hopkins_2008}. Our EISERS sample is, of course, too small to draw general conclusions on this matter. Signatures of mergers are also visible in LS images of the two EISERS J1031-2228 and J1209-2032 in the lower part of Table \ref{tab:redshifts}. We would nevertheless like to point out that  these results are consistent with other findings indicating that compact radio galaxies with peaked radio spectra have a high proportion of galaxy mergers \citep{Odea_2021,Meusinger_2024,Stanghellini_2025, Fida_2026}.

\section{Discussion} \label{sec:discussion}
In this paper, we have modelled the radio continuum SED of a sample of peak spectrum EISERS using well-sampled frequencies over a broad range between 0.15 and 8\,GHz. This allows us to sample both the optically thick and thin regime of the emission, on both sides of the turnover frequency, and model their SEDs by synchrotron self-absorbed medium that is further absorbed by a free--free emitting and absorbing screen, SSA+FFA model described in Section~\ref{sec:SSA+FFA}. These high quality data therefore allowed us to decompose the relative contribution of each of the mechanisms, i.e., SSA and FFA, and estimate various physical parameters presented in Section~\ref{sec:derived_par}.
The SSA+FFA model parameters listed in Table~\ref{tab:best_fit_parameters} reveal significant diversity in both the spectral turnover mechanisms and the physical environments of the compact radio sources in our sample. Rather than forming a homogeneous class, EISERS span a wide range of turnover frequencies,
optical depths, and spectral slopes, indicating differing physical conditions within the emitting regions and along the line of sight.

\subsection{Reliability of extreme spectral indices}
The EISERS sample was initially selected using two-point spectral indices derived from non-simultaneous survey measurements between TGSS and WENSS, i.e., between 0.15 and 0.325\,GHz. These sources appeared to exhibit optically thick spectral indices that seemingly exceeded the SSA limit of $\alpha_{\rm thick} \simeq +2.5$ expected for optically thick, homogeneous synchrotron emitting sources.

However, our new quasi-simultaneous uGMRT observations with improved frequency coverage, uniform calibration, and internally consistent flux-density scaling, show that a significant fraction of the sources do not exhibit such extreme spectral indices. The two-point spectral index $\alpha_{\rm 0.150~GHz}^{\rm 0.325 ~GHz}$, between the same frequencies obtained with our data are listed in Table~\ref{tab:derived_parameters}, and except for J0858$+$7501, all sources have $\alpha_{\rm 0.150~GHz}^{\rm 0.325 ~GHz} < +2.5$. This highlights that a sample selection based on spectral indices between two different surveys, especially at low frequencies, could be significantly affected by systematic calibration differences, resolution effects, flux-scale inconsistencies, and possible variability between survey epochs. Nevertheless, a subset of sources in our sample confirms that the EISERS can have unusually steep optically thick spectra.

Interestingly, from modelling of the broadband radio continuum SED of EISERS in Section~\ref{sec:modelling}, we find the best-fit $\alpha_{\rm thick}$ of the optically thick synchrotron emission listed in Table~\ref{tab:best_fit_parameters} to be significantly lower than $\alpha_{\rm 0.150 ~GHz}^{\rm 0.325 ~GHz}$. This indicates that for EISERS, FFA plays an important role in shaping the low-frequency spectral slope below $\nup$.

\subsection{Properties of peak spectrum sources and EISERS}
Using the best-fit parameters of the SSA+FFA model for the radio SEDs of EISERS, we derived the magnetic field strengths ($B_{\rm SSA}$) from the peak of the SSA component alongside EM and $n_{\rm e}$ from the free--free optical depth. The estimated $B_{\rm SSA}$ for the EISERS in our sample span over 4 orders of magnitude ranging between few tens of milliGauss to a few Gauss. These are largely similar to the magnetic field strengths reported for typical peaked-spectrum (PS) sources, i.e., gigahertz peaked spectrum (GPS) and/or compact steep spectrum (CSS) sources \citep{Odea1998, Orienti2016, Odea_2021}. In fact, the strongest $B_{\rm SSA}$ value in our sample, for the high-redshift QSO J1549+5038, is also accompanied by a large Faraday rotation measure of  $1400 \pm 500\,\mathrm{rad\,m^{-2}}$ \citep{Pasetto2016}, indicative of a strongly magnetised and dense ionised region.

The estimated $\rm EM = \mathcal{O}(10^5\,cm^{-6}\,pc)$ for the EISERS, and subsequently derived $n_{\rm e}$ in the range 30 to $100\,\rm cm^{-3}$, are comparable to those inferred for dense circumnuclear ionised media in compact radio AGN associated with GPS and CSS sources \citep{Bicknell1997,Kameno2000}. This points to the fact that the associated free--free absorption in EISERS occurs in the local vicinity of the compact, optically thick synchrotron emission, likely, close to the black hole. Despite their extreme low-frequency spectra, the physical conditions inferred for the EISERS remain broadly similar to those of other classes of compact PS radio sources. 

\subsection{Role of absorption mechanisms} 
The physical parameters derived from detailed modelling of the densely sampled spectra of EISERS, selected on the basis of the steepness of their optically thick radio spectra, are remarkably similar to those of other PSS selected primarily from the overall shape of their radio SEDs. This is especially true for the magnetic field strengths, whose estimates depend primarily on the turnover frequency, peak flux density, and angular size of the source, rather than directly on the unusually steep optically thick spectral indices observed in several EISERS. In order to understand the relative efficacy of SSA and FFA mechanism in shaping the low frequency part of the spectrum, in Table~\ref{tab:derived_parameters} we present the optical depths $\tauSSA(\nup)$ and $\tauFFA(\nup)$ at the frequency of the peak of the overall spectrum ($\nup$), derived using eqs.~\ref{eq:tau_SSA} and \ref{eq:ffa-k2}, respectively. It is evident that at $\nup$, $\tauSSA \gg \tauFFA$, such that, the spectral turnover is mostly governed by the SSA mechanism internal to the emitting plasma. As a consequence, the frequency of the peak synchrotron emitting component $\nu_{\rm p,SSA} \approx \nup$. Since the estimate of magnetic field strength for optically thick synchrotron emission is more sensitive to $\nu_{\rm p,SSA}$ and depends as $\nu_{\rm p,SSA}^5$ (see eq.~\ref{eq:B_mag_field}), the dominating $\tauSSA$ in EISERS, and likely in other classes of PS sources, provides similar estimates on magnetic fields.

Although the inclusion of an external FFA component in front of an SSA component improves the spectral fits in EISERS, the derived $\tauFFA(\nup)$ remain comparatively modest. This suggests that external ionised gas contributes mainly as a secondary modification to the low frequency spectral curvature rather than being the principal origin of the turnover. Our results suggest that PS in extragalactic sources is largely caused by SSA mechanism.

While the radio continuum SEDs of EISERS are well represented by the SSA+FFA model, in Section~\ref{sec:ancillary_models} we also compared a simpler SSA-only model and an alternate geometrical configuration, FFA+SSA. As per the BIC values, for the majority of the EISERS, the SSA+FFA model is preferred over the alternative absorption models, although for a few sources this preference is only marginal (see Table~\ref{tab:best_fit_parameters}). Nevertheless, the SSA+FFA model provides a convenient framework for quantifying the possible contribution of FFA while remaining consistent with previous studies that have described peaked-spectrum sources in general using SSA-only or FFA-only models \citep[e.g.][]{Bicknell1997, Tingay2003, Callingham2015, Mhaskey2019b}. It should be noted that previous studies modelled PS sources using limited frequency coverage that did not homogeneously cover both the optically thin and thick regime, and mostly using large sky-area survey data.

We further emphasise that SSA+FFA model remains contingent on the assumption that the underlying physical parameters of the synchrotron emitting component are effectively homogeneous. Inhomogeneous synchrotron sources generally exhibit flatter optically thick spectra because different emitting regions become optically thick at different frequencies. The proximity of several EISERS to the homogeneous SSA limit therefore lends support to the adopted description, but does not uniquely establish it. More complex source structures, such as core--jet systems (e.g., J1549+5038) or different spatial arrangements of the synchrotron-emitting and thermal absorbing plasma, may produce similar broadband spectra. Thus, the radio continuum SED alone cannot uniquely determine the physical geometry of the emitting and absorbing media, highlighting the importance of ancillary constraints such as VLBI morphology and multi-wavelength observations.

\subsection{Optical hosts of EISERS}

As shown in Section~\ref{subsec:SED}, the optical and infrared SEDs reveal  a clear difference between the activity types (SB or AGN) of the host galaxies in the EISERS sample and those in the two comparison samples.
In the comparison samples, most objects classified as galaxies exhibit SEDs corresponding to those of normal galaxies, although the proportion of (normal) star-forming galaxies is higher in the PSS-CSO sample than in the ECS sample. For the EISERS, on the other hand, there is no normal galaxy among the best-matching templates and only one (a late-type spiral galaxy) among the second-best matches, which, however, can be regarded as unlikely (see  Sect.\,\ref{subsec:SED}). 

In the subsequent discussion, we exclude the bright QSOs from the statistics, as they outshine their host galaxies. When we take the average of the best and second-best matches, we find that 70\% of the EISERS host galaxies harbour an SB and/or a reddened AGN (classes 3, 4, or 5), compared to only 16\% for the PSS/CSO  and 14\% for the ECS sample. According to the one-tailed two-sample Z test of proportions, the difference between the EISERS sample and the two other samples is highly significant ($p < 10^{-5}$).  We cannot rule out the possibility that the five EISERS for which we do not have a reliable WBSED are hosted by normal galaxies. However, even in this case, the difference compared with the other samples would still be statistically significant ($p = 0.003$).
Our analysis therefore suggests that the central engines of EISERS host galaxies may be embedded in gas-rich environments. 
However, these results must be verified using a larger EISERS sample and deeper optical observations.

The optical and infrared properties of the EISERS support the idea that these sources preferentially reside in dusty and gas-rich host galaxies. Most sources are either faint or undetected in the optical bands, while infrared counterparts are commonly detected. The SED fitting suggests that the hosts are predominantly composite SB--AGN systems, or reddened AGN. Such environments naturally provide large gas reservoirs capable of producing both free--free absorption and strong jet--ISM interactions. Similar associations between compact PS radio sources and dense circumnuclear environments have been reported for GPS/CSS populations \citep{Holt2008,Morganti2013,Tadhunter2016}. The inferred presence of dense ionised gas responsible for free–free absorption based on SSA+FFA model is consistent with the host galaxy properties, suggesting a gas-rich environment that can confine or interact with the radio jets.

The EISERS hosts are predominantly associated with dusty starburst galaxies, composite systems, and obscured AGN. This suggests that EISERS may not necessarily represent a fundamentally different class of peaked spectrum sources, but could instead be compact radio AGN embedded in denser and more gas-rich circumnuclear environments. Such environmental differences may contribute to the extreme low-frequency spectra, although the present observations do not allow us to establish a unique physical origin.

\section{Summary} \label{sec:summary}

The summary of results obtained with quasi-simultaneous observations between 0.15 and 8\,GHz of a sample of extremely inverted spectrum sources, called EISERS, is listed here.

\begin{itemize}
\item[--] We presented a sample of extremely inverted spectrum radio galaxies identified based on their unusually steep low-frequency spectral indices between the TGSS-ADR1 and WENSS surveys at 0.150 and 0.325 GHz, respectively. Subsequently characterised using broadband quasi-simultaneous uGMRT and VLA observations covering $0.15$--$8$\,GHz.

\item[--] Several survey-selected candidates do not retain extremely inverted spectra in the follow-up observations, demonstrating that two-point spectral-index selection can be affected by flux-scale offsets, spectral curvature near the turnover, variability, and resolution effects.

\item[--] The SSA+FFA radiative-transfer modelling successfully reproduces the observed spectral turnovers, with the inferred SSA optical depths generally exceeding the corresponding FFA optical depths at the turnover frequency.

\item[--] The derived physical parameters, including magnetic field strengths ($\sim$0.02--8~G) and electron densities ($\sim$30--100 cm$^{-3}$), imply modest physical conditions on parsec scales, consistent with the presence of ionised circumnuclear gas surrounding the compact radio source. These parameters are generally not large enough for FFA to dominate the observed
spectral turnover and are broadly consistent with those reported for compact GPS/CSS radio sources, despite the unusually steep optically thick spectra of the EISERS.

\item[--] Optical and infrared analysis support the presence of gas-rich host environments: optical counterparts are generally faint or undetected, whereas infrared detections are common. SED fitting indicates that the hosts are predominantly starburst galaxies, composite starburst–AGN systems, or reddened AGN, with a marked absence of normal early-type galaxies. This seems to support a scenario in which at least a fraction of these sources represent young or environmentally confined radio AGN embedded in dense circumnuclear media. However, given the small size of our sample, these results should be regarded as an indication rather than as proof.

\end{itemize}
\section*{Data availability}
Tables \ref{app:sed},\ref{tab:data-validation-I} and \ref{tab:fluxJ0858} are only available in electronic form at the CDS via anonymous ftp to cdsarc.u-strasbg.fr (130.79.128.5) or via http://cdsweb.u-strasbg.fr/cgi-bin/qcat?J/A+A/.

\begin{acknowledgements}
We thank the anonymous referee for their constructive comments. AB gratefully acknowledges the financial support provided by the German Federal Ministry of Research, Technology and Space (BMFTR) in the framework of the Knowledge creates perspectives for the region!, for the project StStG – DZA – Aufbauphase: Deutsches Zentrum f\"ur Astrophysik, Gro{\ss}forschungszentrum in der s\"achsischen Lausitz: Aufbauphase 2026, grant number 03WSP1746.
This research has made use of data products from the Legacy Surveys, including the Dark Energy Camera Legacy Survey (DECaLS), the Beijing-Arizona Sky Survey (BASS), and the Mayall z-band Legacy Survey (MzLS). The Legacy Surveys imaging of the DESI footprint is supported by the U.S. Department of Energy and the U.S. National Science Foundation. The Legacy Survey team also makes use of data products from NEOWISE, a NASA-funded project of the Jet Propulsion Laboratory/California Institute of Technology.
This publication has made use of the VizieR catalogue access tool, CDS, Strasbourg, France, and of the NASA/IPAC Infrared Science Archive (IRSA), operated by the Jet Propulsion Laboratories/California Institute of Technology, founded by the National Aeronautic and Space Administration. In particular, this publication makes use of data products from the Wide-field Infrared Survey Explorer, which is a joint project of the University of California, Los Angeles, and the Jet Propulsion Laboratory/California Institute of Technology, funded by the National Aeronautics and Space Administration. 

\end{acknowledgements}

\bibliographystyle{aa} 
\bibliography{literature}

\clearpage
\newpage
\onecolumn

\begin{appendix}
\section{Composite radio images of the EISERS}
\label{appendix:ror}
This appendix presents the composite radio images of the EISERS where we show the observed low frequency uGMRT radio map with the high frequency VLA contours along with the VLBI core. This appendix is a continuation of Fig. \ref{fig:Radiomap_J0858+7501}.
\begin{figure*} [htp!]
    \centering
    \begin{subfigure}{0.32\textwidth}
        \centering
        \includegraphics[width=\textwidth]{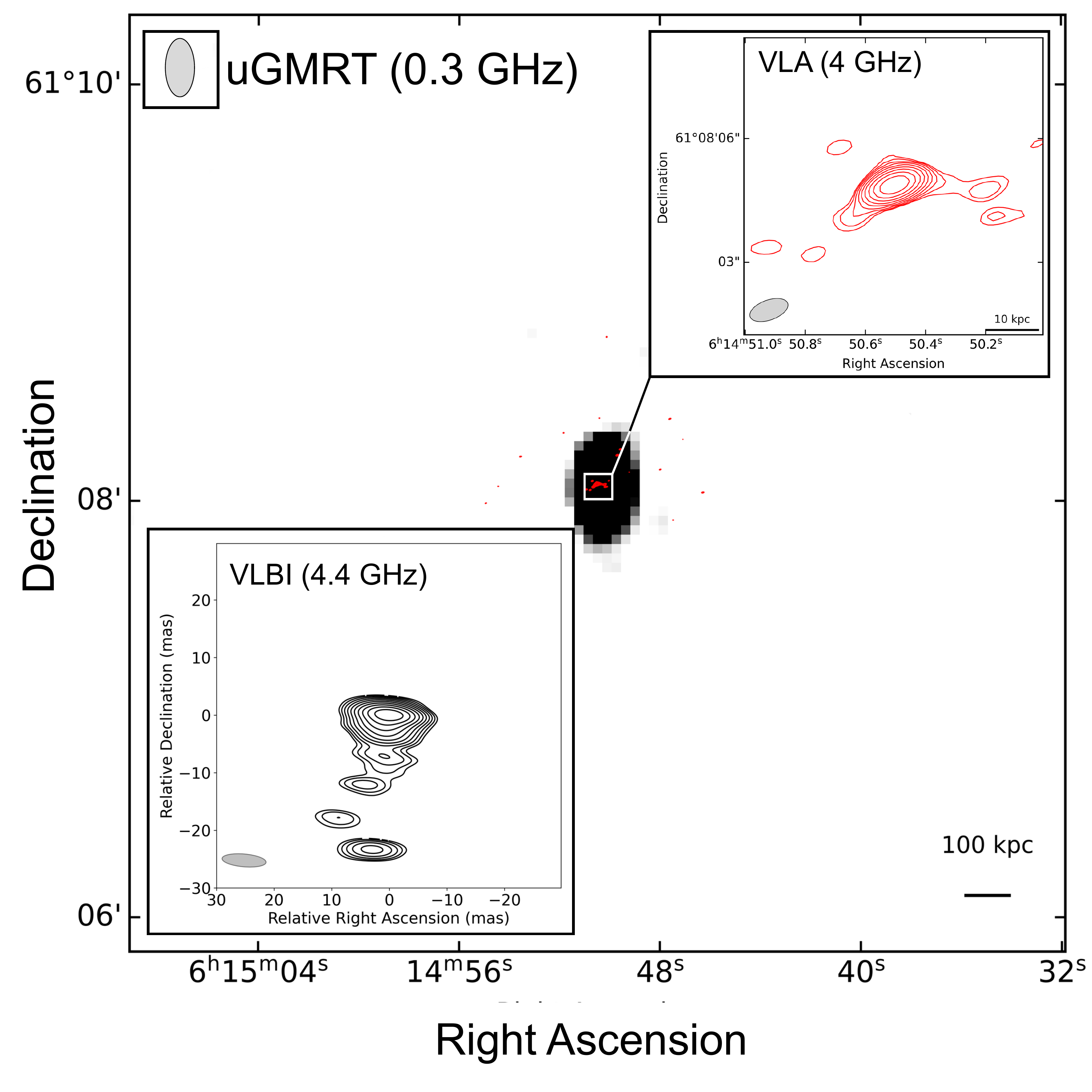}
        \caption{J0614+6108}
        \label{fig:image1}
    \end{subfigure}
    \hfill
    \begin{subfigure}{0.32\textwidth}
        \centering
        \includegraphics[width=\textwidth]{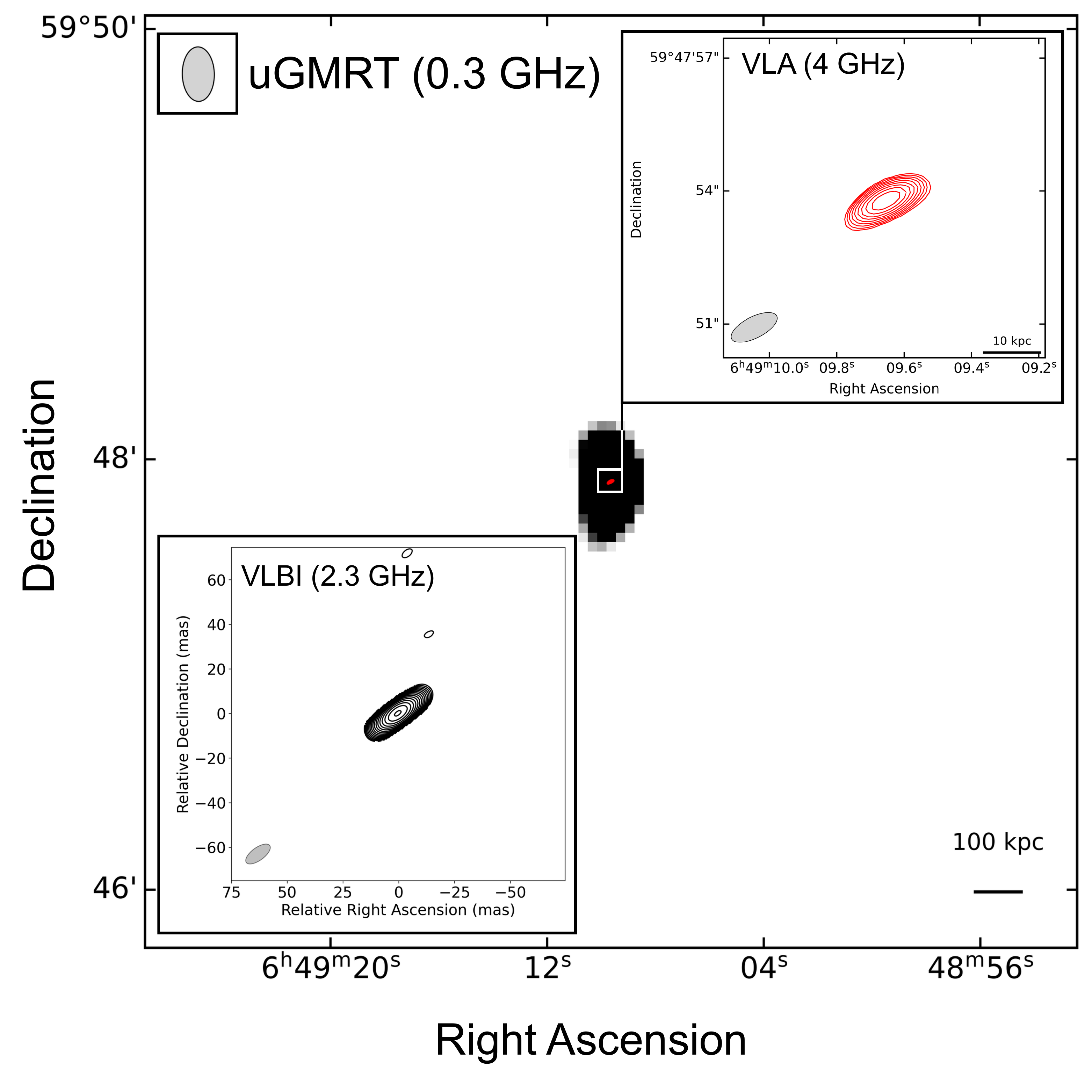}
        \caption{J0649+5947}
        \label{fig:image2}
    \end{subfigure}
    \hfill
    \begin{subfigure}{0.32\textwidth}
        \centering
        \includegraphics[width=\textwidth]{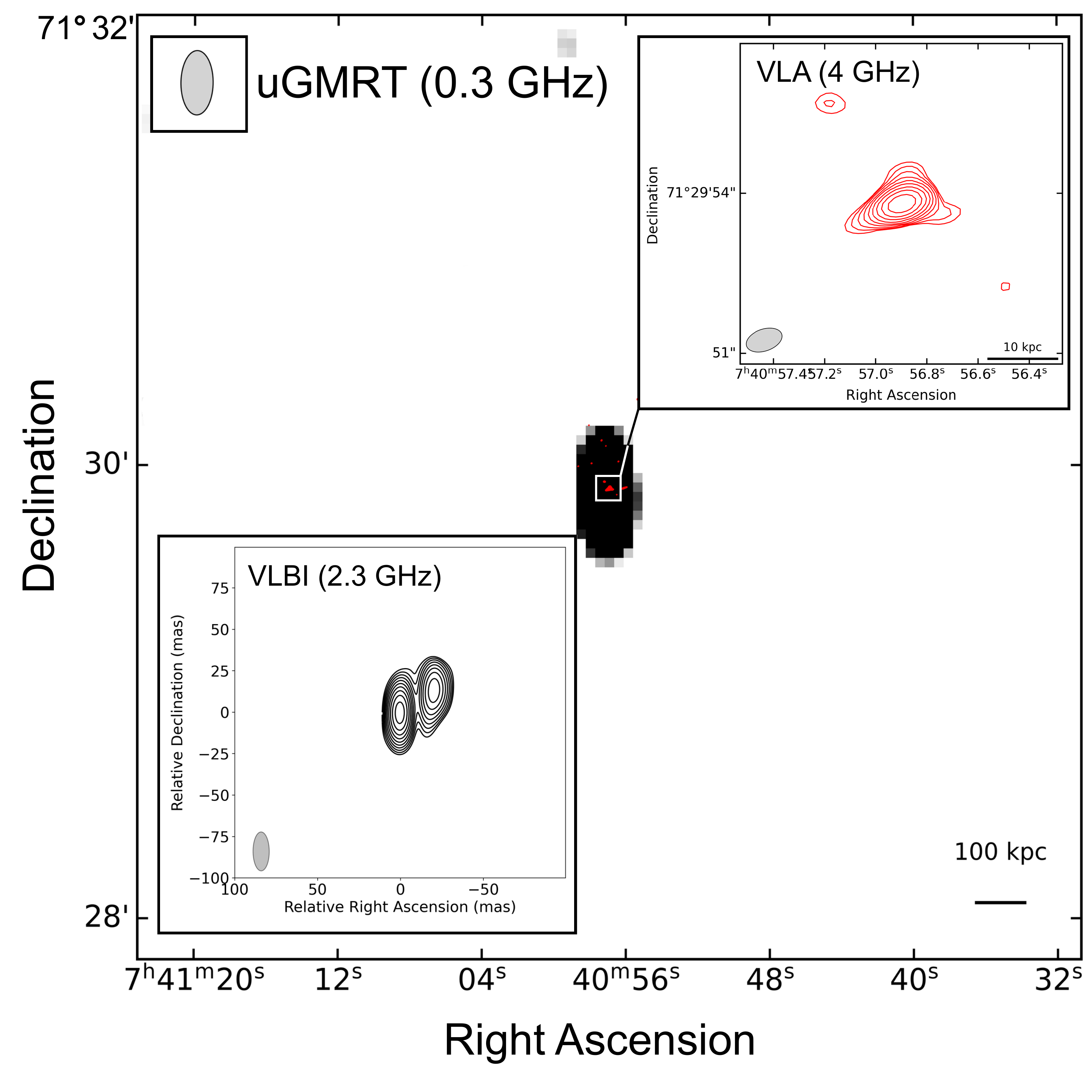}
        \caption{J0740+7129}
        \label{fig:image3}
    \end{subfigure}

    \hfill

    \begin{subfigure}{0.32\textwidth}
        \centering
        \includegraphics[width=\textwidth]{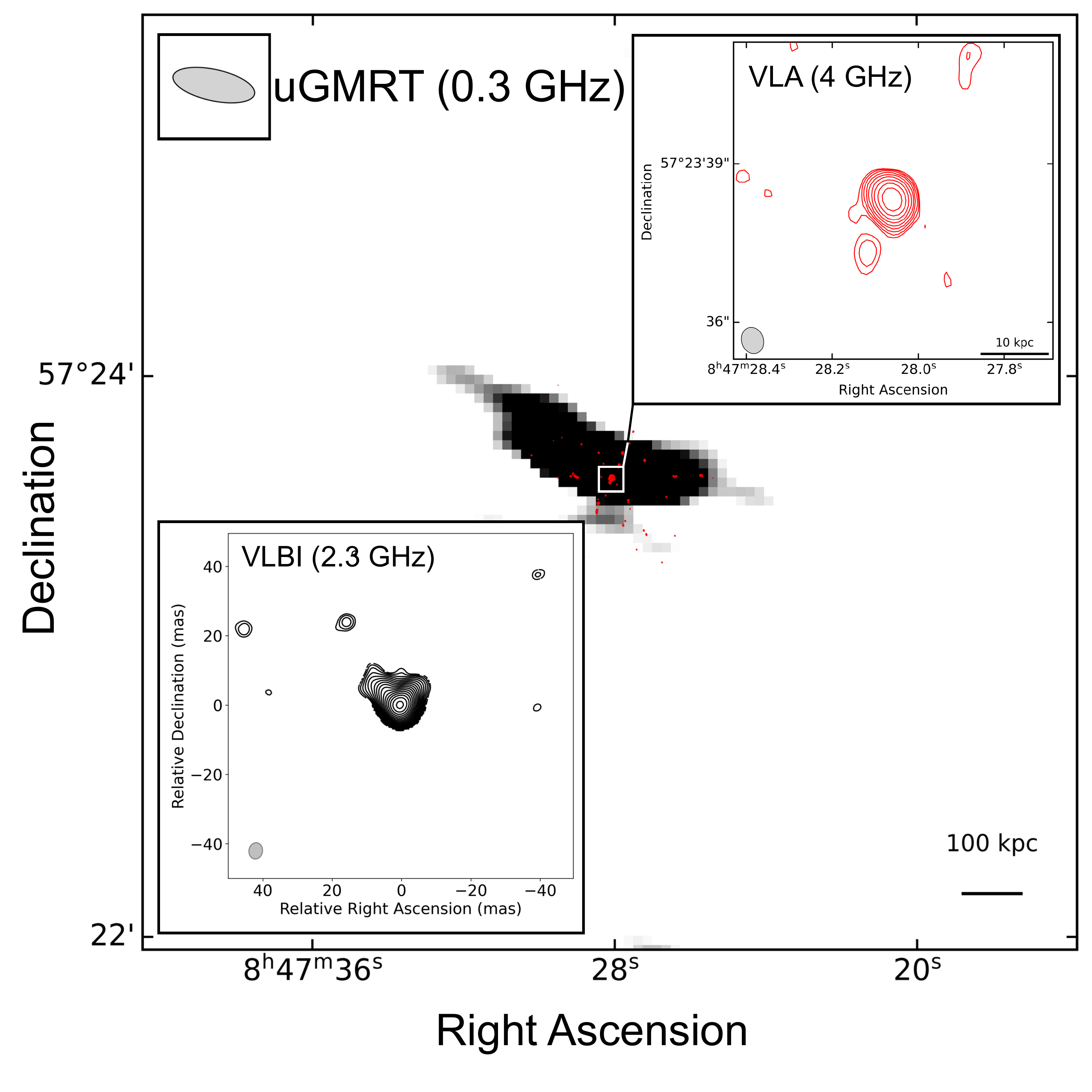}
        \caption{J0847+5723}
        \label{fig:image4}
    \end{subfigure}
    \hfill
    \begin{subfigure}{0.32\textwidth}
        \centering
        \includegraphics[width=\textwidth]{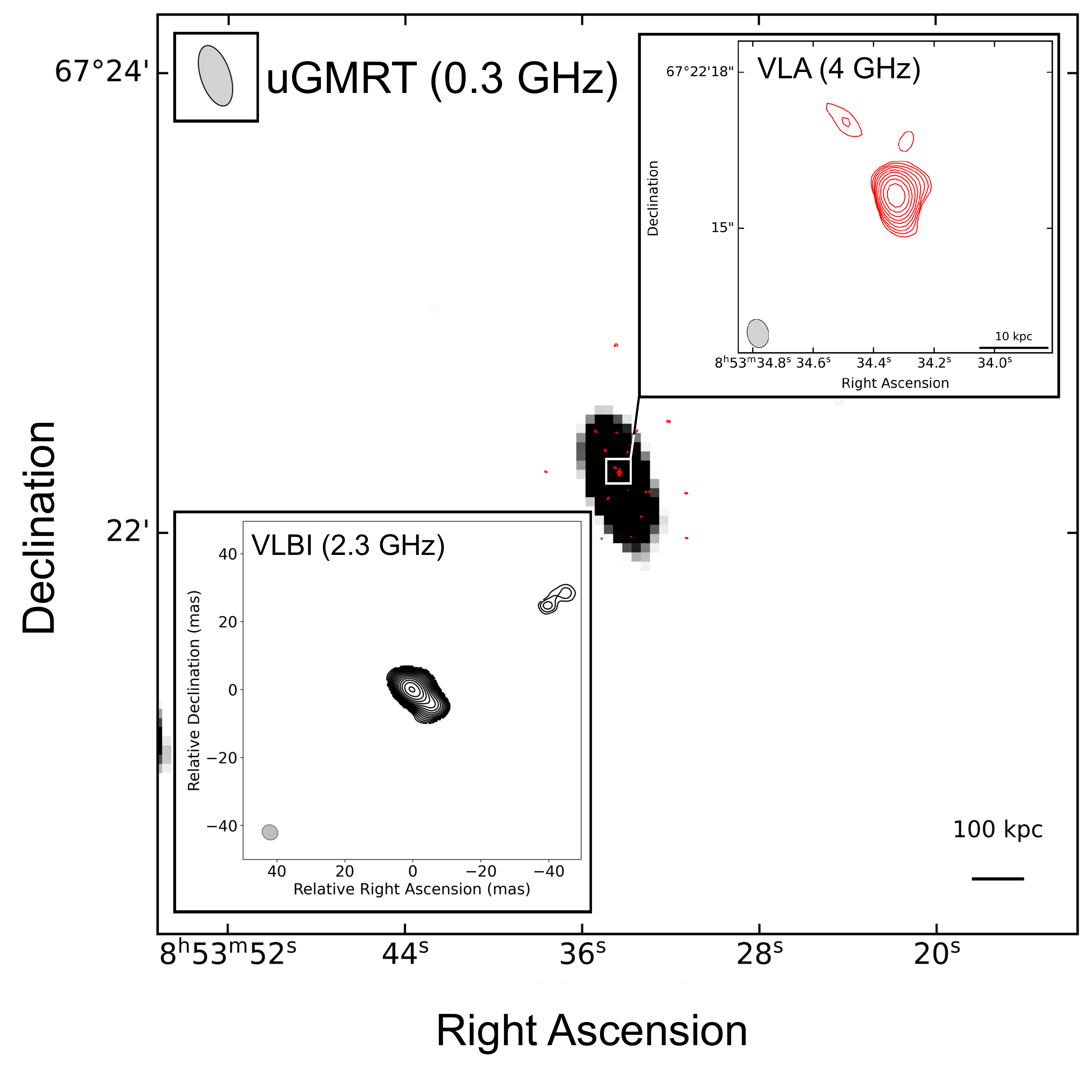}
        \caption{J0853+6722}
        \label{fig:image5}
    \end{subfigure}
    \hfill
    \begin{subfigure}{0.32\textwidth}
        \centering
        \includegraphics[width=\textwidth]{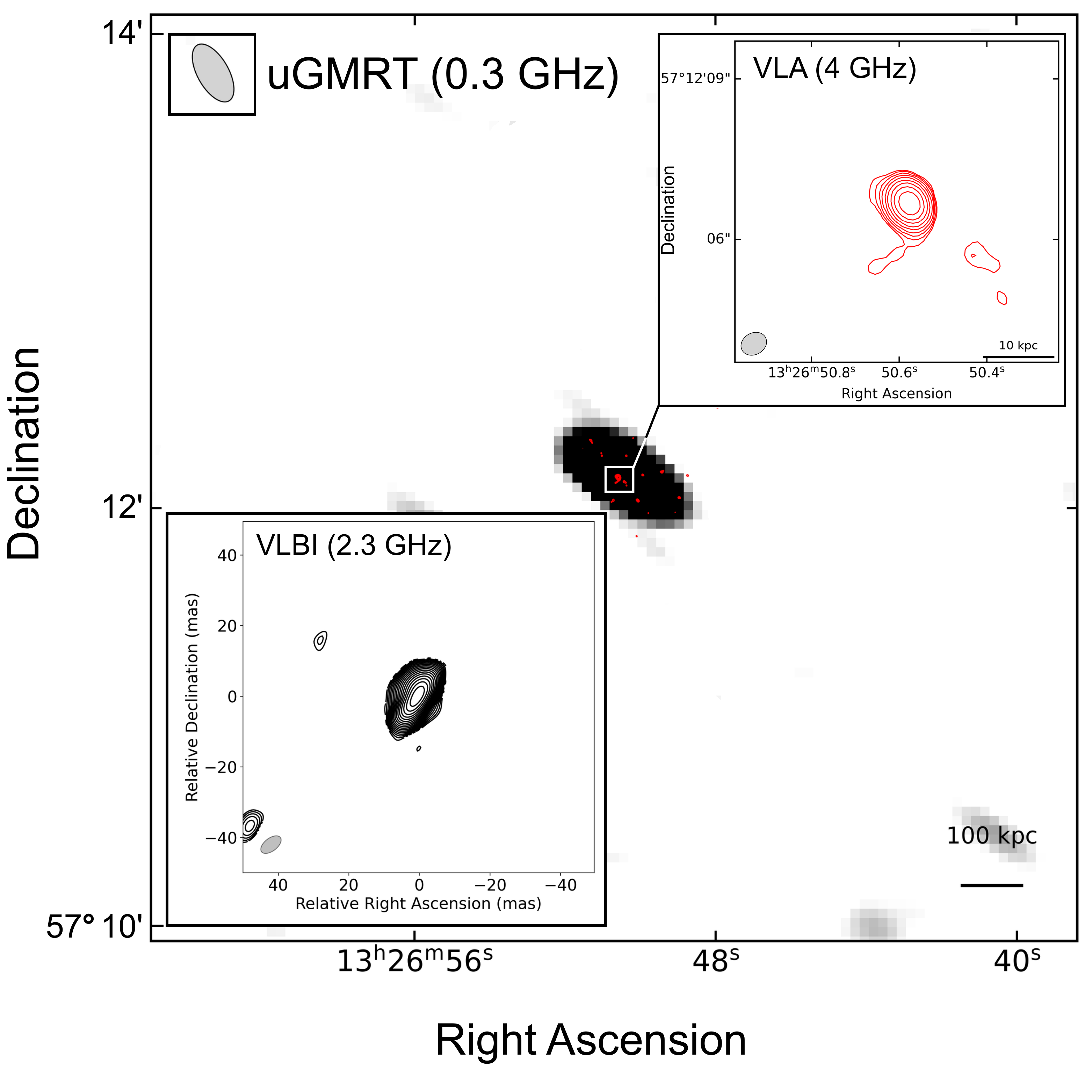}
        \caption{J1326+5712}
        \label{fig:image6}
    \end{subfigure}
        \begin{subfigure}{0.32\textwidth}
        \centering
        \includegraphics[width=\textwidth]{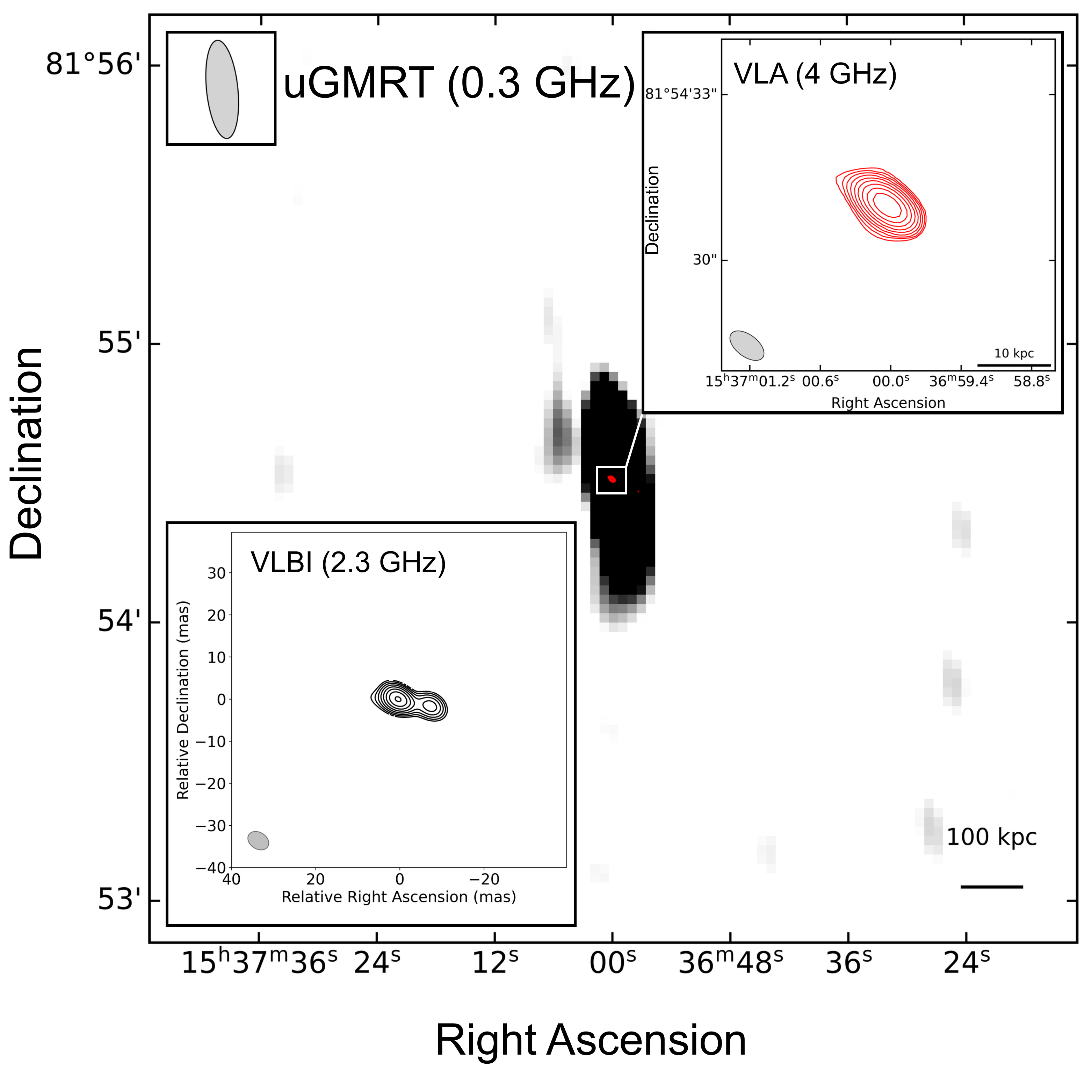}
        \caption{J1536+8154}
        \label{fig:image7}
    \end{subfigure}
    \hfill
    \begin{subfigure}{0.32\textwidth}
        \centering
        \includegraphics[width=\textwidth]{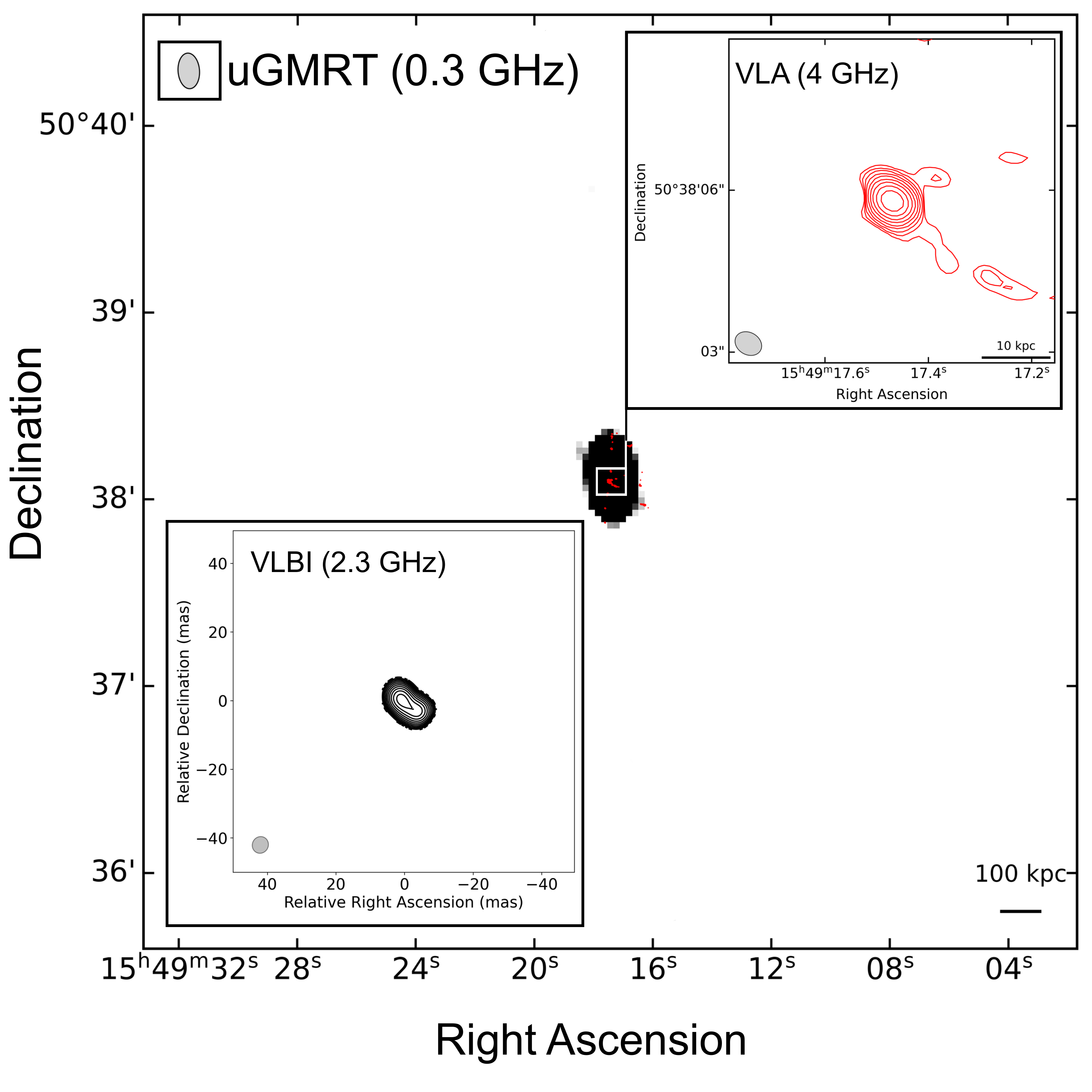}
        \caption{J1549+5038}
        \label{fig:image8}
    \end{subfigure}
    \hfill
    \begin{subfigure}{0.32\textwidth}
        \centering
        \includegraphics[width=\textwidth]{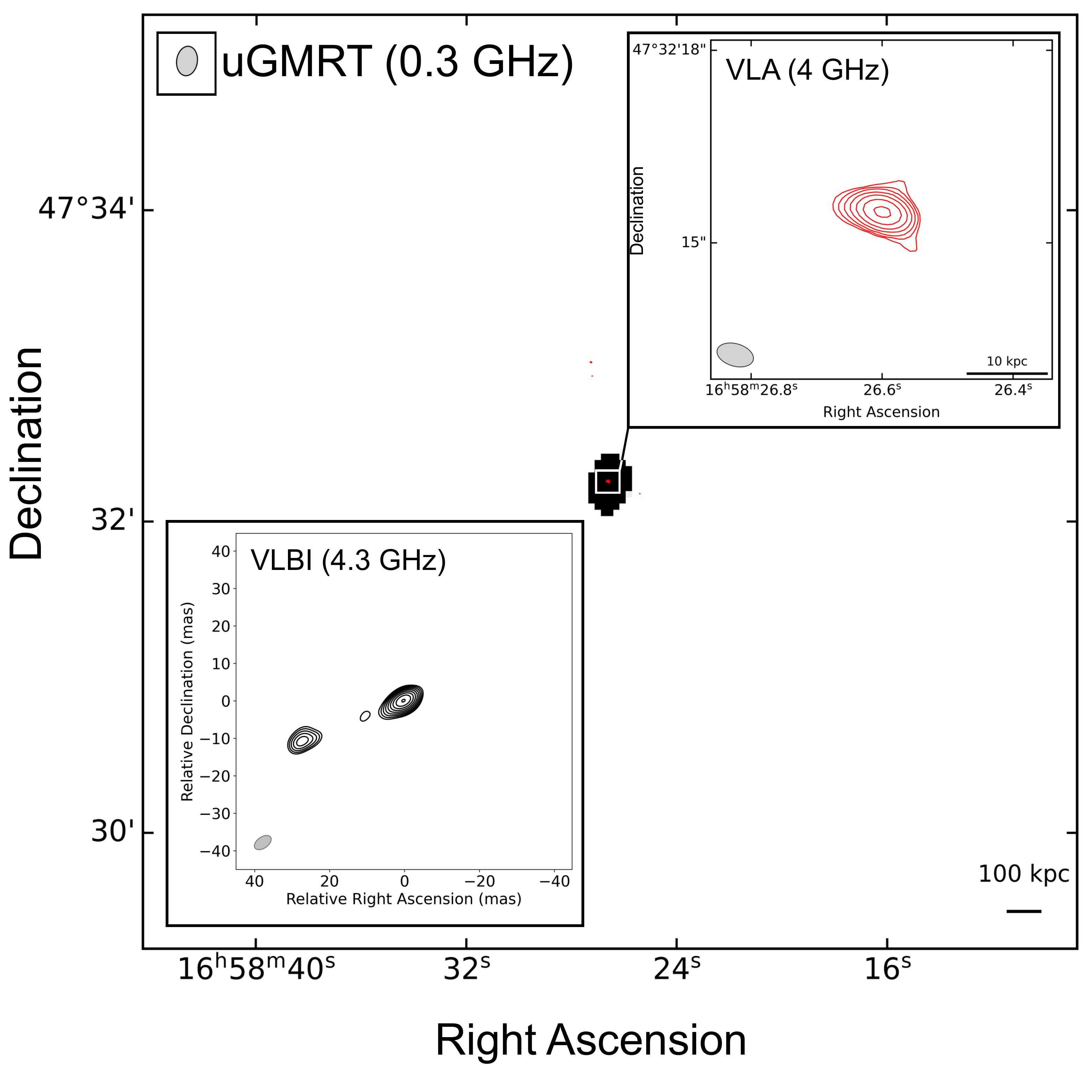}
        \caption{J1658+4732}
        \label{fig:image9}
    \end{subfigure}
    \caption{Composite radio images of the remaining nine EISERs. The raster image is uGMRT Band-3 (0.3 GHz) and VLA C-band (4 GHz) contours are overlaid. The VLBI contour maps are plotted in the inset. All contours are plotted starting at the respective $5\sigma$ and subsequent levels increase by factor of $\sqrt2$.}
\label{fig:RoR}
\end{figure*}

\section{Flux densities, radio properties, and SEDs of the EISERS sample} \label{app:sed}
This appendix presents supplementary observational results related to variability, flux-density validation, and broadband spectral properties of the EISERS sample. The tables provide flux-density measurements from different observing epochs and surveys, while the figures show the corresponding broadband radio spectra and SSA+FFA model fit discussed in the main text. \\ \\
The tables included here correspond to the source J0858+7501 as a representative example; the equivalent tables for the remaining nine EISERS are available as online supplementary material.

\setlength{\tabcolsep}{4pt}
\begin{longtable}{lccccc}
\caption{Flux densities of the ten brightest unresolved sources in the J0858+5701 field used to assess possible variability and consistency between observing epochs. (A sample of the table is given here. The complete table is available in electronic form). } 
\label{table:variability}\\
\hline
Source & Band & Epoch & $\nu$ (GHz) & Flux (mJy), $S_\nu$ & $\Delta\,f_\nu$ (\%) \\
\hline
\hline \addlinespace
J0903+7518 & 03 & 01 & 0.322 & 252.0$\pm$37.8 & 8.6$\pm$23.1 \\
           &    & 02 & 0.322 & 273.7$\pm$41.1 &  \\

J0906+7437 & 03 & 01 & 0.322 & 197.5$\pm$29.6 & 0.71$\pm$21.0 \\
           &    & 02 & 0.322 & 196.1$\pm$29.4 &  \\

J0854+7423 & 03 & 01 & 0.322 & 114.8$\pm$17.2 & 0.52$\pm$21.1 \\
           &    & 02 & 0.322 & 114.2$\pm$17.1 &  \\
\hline
\end{longtable}
\tablefoot{$\Delta\,f_\nu = |\Delta\,S_\nu|/\langle S_\nu \rangle$, is the degree of variability and expressed in percentage, where $|\Delta\,S_\nu|$ is the absolute difference of the flux densities measured between the two epochs, and $\langle S_\nu\rangle$ is the mean. For both EISERS and the field sources, we find a typical $\Delta\,f_\nu \approx 15\%$.\\
}

\begin{longtable}{@{}lcccccc@{}}

\caption{Flux-density validation through comparison of the uGMRT and VLA measurements (in mJy) with standard radio surveys, including LoTSS \citep{Shimwell2022,Shimwell2026}, FIRST \citep{Becker1995}, and NVSS \citep{Condon1998}. (A sample of the table is given here. The complete table is available in electronic form).}
\label{tab:data-validation-I}
\\

\hline\hline
\multicolumn{7}{c}{Part A: EISERS sample}\\
\hline

Source & \multicolumn{3}{c}{0.145 GHz} & \multicolumn{3}{c}{1.4 GHz} \\
\cmidrule(lr){2-4} \cmidrule(lr){5-7}

 & \multicolumn{2}{c}{uGMRT} & LoTSS & VLA & FIRST/ & uGMRT \\
 & Epoch\,1 & Epoch\,2 &  &  & NVSS & e1 \\
 &  &   &   &   & $(\star/\dagger)$ & \\

\midrule
\endfirsthead

\multicolumn{7}{c}{Table \thetable\ - continued from the previous page}\\
\hline
\hline
\endhead

J0614+6108 & -- &
\begin{tabular}[t]{@{}c@{}}28.3\\$\pm$4.2\end{tabular} &
\begin{tabular}[t]{@{}c@{}}24.0\\$\pm$2.4\end{tabular} &
\begin{tabular}[t]{@{}c@{}}254.5\\$\pm$12.7\end{tabular} &
\begin{tabular}[t]{@{}c@{}}259.3$^\dagger$\\$\pm$7.8\end{tabular} &
-- \\

J0649+5947 & -- &
\begin{tabular}[t]{@{}c@{}}28.0\\$\pm$4.2\end{tabular} &
\begin{tabular}[t]{@{}c@{}}17.0\\$\pm$1.7\end{tabular} &
\begin{tabular}[t]{@{}c@{}}126.0\\$\pm$6.3\end{tabular} &
\begin{tabular}[t]{@{}c@{}}128.3$^\dagger$\\$\pm$4.6\end{tabular} &
-- \\

J0740+7129 & -- &
\begin{tabular}[t]{@{}c@{}}8.0\\$\pm$0.8\end{tabular} &
\begin{tabular}[t]{@{}c@{}}14.0\\$\pm$1.4\end{tabular} &
\begin{tabular}[t]{@{}c@{}}243.2\\$\pm$12.2\end{tabular} &
\begin{tabular}[t]{@{}c@{}}254.1$^\dagger$\\$\pm$7.6\end{tabular} &
-- \\

J0847+5723 & -- &
\begin{tabular}[t]{@{}c@{}}35.6\\$\pm$5.3\end{tabular} &
\begin{tabular}[t]{@{}c@{}}33.0\\$\pm$3.3\end{tabular} &
\begin{tabular}[t]{@{}c@{}}363.5\\$\pm$18.2\end{tabular} &
\begin{tabular}[t]{@{}c@{}}363.0$^\star$\\$\pm$36.3\end{tabular} &
\begin{tabular}[t]{@{}c@{}}349.0\\$\pm$52.4\end{tabular} \\

J0853+6722 & -- &
\begin{tabular}[t]{@{}c@{}}18.4\\$\pm$2.8\end{tabular} &
\begin{tabular}[t]{@{}c@{}}18.0\\$\pm$1.8\end{tabular} &
\begin{tabular}[t]{@{}c@{}}384.7\\$\pm$19.2\end{tabular} &
\begin{tabular}[t]{@{}c@{}}342.1$^\dagger$\\$\pm$10.3\end{tabular} &
-- \\

J0858+7501 & -- &
\begin{tabular}[t]{@{}c@{}}18.4\\$\pm$2.8\end{tabular} &
\begin{tabular}[t]{@{}c@{}}15.0\\$\pm$1.5\end{tabular} &
\begin{tabular}[t]{@{}c@{}}823.6\\$\pm$41.2\end{tabular} &
\begin{tabular}[t]{@{}c@{}}947.4$^\dagger$\\$\pm$28.4\end{tabular} &
-- \\

J1326+5712 &
\begin{tabular}[t]{@{}c@{}}13.6\\$\pm$2.0\end{tabular} &
\begin{tabular}[t]{@{}c@{}}9.1\\$\pm$1.4\end{tabular} &
\begin{tabular}[t]{@{}c@{}}13.9\\$\pm$1.4\end{tabular} &
\begin{tabular}[t]{@{}c@{}}532.9\\$\pm$26.6\end{tabular} &
\begin{tabular}[t]{@{}c@{}}528.5$^\star$\\$\pm$52.9\end{tabular} &
\begin{tabular}[t]{@{}c@{}}520.0\\$\pm$78.0\end{tabular} \\

J1536+8154 & -- &
\begin{tabular}[t]{@{}c@{}}42.7\\$\pm$6.4\end{tabular} &
\begin{tabular}[t]{@{}c@{}}21.8\\$\pm$2.2\end{tabular} &
\begin{tabular}[t]{@{}c@{}}422.6\\$\pm$21.1\end{tabular} &
\begin{tabular}[t]{@{}c@{}}432.3$^\dagger$\\$\pm$13.0\end{tabular} &
\begin{tabular}[t]{@{}c@{}}380.7\\$\pm$57.1\end{tabular} \\

J1549+5038 &
\begin{tabular}[t]{@{}c@{}}23.7\\$\pm$3.6\end{tabular} &
\begin{tabular}[t]{@{}c@{}}43.9\\$\pm$6.6\end{tabular} &
\begin{tabular}[t]{@{}c@{}}34.6\\$\pm$3.5\end{tabular} &
\begin{tabular}[t]{@{}c@{}}621.1\\$\pm$31.1\end{tabular} &
\begin{tabular}[t]{@{}c@{}}701.4$^\star$\\$\pm$70.1\end{tabular} &
-- \\

J1658+4732 &
\begin{tabular}[t]{@{}c@{}}23.5\\$\pm$3.5\end{tabular} &
-- &
\begin{tabular}[t]{@{}c@{}}11.1\\$\pm$1.1\end{tabular} &
\begin{tabular}[t]{@{}c@{}}267.4\\$\pm$13.4\end{tabular} &
\begin{tabular}[t]{@{}c@{}}312.4$^\dagger$\\$\pm$11.0\end{tabular} &
\begin{tabular}[t]{@{}c@{}}310.6\\$\pm$46.6\end{tabular} \\
\addlinespace
\hline

\multicolumn{7}{r}{Continued on next page} \\
\endfoot

\hline
\endlastfoot

\pagebreak

\hline
\multicolumn{7}{c}{Part B: Sources in the J0858+7501 field}\\
\hline

Source & \multicolumn{3}{c}{0.145 GHz} & \multicolumn{3}{c}{1.4 GHz} \\
\cmidrule(lr){2-4} \cmidrule(lr){5-7}

 & \multicolumn{2}{c}{uGMRT} & LoTSS & VLA & FIRST/ & uGMRT \\
 & Epoch\,1 & Epoch\,2 &  &  & NVSS & Epoch\,1 \\
 &  &   &   &   & $(\star/\dagger)$ & \\

\midrule

J0858+7510 &
\begin{tabular}[t]{@{}c@{}}--\\\end{tabular} &
\begin{tabular}[t]{@{}c@{}}--\\\end{tabular} &
\begin{tabular}[t]{@{}c@{}}27.2\\$\pm$2.7\end{tabular} &
\begin{tabular}[t]{@{}c@{}}12.4\\$\pm$1.2\end{tabular} &
\begin{tabular}[t]{@{}c@{}}13.8$^\dagger$\\$\pm$1.4\end{tabular} &
\begin{tabular}[t]{@{}c@{}}17.9\\$\pm$2.7\end{tabular} \\
\addlinespace

J0856+7507 &
\begin{tabular}[t]{@{}c@{}}--\\\end{tabular} &
\begin{tabular}[t]{@{}c@{}}--\\\end{tabular} &
\begin{tabular}[t]{@{}c@{}}10.9\\$\pm$1.1\end{tabular} &
\begin{tabular}[t]{@{}c@{}}2.9\\$\pm$0.3\end{tabular} &
\begin{tabular}[t]{@{}c@{}}3.6$^\dagger$\\$\pm$0.4\end{tabular} &
\begin{tabular}[t]{@{}c@{}}4.6\\$\pm$0.7\end{tabular} \\
\addlinespace

J0855+7506 &
\begin{tabular}[t]{@{}c@{}}--\\\end{tabular} &
\begin{tabular}[t]{@{}c@{}}--\\\end{tabular} &
\begin{tabular}[t]{@{}c@{}}4.5\\$\pm$0.4\end{tabular} &
\begin{tabular}[t]{@{}c@{}}9.9\\$\pm$1.0\end{tabular} &
\begin{tabular}[t]{@{}c@{}}--\\\end{tabular} &
\begin{tabular}[t]{@{}c@{}}10.1\\$\pm$1.5\end{tabular} \\
\addlinespace

J0859+7454 &
\begin{tabular}[t]{@{}c@{}}--\\\end{tabular} &
\begin{tabular}[t]{@{}c@{}}--\\\end{tabular} &
\begin{tabular}[t]{@{}c@{}}21.0\\$\pm$2.1\end{tabular} &
\begin{tabular}[t]{@{}c@{}}6.14\\$\pm$0.6\end{tabular} &
\begin{tabular}[t]{@{}c@{}}5.7$^\dagger$\\$\pm$0.6\end{tabular} &
\begin{tabular}[t]{@{}c@{}}5.6\\$\pm$0.8\end{tabular} \\
\addlinespace

J0856+7454 &
\begin{tabular}[t]{@{}c@{}}--\\\end{tabular} &
\begin{tabular}[t]{@{}c@{}}--\\\end{tabular} &
\begin{tabular}[t]{@{}c@{}}18.0\\$\pm$1.8\end{tabular} &
\begin{tabular}[t]{@{}c@{}}11.1\\$\pm$1.1\end{tabular} &
\begin{tabular}[t]{@{}c@{}}11.2$^\dagger$\\$\pm$1.1\end{tabular} &
\begin{tabular}[t]{@{}c@{}}10.6\\$\pm$1.6\end{tabular} \\
\addlinespace
\hline

\end{longtable}
\tablefoot{The uGMRT measurements correspond to observations taken at two different times labelled as epoch-1 and epoch-2 and described in Sect.~\ref{sec:ugmrt}. Part~A compares the EISERS flux densities with LoTSS at 0.145\,GHz and with FIRST/NVSS at 1.4\,GHz, while Part~B presents the same comparison for bright unresolved sources in the J0858+7501 field.}

\begin{longtable}{ccccccccc}
\caption{Flux densities for the source J0858+7501 measured at different observing frequencies using uGMRT (epoch-1 and 2) and VLA. (A sample of the table is given here. The complete table is available in electronic form).}
\label{tab:fluxJ0858}
\\

\hline
\hline

Source & \multicolumn{5}{c}{uGMRT} & \multicolumn{3}{c}{VLA} \\
\cmidrule(rl){2-6} \cmidrule(rl){7-9}

 & Band & Epoch & $\nu$ & Flux,$S_\nu$ & $\Delta\,f_\nu$ & Band & $\nu$ & Flux \\
 &       &       & (GHz) & (mJy) & (\%) &      & (GHz) & (mJy) \\

\hline
\endfirsthead

\addlinespace
J0858+7501 & 2 & 2 & 0.149 & 18.4 $\pm$ 2.8 &  & L & 1.039 & 890.6 $\pm$ 89.1 \\
 &  & 2 & 0.198 & 39.6 $\pm$ 5.9 &  &  & 1.104 & 925.0 $\pm$ 92.5 \\
 &  & 2 & 0.215 & 50.9 $\pm$ 7.6 &  &  & 1.168 & 900.0 $\pm$ 90.0 \\
 &  & 2 & 0.231 & 78.0 $\pm$ 11.7 &  &  & 1.360 & 859.4 $\pm$ 85.9 \\
 & 3 & 1 & 0.316 & 166.1 $\pm$ 24.9 & 0.0 $\pm$ 21.2 &  & 1.680 & 776.2 $\pm$ 77.6 \\
 &  & 2 & 0.316 & 166.0 $\pm$ 24.9 &  & S & 2.051 & 632.6 $\pm$ 63.3 \\ \rule{0pt}{12pt}
 &  & 1 & 0.322 & 145.0 $\pm$ 21.8 &  &  & 2.563 & 513.2 $\pm$ 51.3 \\
 &  & 1 & 0.348 & 232.1 $\pm$ 34.8 & 2.3 $\pm$ 21.7 & C & 4.039 & 272.7 $\pm$ 27.3 \\
 &  & 2 & 0.349 & 237.4 $\pm$ 35.6 &  &  & 6.039 & 141.7 $\pm$ 14.2 \\ \rule{0pt}{12pt}
 &  & 1 & 0.385 & 288.8 $\pm$ 43.3 & 13.1 $\pm$ 24.0 &  & 7.800 & 67.3 $\pm$ 6.7 \\
 &  & 2 & 0.385 & 326.8 $\pm$ 49.0 &  &  &  &  \\ \rule{0pt}{12pt}
 &  & 1 & 0.417 & 451.0 $\pm$ 67.6 & 18.5 $\pm$ 17.3 &  &  &  \\
 &  & 2 & 0.417 & 367.3 $\pm$ 55.1 &  &  &  &  \\ \rule{0pt}{12pt}
 &  & 1 & 0.450 & 427.0 $\pm$ 64.0 & 2.4 $\pm$ 21.7 &  &  &  \\
 &  & 2 & 0.450 & 437.2 $\pm$ 65.6 &  &  &  &  \\ \rule{0pt}{12pt}
 &  & 1 & 0.482 & 514.2 $\pm$ 77.1 & 12.8 $\pm$ 23.9 &  &  &  \\
 &  & 2 & 0.480 & 580.2 $\pm$ 87.0 &  &  &  &  \\ \rule{0pt}{12pt}
 & 4 & 2 & 0.579 & 577.6 $\pm$ 86.6 &  &  &  &  \\
 &  & 2 & 0.626 & 633.3 $\pm$ 95.0 &  &  &  &  \\
 &  & 2 & 0.675 & 684.5 $\pm$ 102.7 &  &  &  &  \\
 &  & 2 & 0.725 & 731.2 $\pm$ 109.7 &  &  &  &  \\
 & 5 & 1 & 1.095 & 1041.0 $\pm$ 156.2 &  &  &  &  \\
 &  & 1 & 1.157 & 1003.3 $\pm$ 150.5 &  &  &  &  \\
 &  & 1 & 1.230 & 935.5 $\pm$ 140.3 &  &  &  &  \\
 &  & 1 & 1.299 & 859.1 $\pm$ 128.9 &  &  &  &  \\
 &  & 1 & 1.360 & 913.4 $\pm$ 137.0 &  &  &  &  \\
 &  & 1 & 1.392 & 851.8 $\pm$ 127.8 &  &  &  &  \\
 &  & 1 & 1.423 & 832.1 $\pm$ 124.8 &  &  &  &  \\ \addlinespace
\hline

\end{longtable}
\tablefoot{$\Delta\,f_\nu = |\Delta\,S_\nu|/\langle S_\nu \rangle$, is the degree of variability and expressed in percentage, where $|\Delta\,S_\nu|$ is the absolute difference of the flux densities measured between the two epochs, and $\langle S_\nu \rangle$ is the mean. For both EISERS and the field sources, we find a typical $\Delta\,f_\nu \approx 15\%$.}

\section{Radio SED Modelling of the Remaining EISERS: SSA, SSA+FFA, and FFA+SSA}
\label{appendix:radiosed}
In this appendix, we present the radio spectral energy distributions (SEDs) of the remaining nine EISERs, together with the corresponding SSA, SSA+FFA, and FFA+SSA model fits. The best-fit parameters of the SSA+FFA model, listed in Table \ref{tab:best_fit_parameters}, are used to derive the FFA optical depth, magnetic field strength, emission measure, and electron number density ($n_{\rm e}$), which are presented in Table~\ref{tab:derived_parameters}.

\begin{figure}[htp!]
\centering

\includegraphics[trim={0cm 0cm 0cm 0cm},clip,width=0.32\columnwidth]{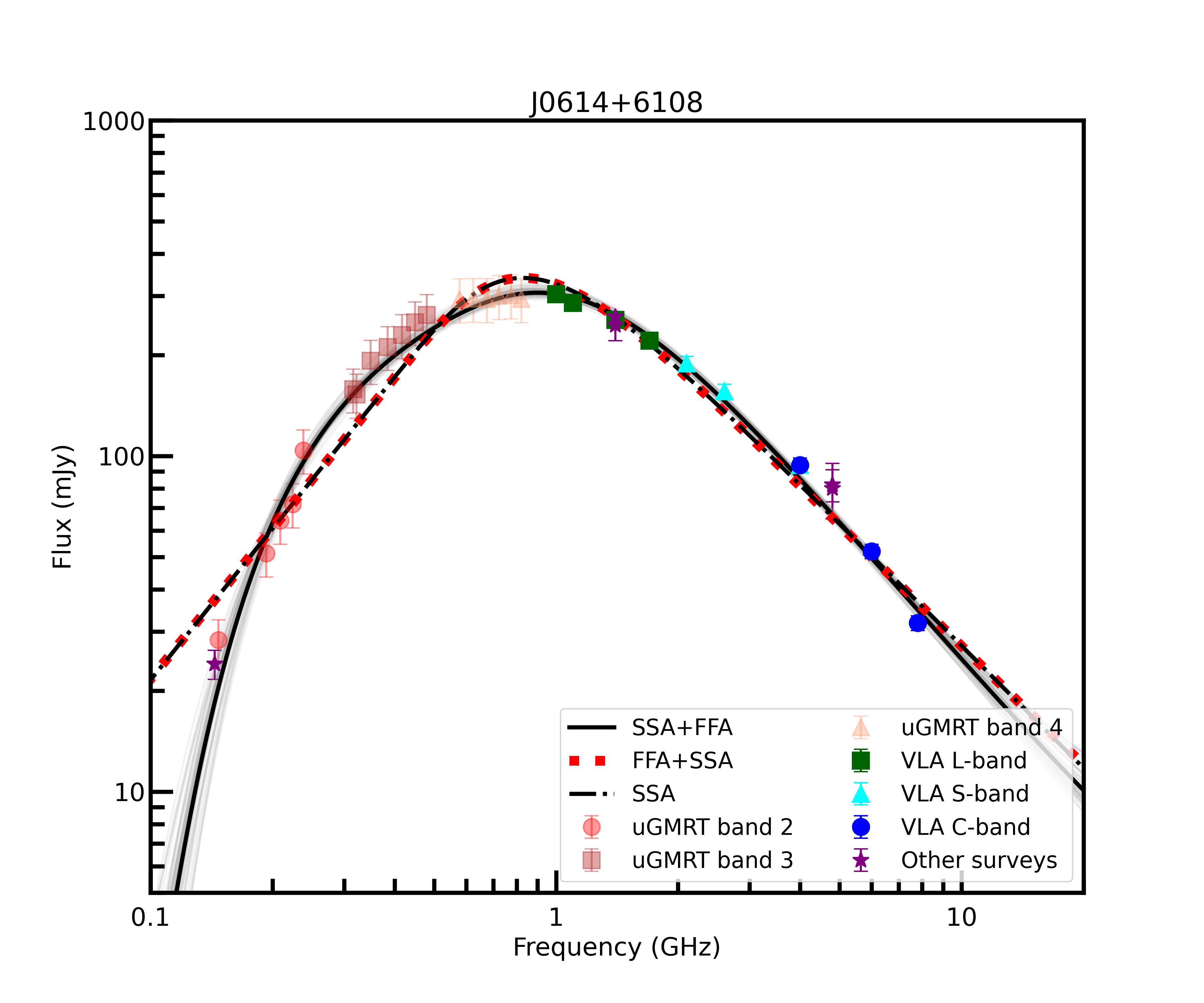}
\includegraphics[trim={0cm 0cm 0cm 0cm},clip,width=0.32\columnwidth]{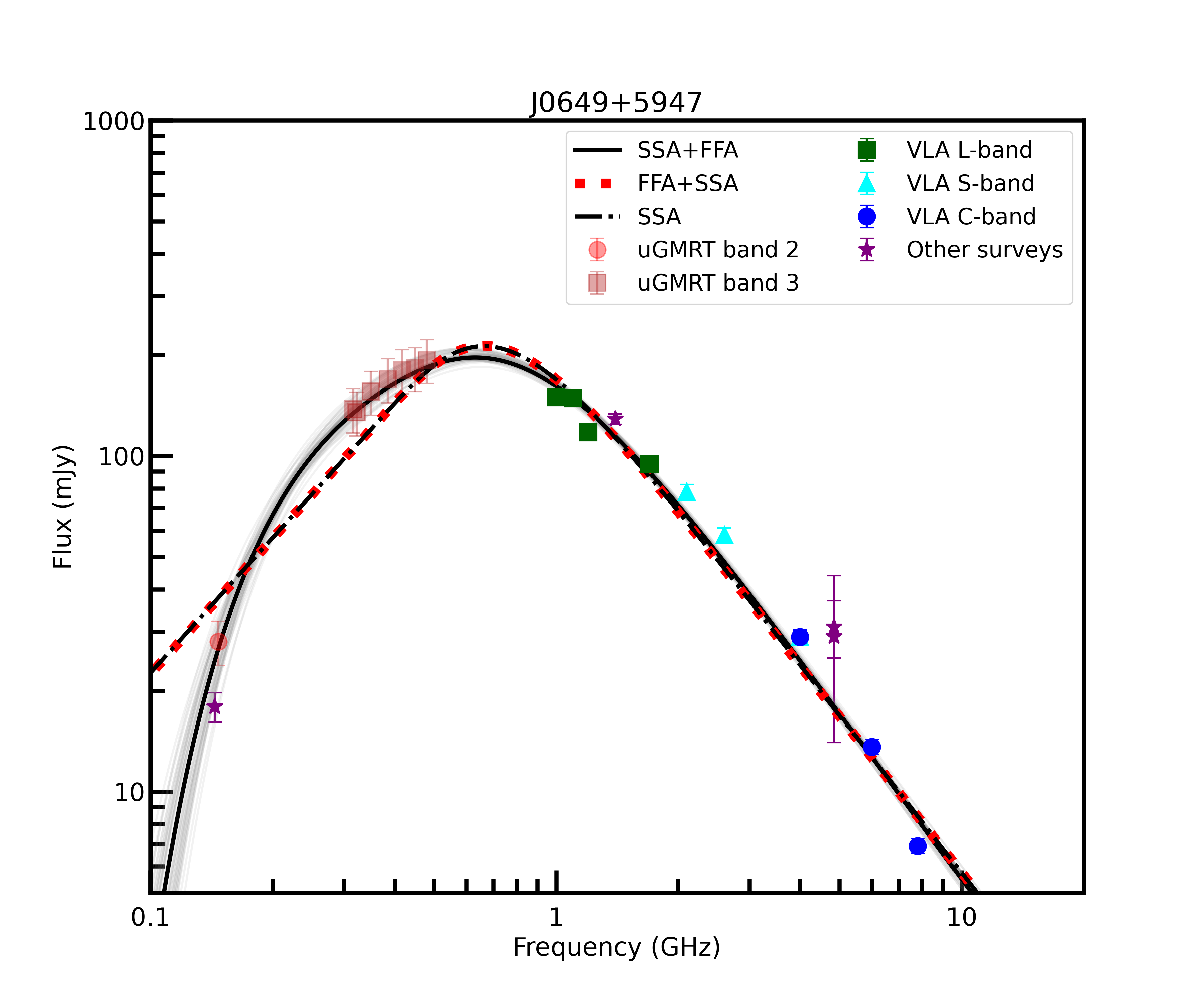}
\includegraphics[trim={0cm 0cm 0cm 0cm},clip,width=0.32\columnwidth]{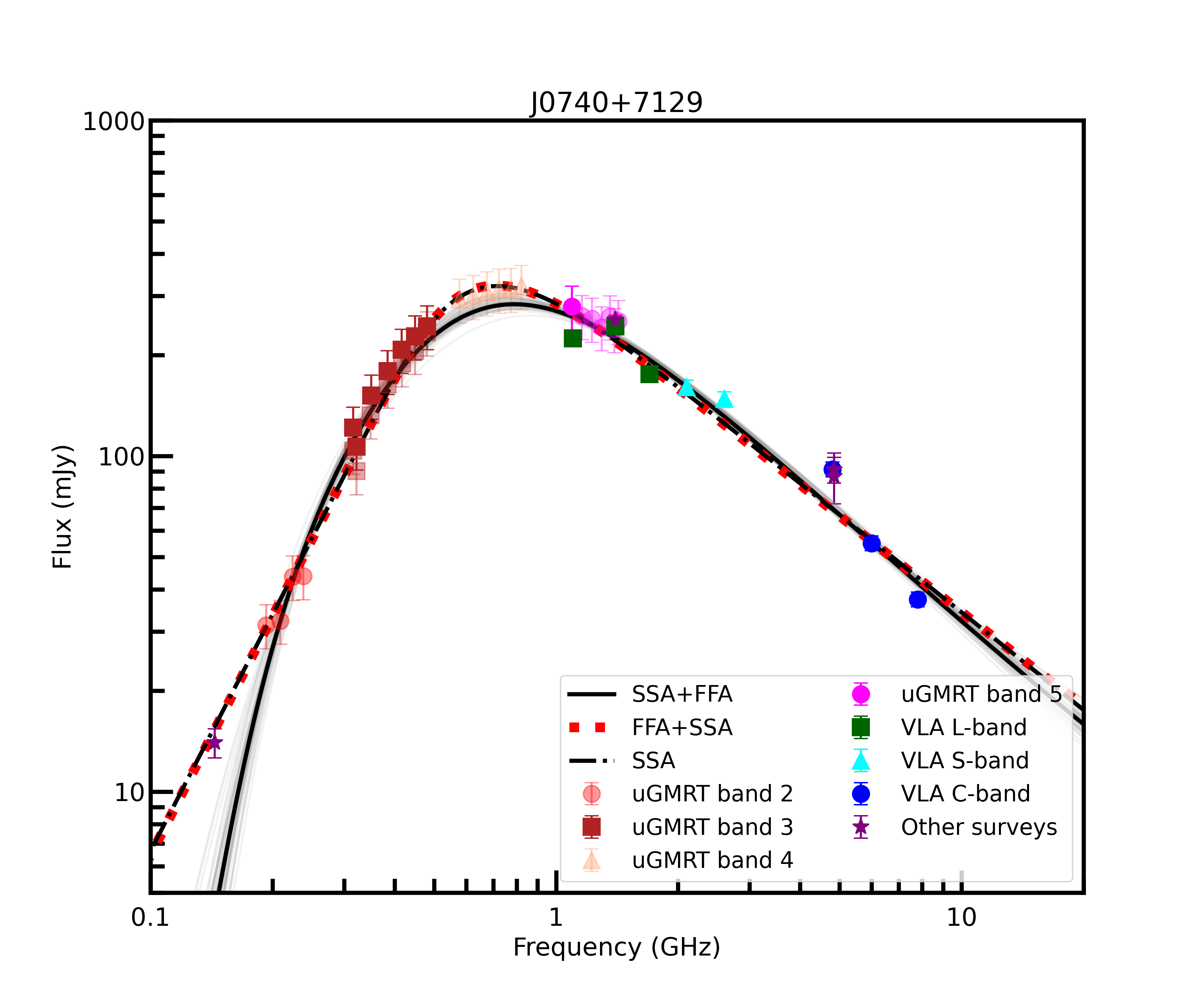}

\vspace{0.3cm}

\includegraphics[trim={0cm 0cm 0cm 0cm},clip,width=0.32\textwidth]
{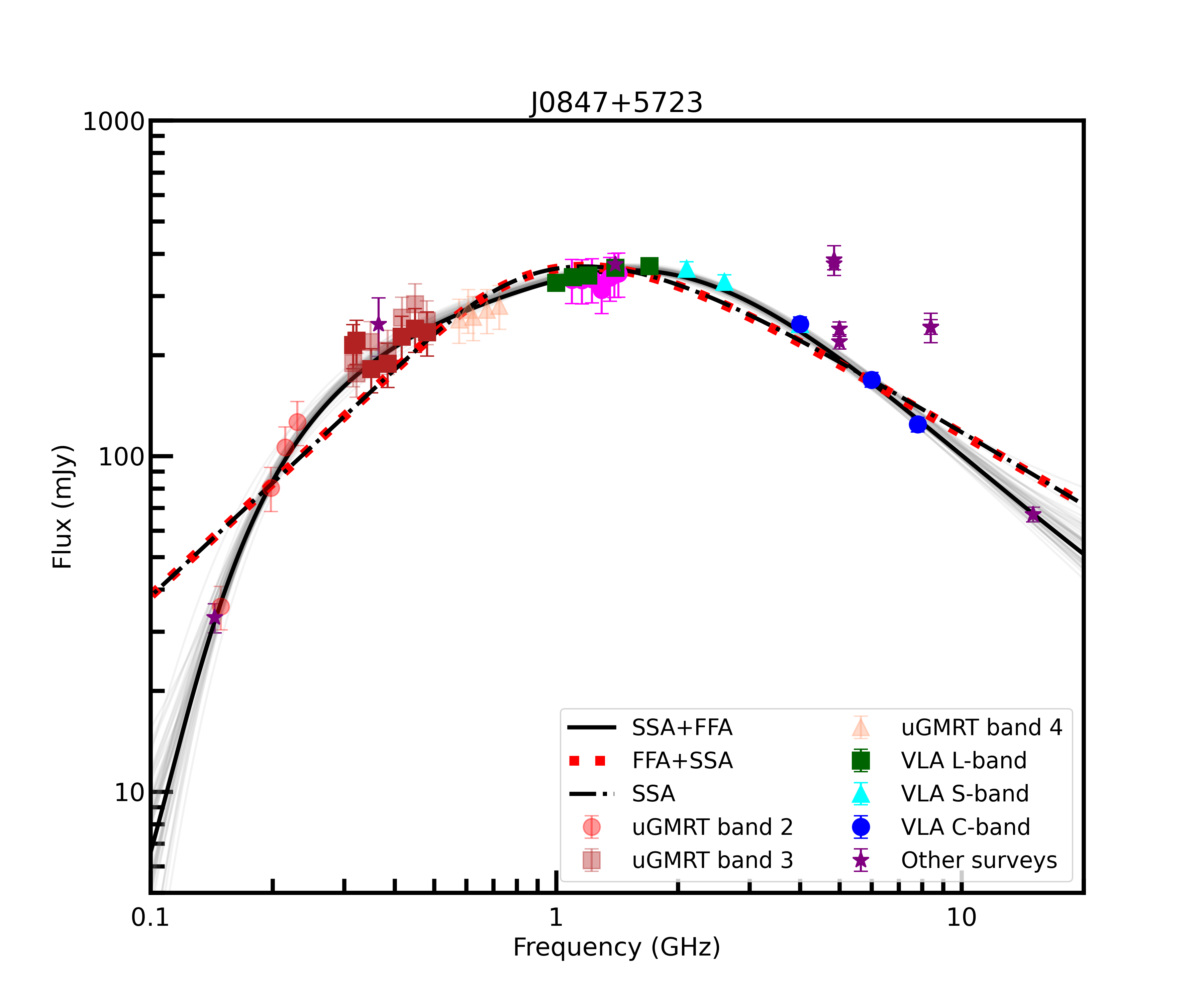}
\includegraphics[trim={0cm 0cm 0cm 0cm},clip,width=0.32\textwidth]
{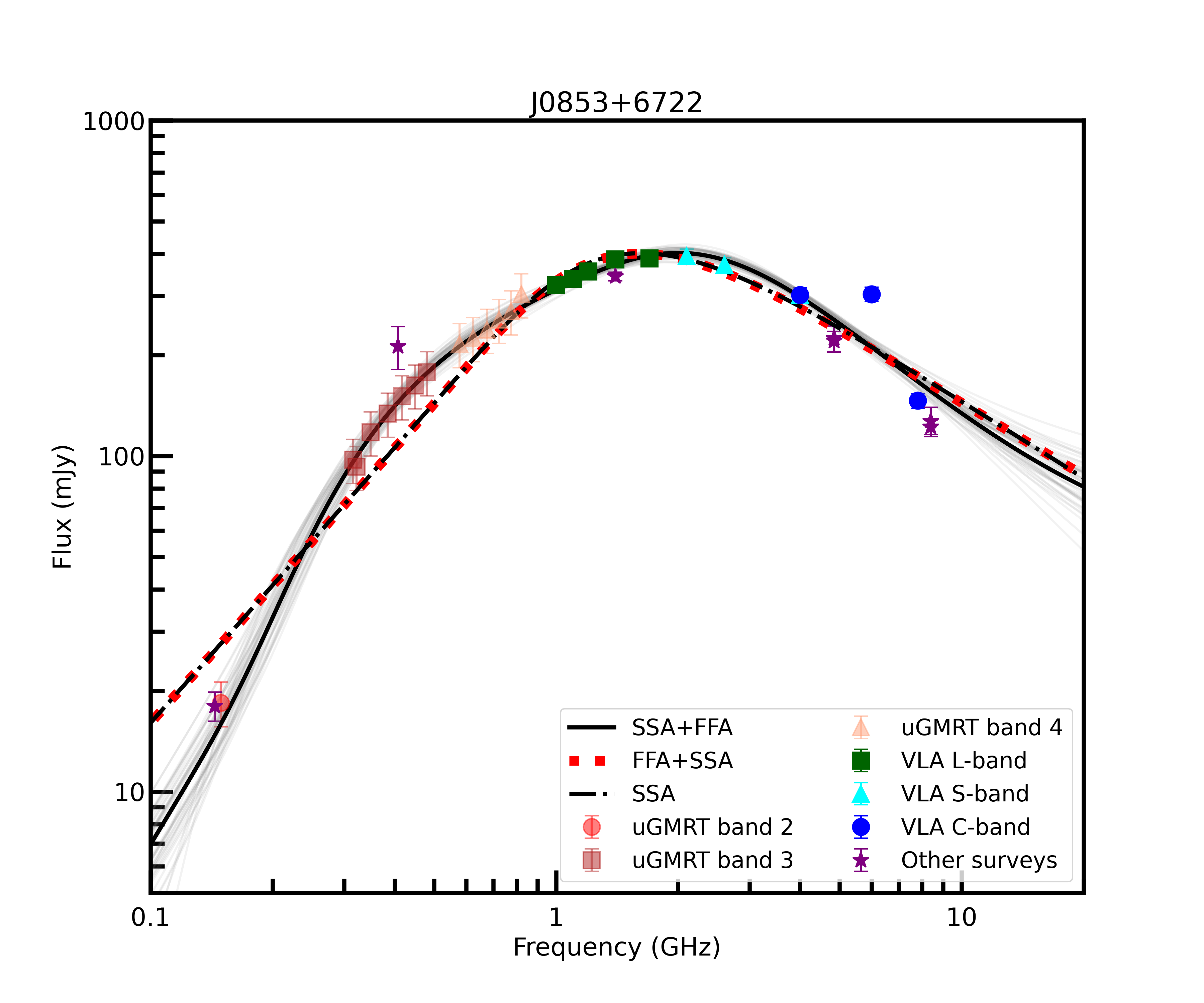}
\includegraphics[trim={0cm 0cm 0cm 0cm},clip,width=0.32\textwidth]
{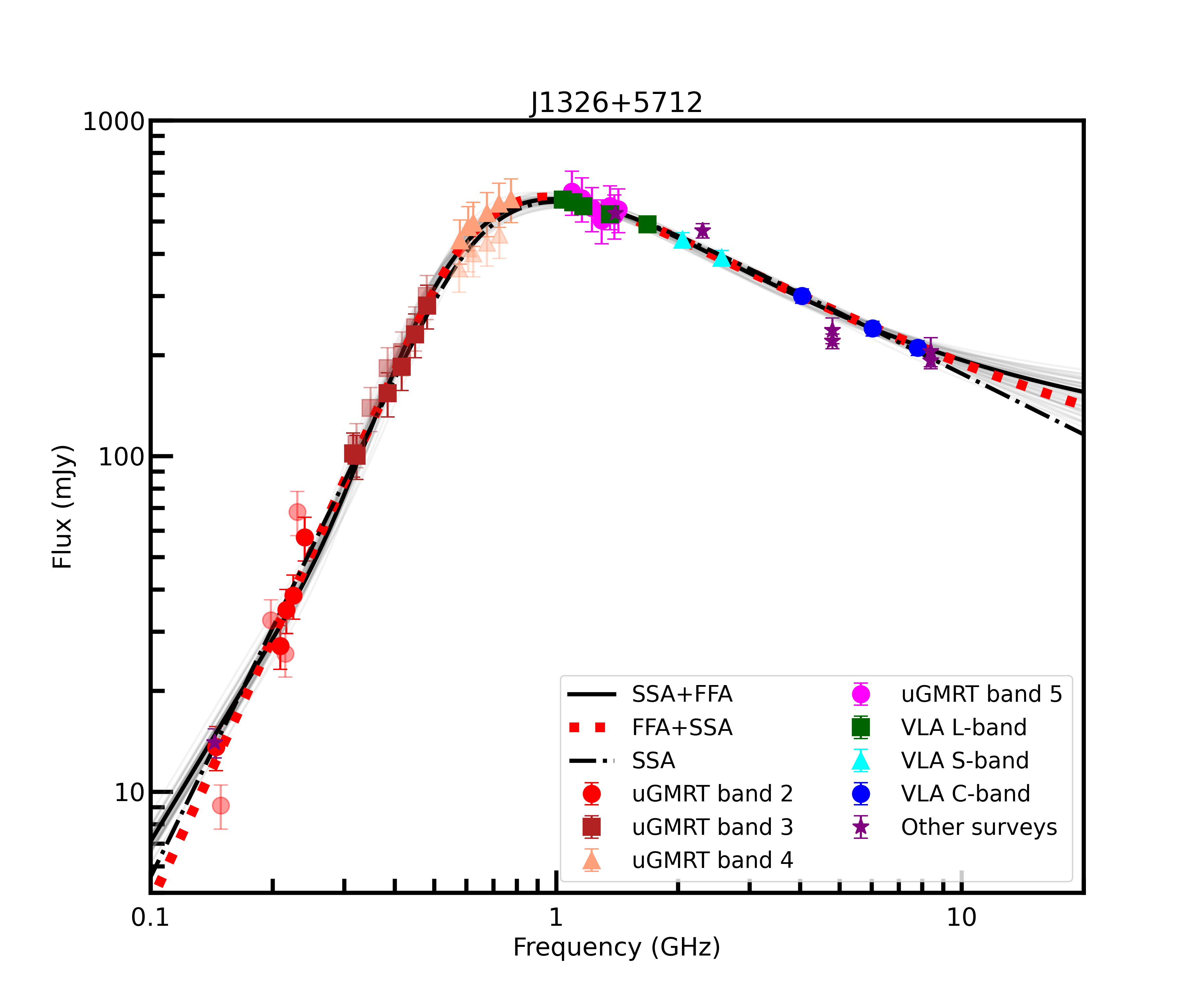}

\vspace{0.3cm}

\includegraphics[trim={0cm 0cm 0cm 0cm},clip,width=0.32\textwidth]
{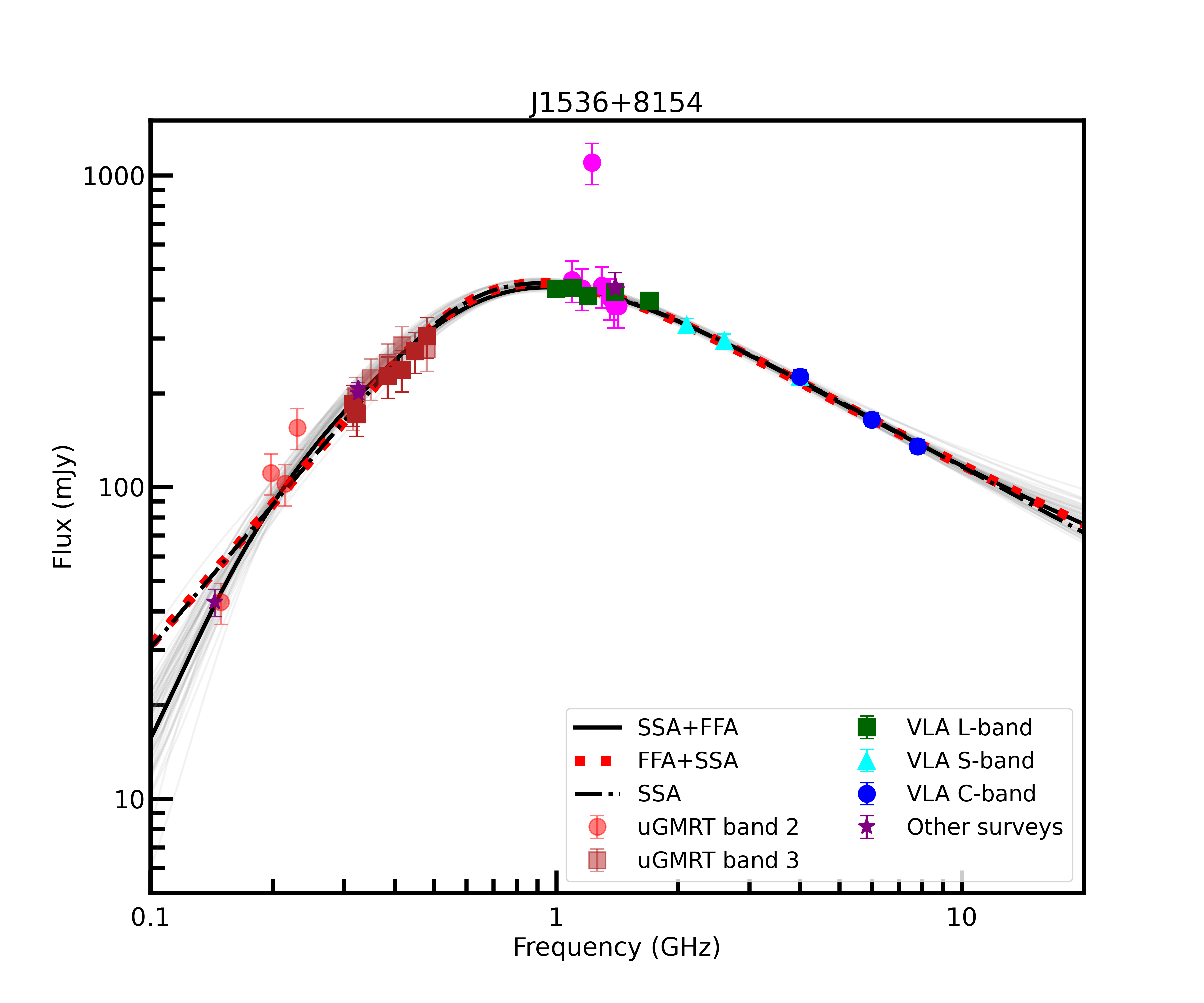}
\includegraphics[trim={0cm 0cm 0cm 0cm},clip,width=0.32\textwidth]
{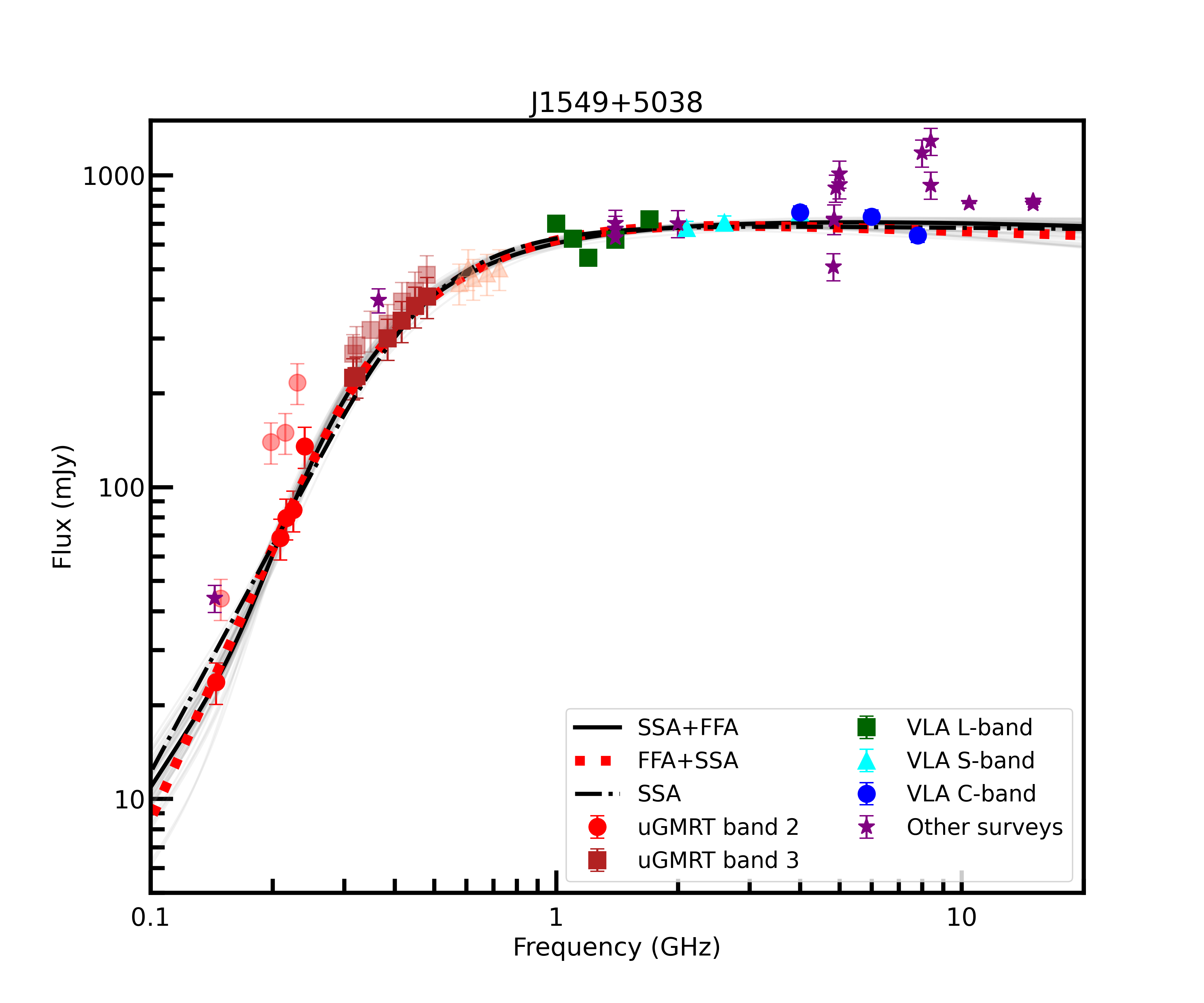}
\includegraphics[trim={0cm 0cm 0cm 0cm},clip,width=0.32\textwidth]
{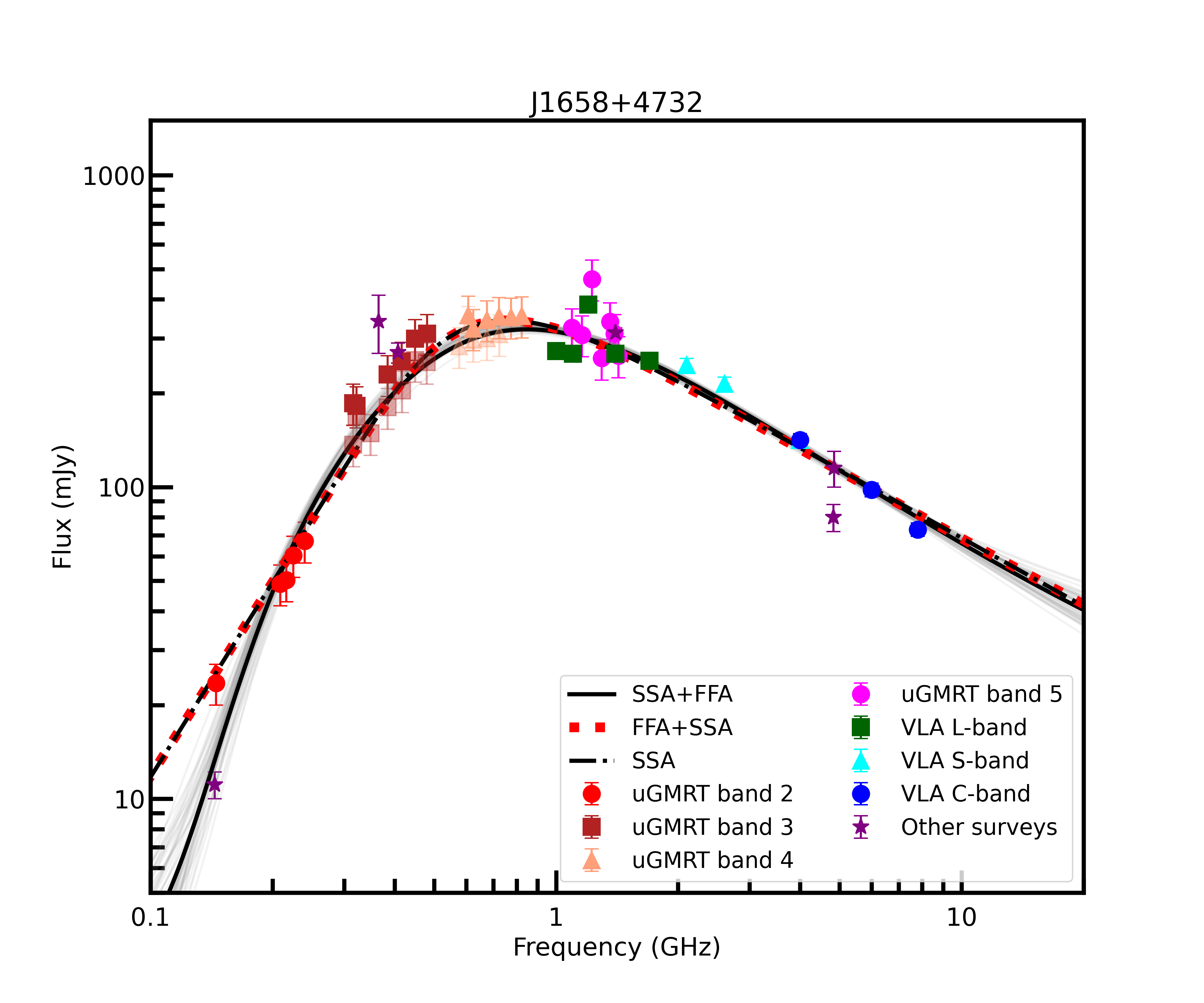}

\caption{SED of the remaining nine EISERS constructed using radio observations from uGMRT and VLA. The different frequency bands covered by the uGMRT and VLA are marked as different colours. The other available radio surveys (eg., LoTSS, FIRST, NVSS, WENSS etc.) are shown under the legend \lq other surveys', which are not included for the fitting. The solid and shaded uGMRT datapoints correspond to Epochs\,1 and 2, respectively. The solid, dotted and dash-dotted lines represent the best fit SSA+FFA, FFA+SSA and pure SSA emission models discussed in Section~\ref{sec:modelling}.}
\label{fig:SED-EISERS}
\end{figure}

\clearpage
\section{WISE SEDs and optical images of the EISERS}
\label{appendix:wisesed}

This appendix presents the optical images and infrared spectral energy distributions (WISE SEDs) of the EISERS sample used to investigate their host-galaxy properties and optical/infrared counterparts discussed in Sect.~\ref{sec:optical_SED}.

\begin{center}
\includegraphics[width=4.5cm]{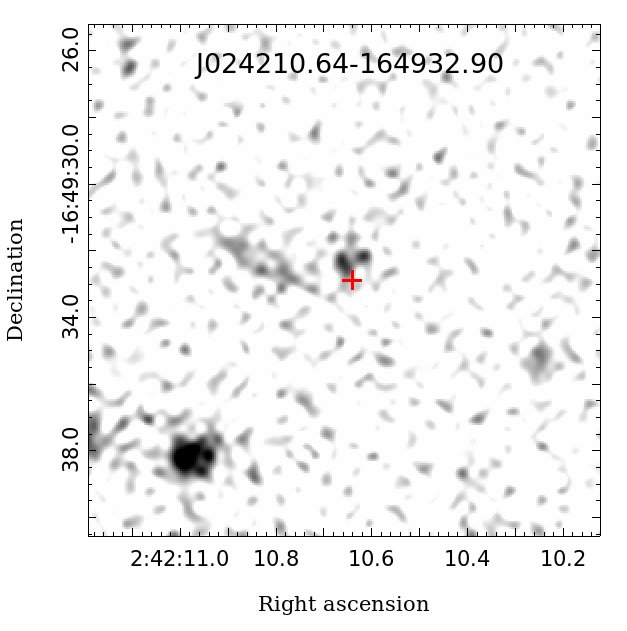}
\hspace{0.1cm}
\includegraphics[width=4.5cm]{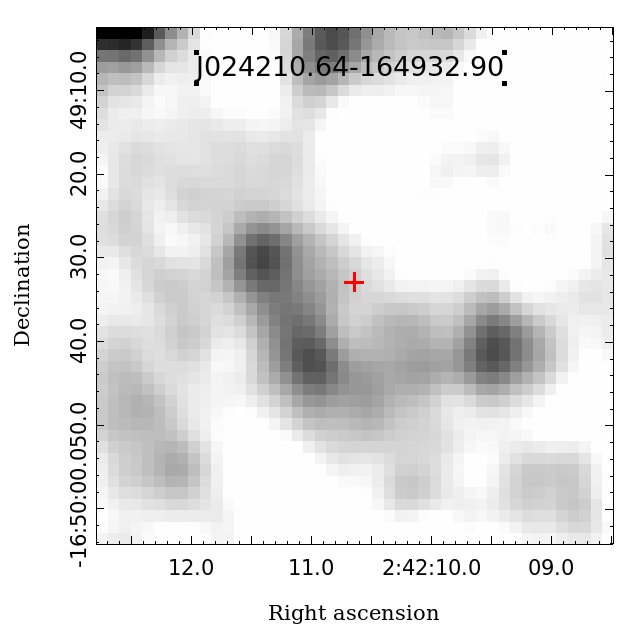}
\hspace{0.1cm}
\includegraphics[width=7.0cm]{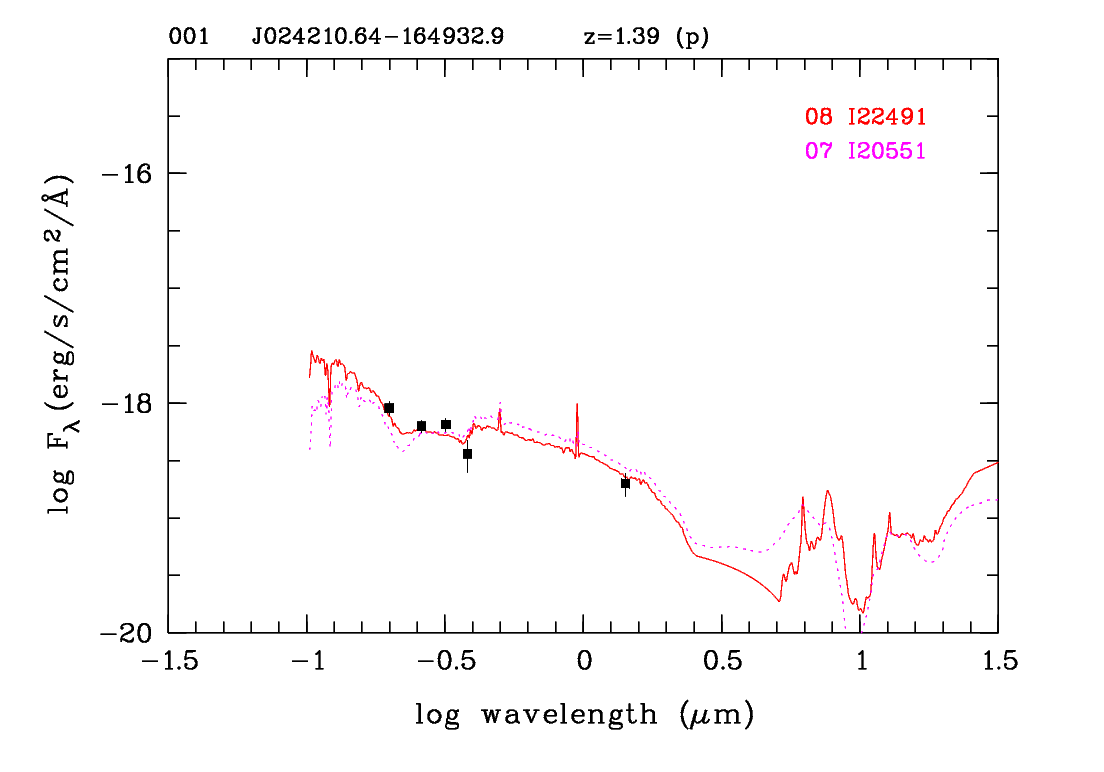}
\end{center}

\begin{center}
\includegraphics[width=4.5cm]{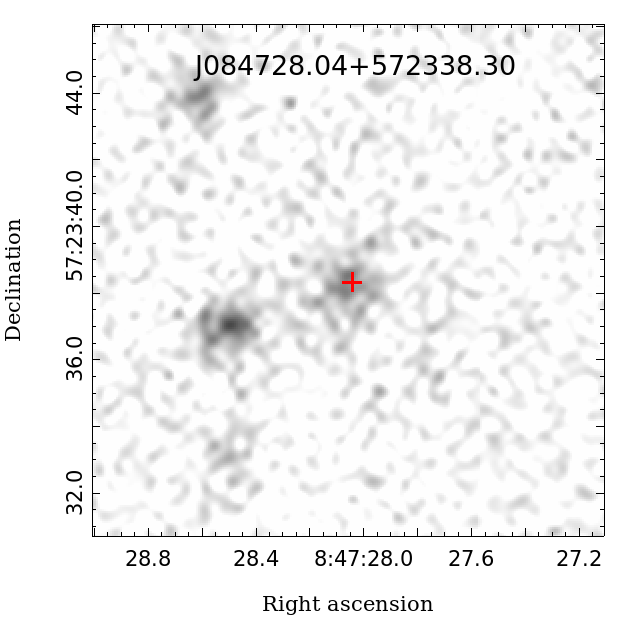}
\hspace{0.1cm}
\includegraphics[width=4.5cm]{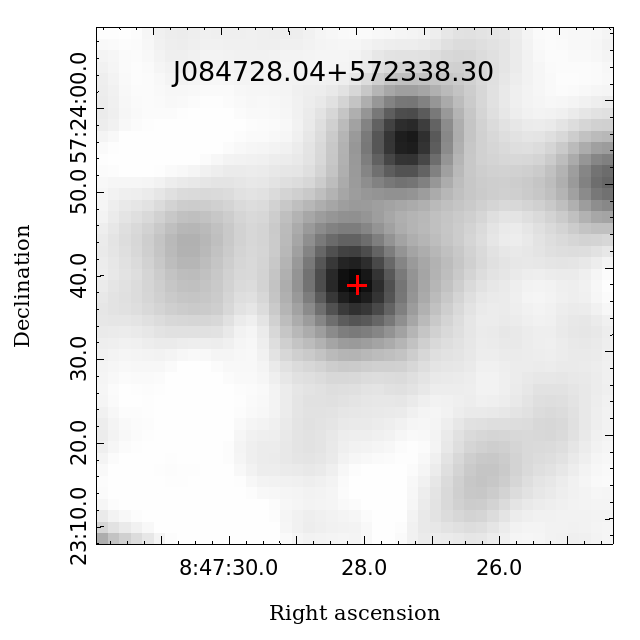}
\hspace{0.1cm}
\includegraphics[width=7.0cm]{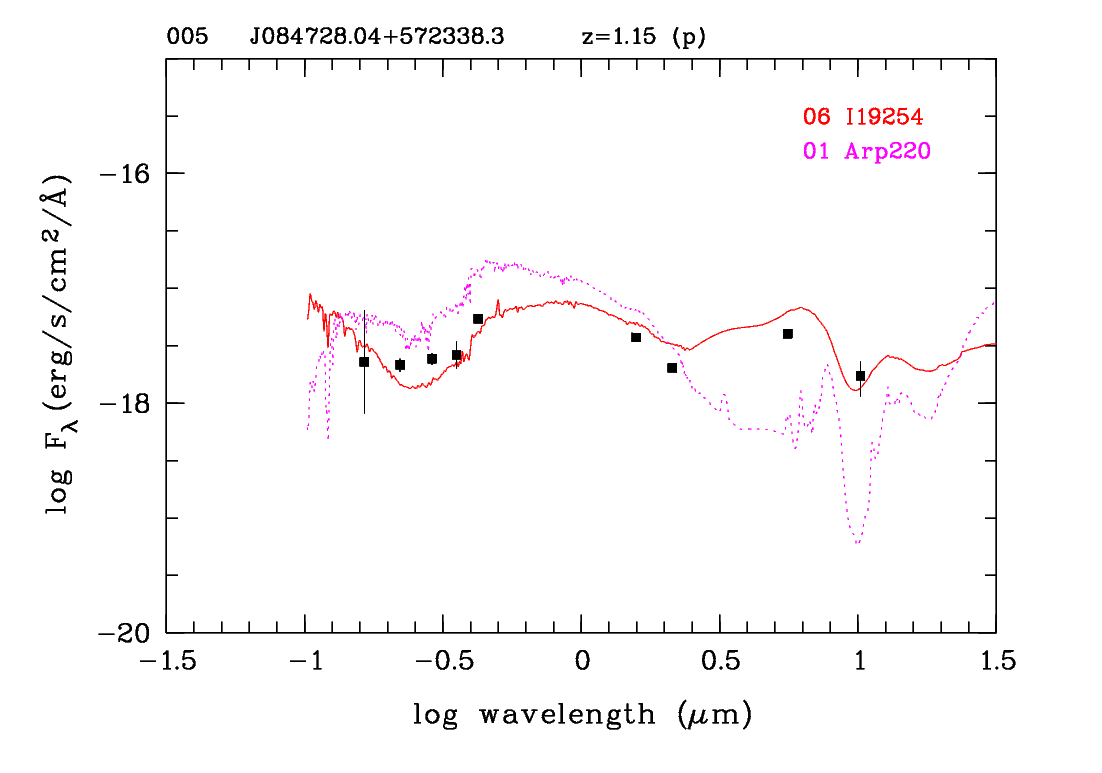}
\end{center}

\begin{center}
\includegraphics[width=4.5cm]{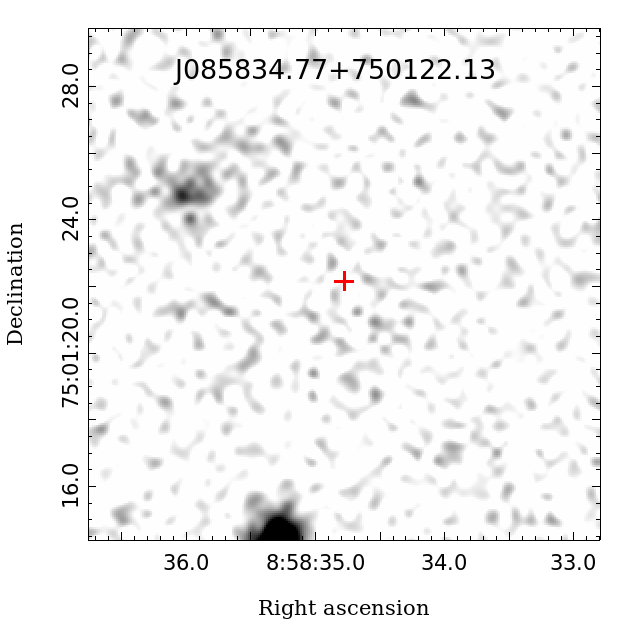}
\hspace{0.1cm}
\includegraphics[width=4.5cm]{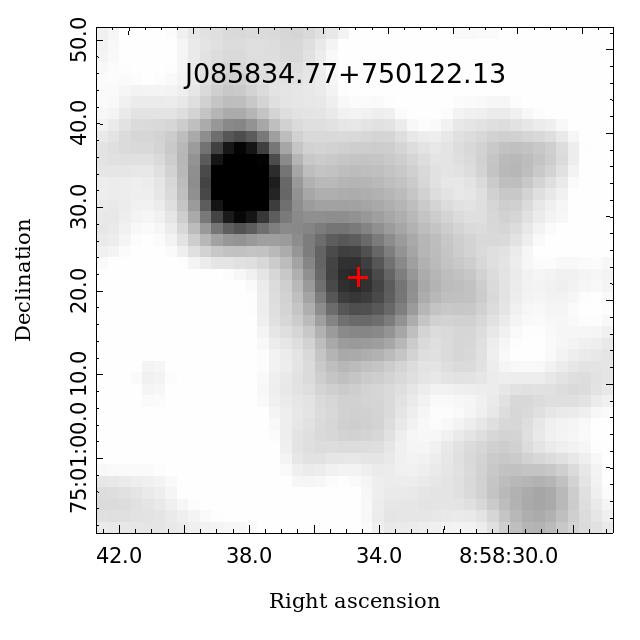}
\hspace{0.1cm}
\includegraphics[width=7.0cm]{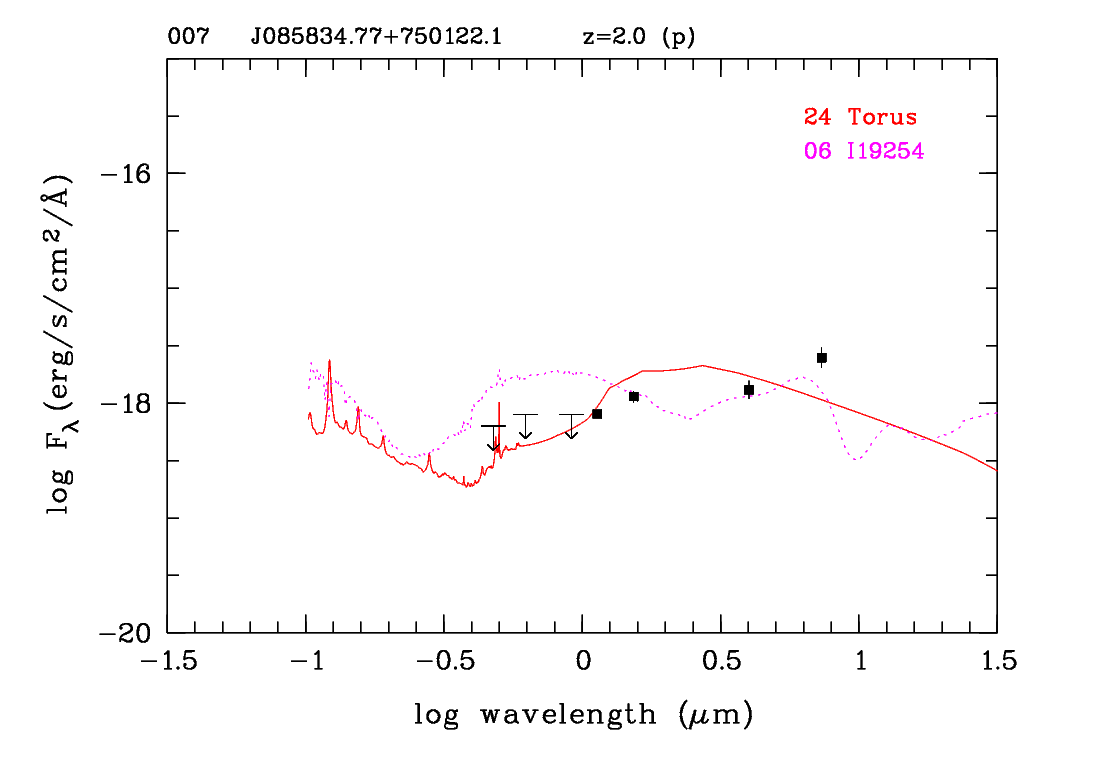}
\end{center}

\begin{center}
\includegraphics[width=4.5cm]{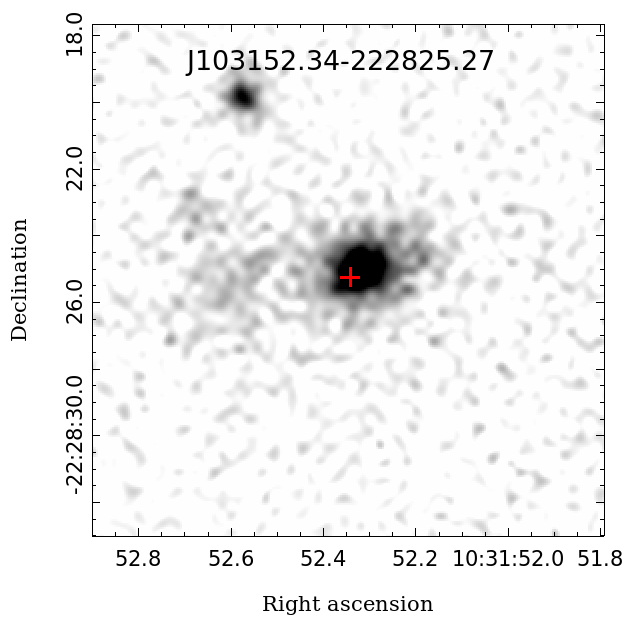}
\hspace{0.1cm}
\includegraphics[width=4.5cm]{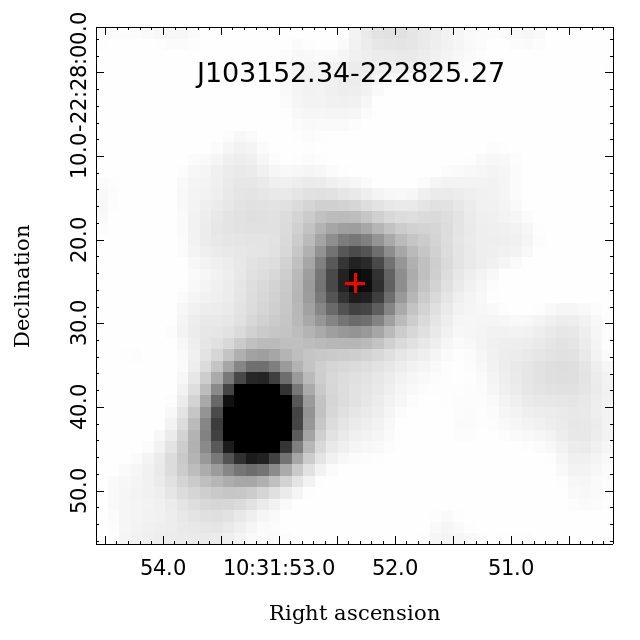}
\hspace{0.1cm}
\includegraphics[width=7.0cm]{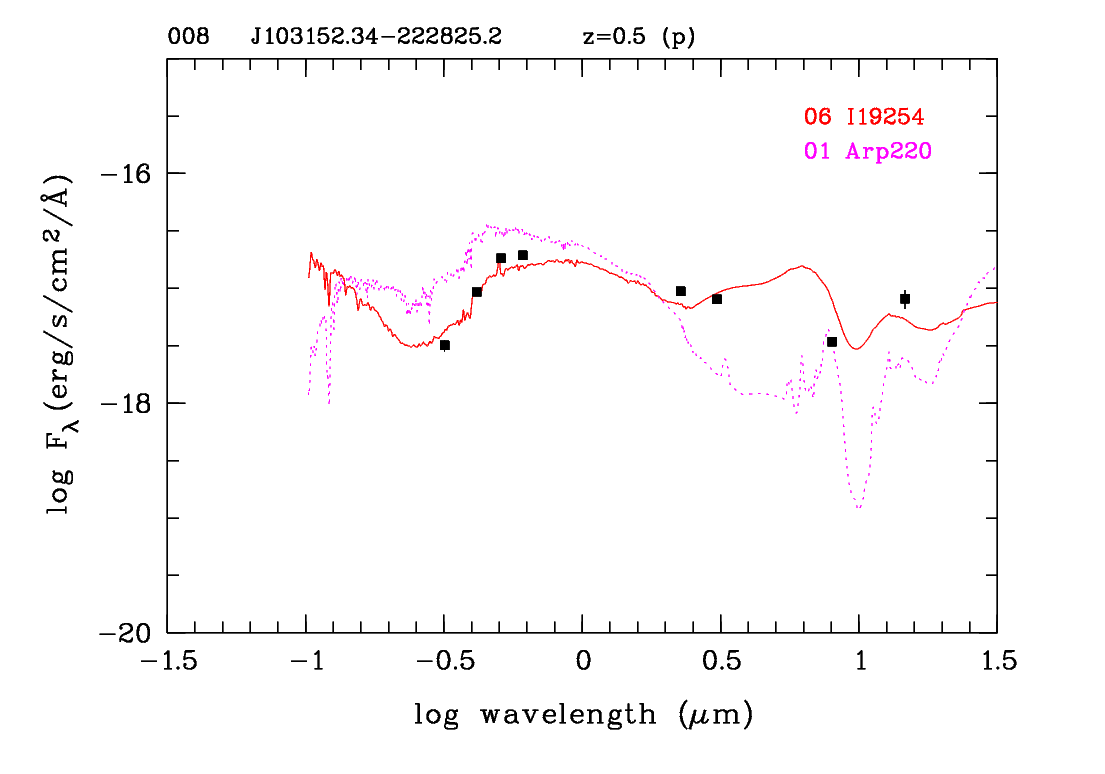}
\end{center}

\captionof{figure}{
EISERS measured in at least four photometric bands: Left and middle: co-added images from the LS (FOV 15 arcsec) and from WISE (FOV 60 arcsec); the position of the radio source is marked by the red plus sign. Right: WBSEDs in the rest frame with best-fitting template (red) and second-best fit (magenta, dotted).
}
\label{fig:MBSED}

\clearpage

\begin{center}
\includegraphics[width=4.5cm]{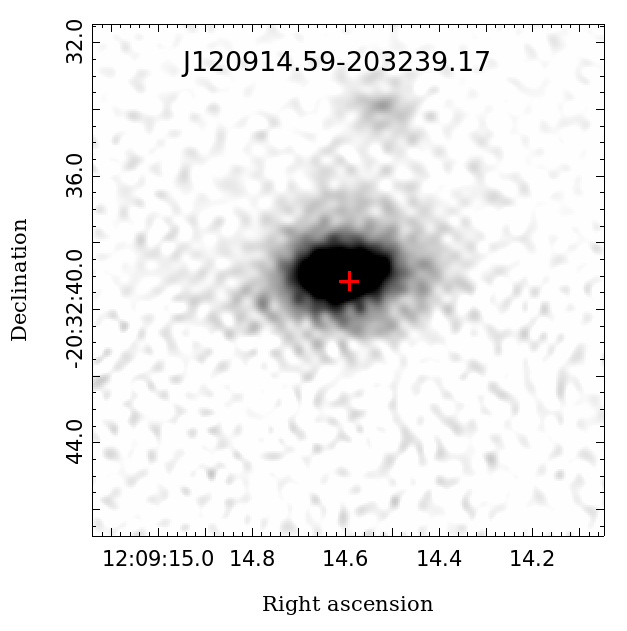}
\hspace{0.1cm}
\includegraphics[width=4.5cm]{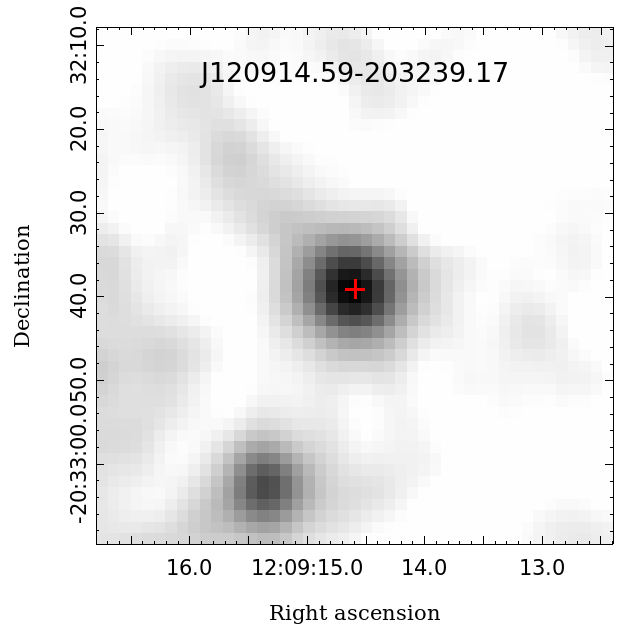}
\hspace{0.1cm}
\includegraphics[width=7.0cm]{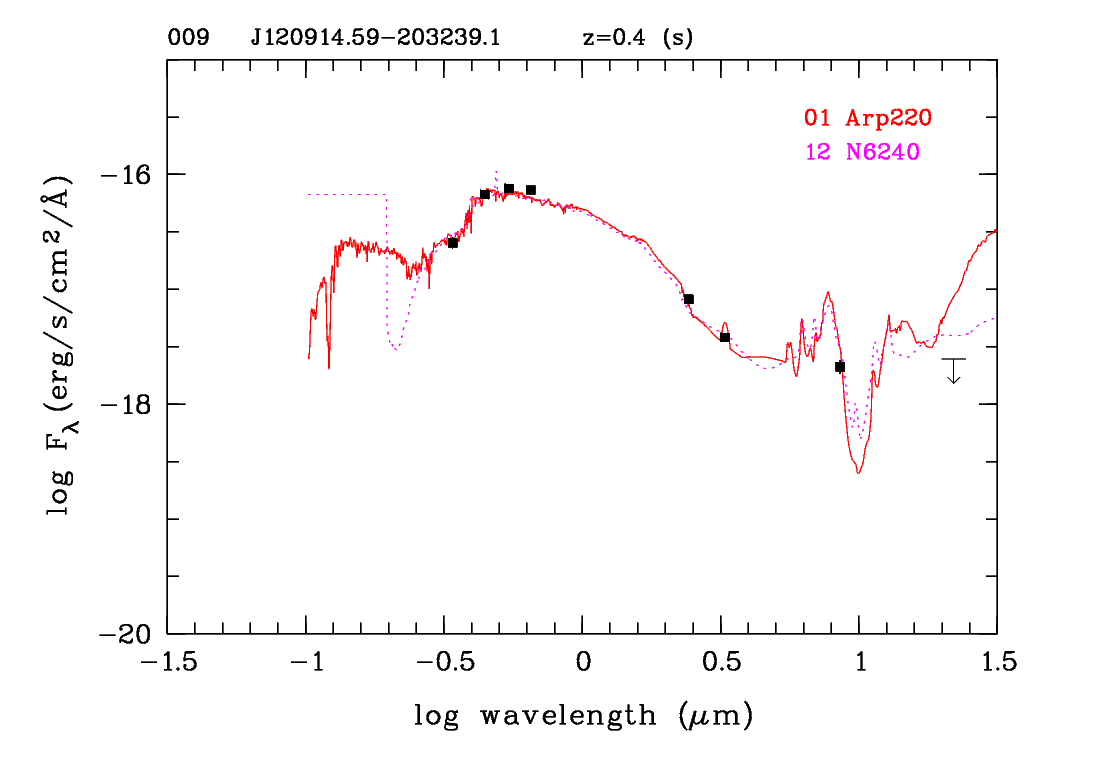}
\end{center}

\begin{center}
\includegraphics[width=4.5cm]{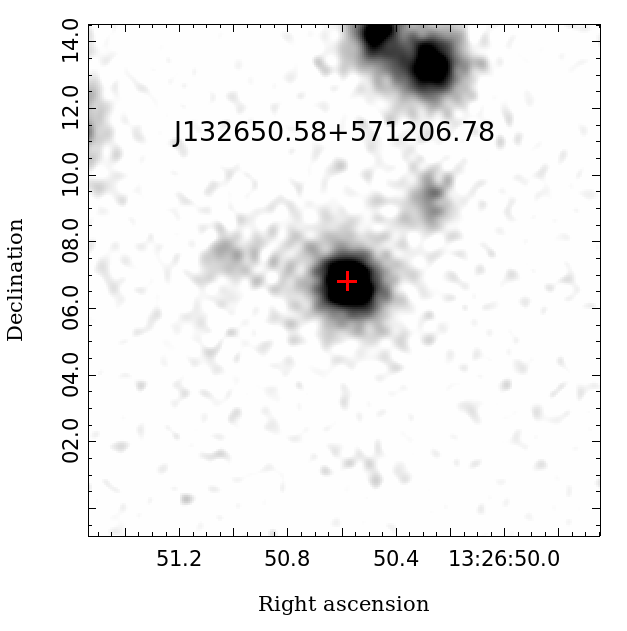}
\hspace{0.1cm}
\includegraphics[width=4.5cm]{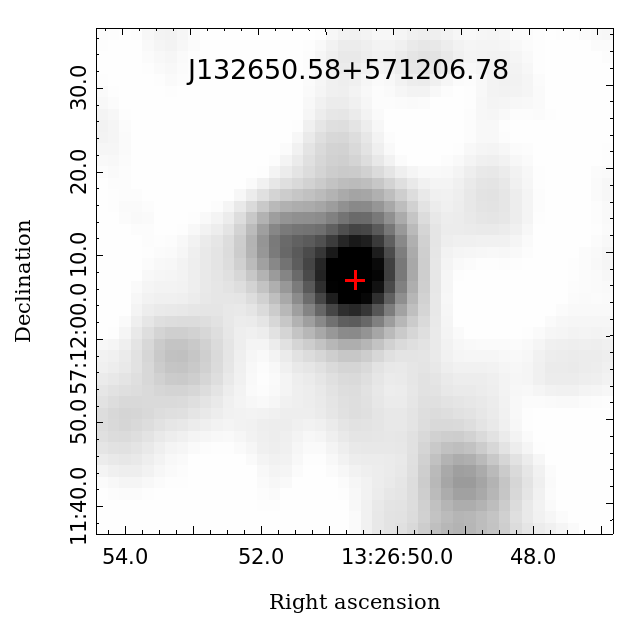}
\hspace{0.1cm}
\includegraphics[width=7.0cm]{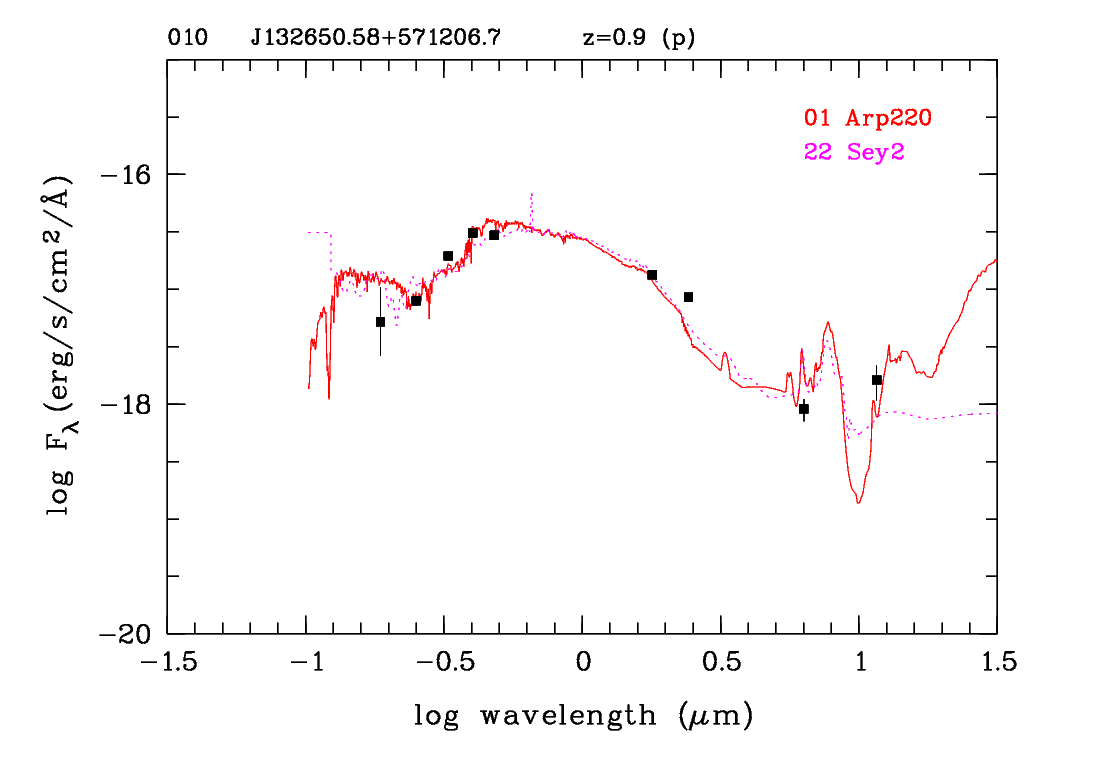}
\end{center}

\begin{center}
\includegraphics[width=4.5cm]{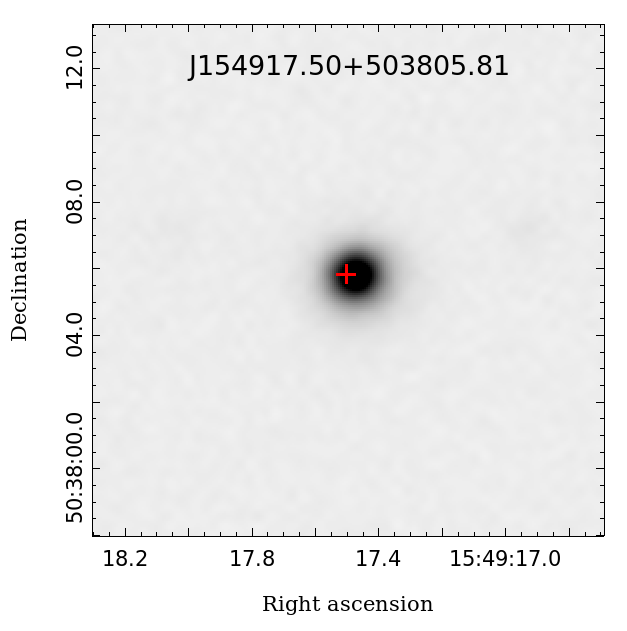}
\hspace{0.1cm}
\includegraphics[width=4.5cm]{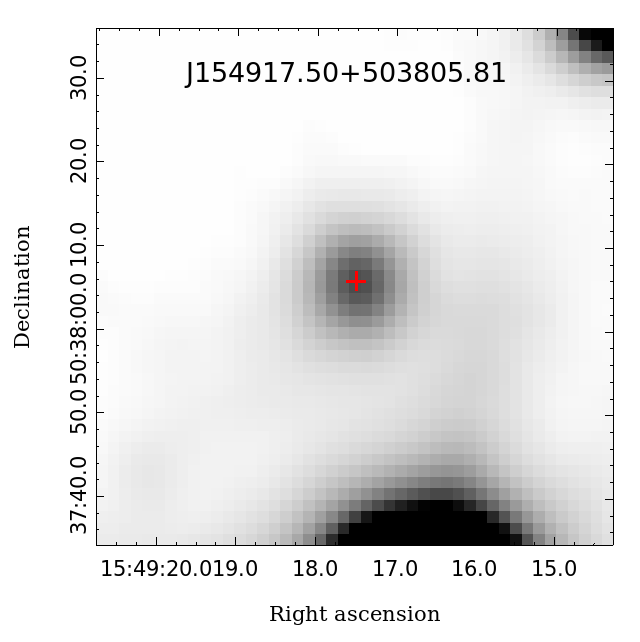}
\hspace{0.1cm}
\includegraphics[width=7.0cm]{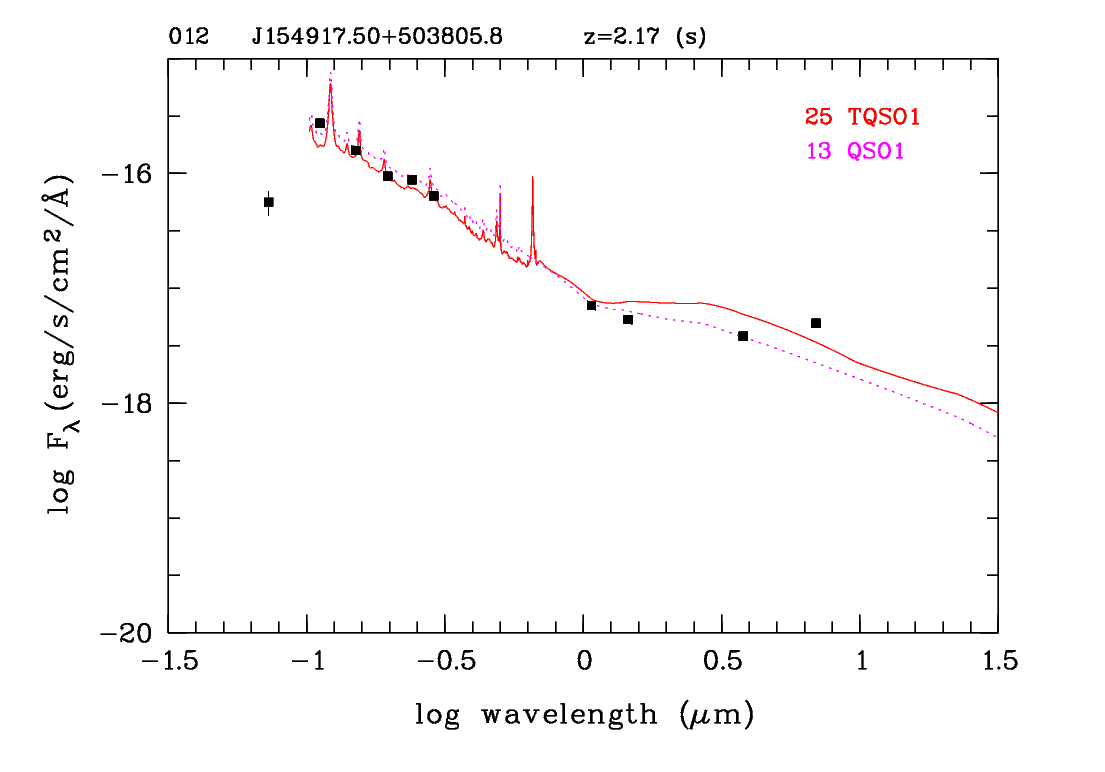}
\end{center}

\begin{center}
\includegraphics[width=4.5cm]{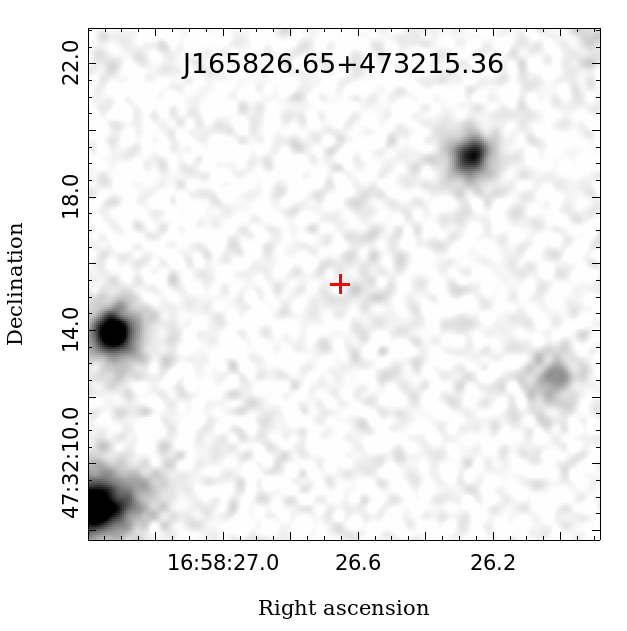}
\hspace{0.1cm}
\includegraphics[width=4.5cm]{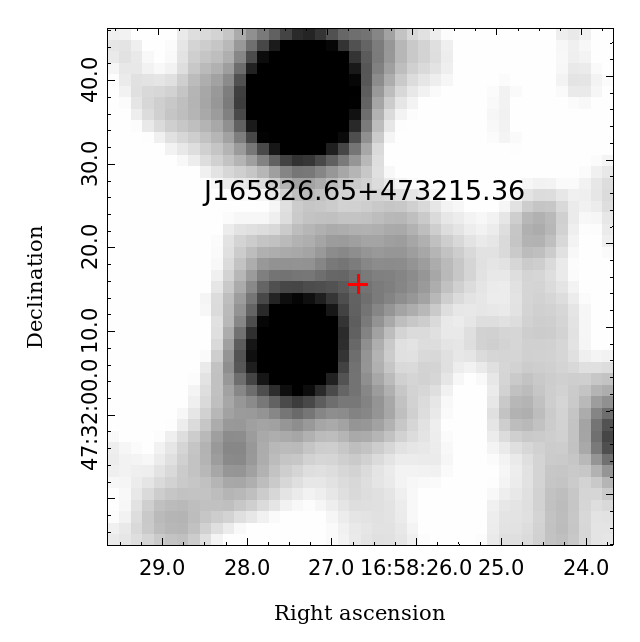}
\hspace{0.1cm}
\includegraphics[width=7.0cm]{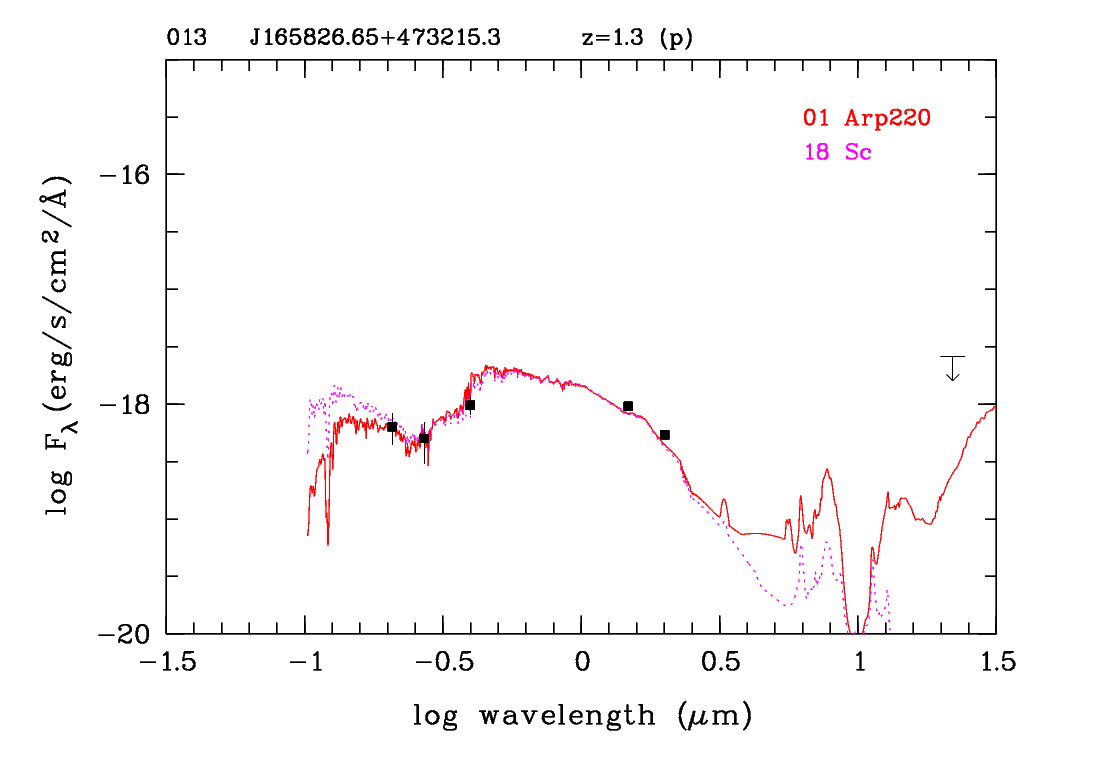}
\end{center}

\caption*{(Fig. D.1. continued)}
\clearpage

\begin{figure*}[htbp]
\includegraphics[viewport=0 -10 640 640,width=4.4cm,angle=0]{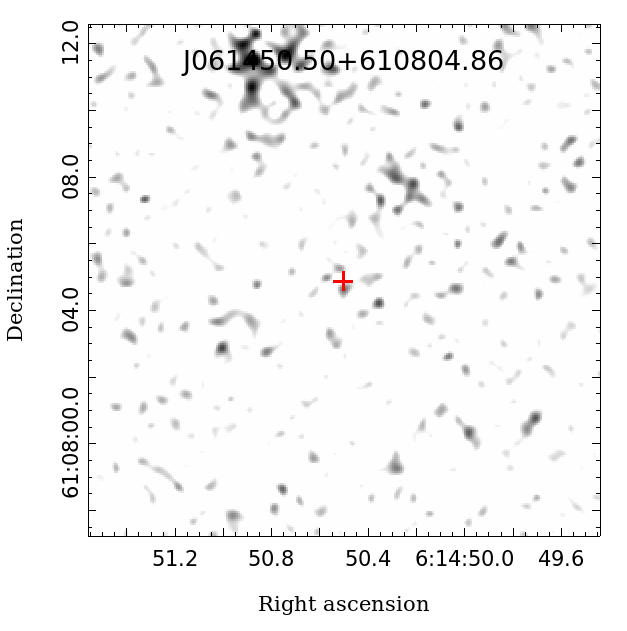}
\includegraphics[viewport=0 0 640 640,width=4.4cm,angle=0]{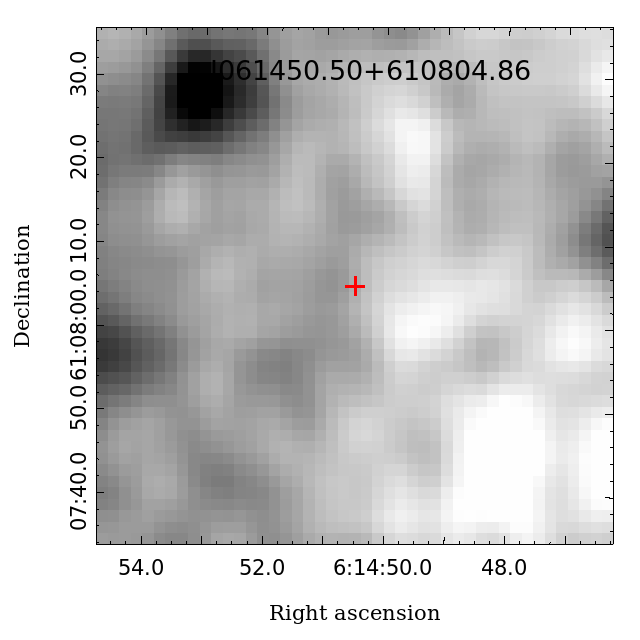}
\hspace{0.5cm}
\includegraphics[viewport=0 -10 640 640,width=4.4cm,angle=0]{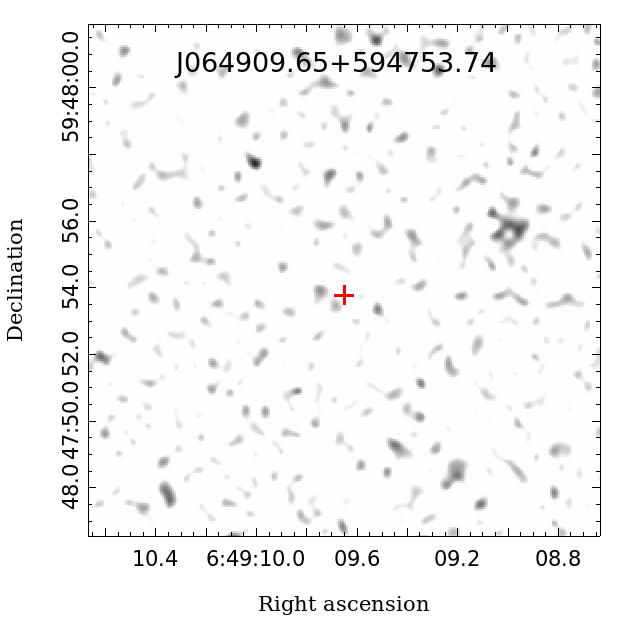}
\includegraphics[viewport=0 0 640 640,width=4.4cm,angle=0]{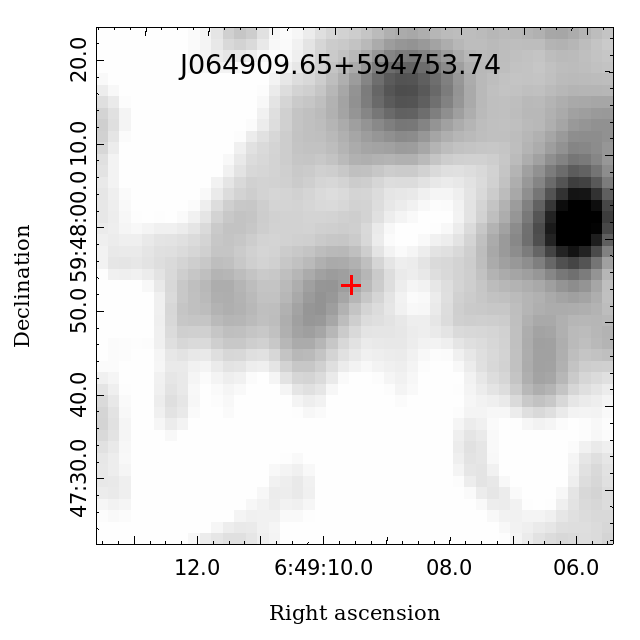}\\

\includegraphics[viewport=0 -10 640 640,width=4.4cm,angle=0]{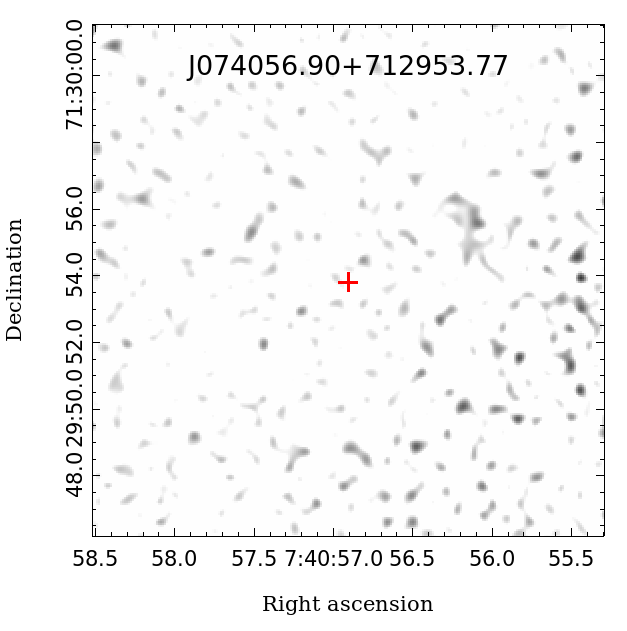}
\includegraphics[viewport=0 0 640 640,width=4.4cm,angle=0]{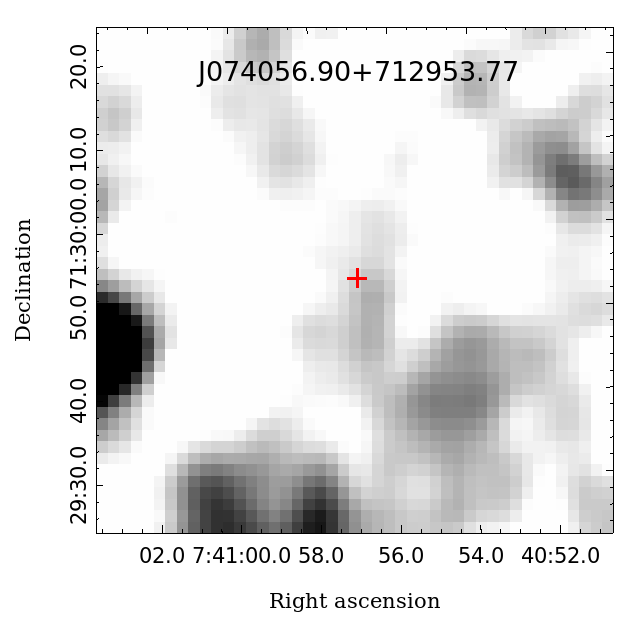}
\hspace{0.5cm}
\includegraphics[viewport=0 -10 640 640,width=4.4cm,angle=0]{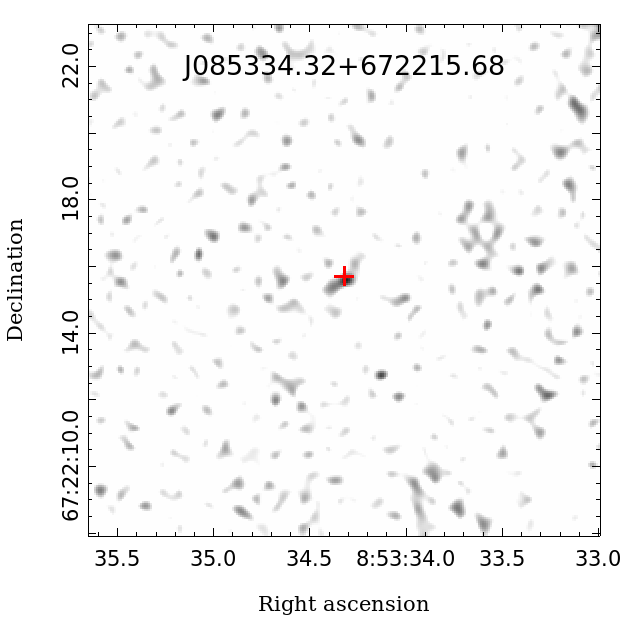}
\includegraphics[viewport=0 0 640 640,width=4.4cm,angle=0]{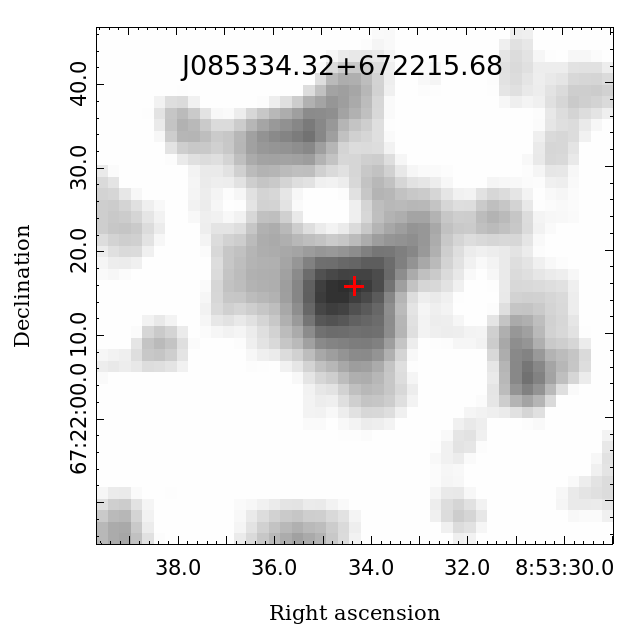}\\

\includegraphics[viewport=0 -10 640 640,width=4.4cm,angle=0]{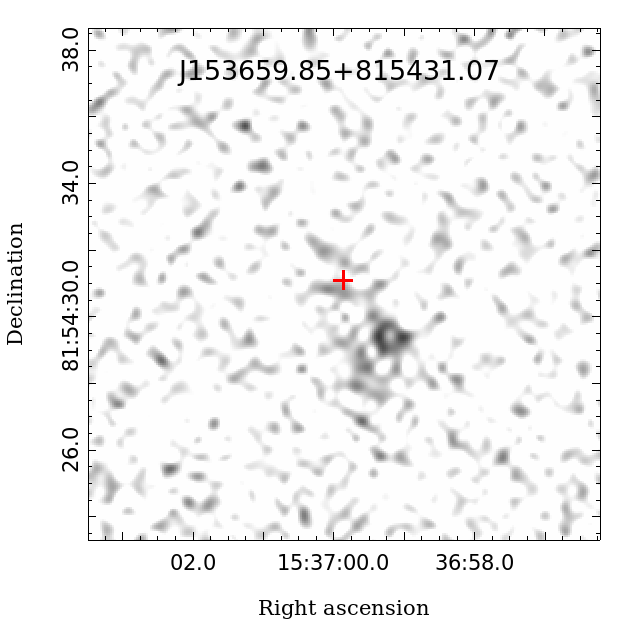}
\includegraphics[viewport=0 0 640 640,width=4.4cm,angle=0]{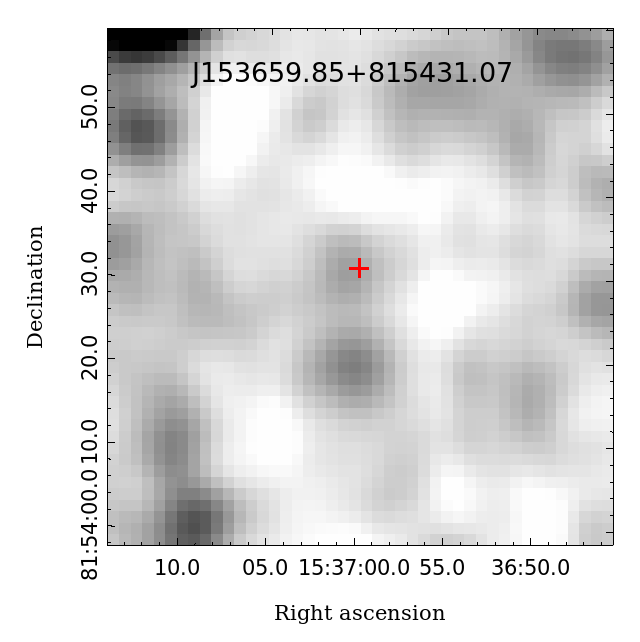}

\caption{Co-added optical images of the EISERS detected in fewer than four photometric bands, processed in the same manner as Fig.~\ref{fig:MBSED}.}
\label{fig:EISERS_wo_MBSED}
\end{figure*}

\clearpage

\end{appendix}

\end{document}